\documentclass[prd,twocolumn,superscriptaddress,amsfonts,amssymb,amsmath,nofootinbib]{revtex4-2}
\usepackage{graphicx} 
\usepackage{dcolumn}
\usepackage{bm} 
\usepackage{accents}
\usepackage{float} 
\usepackage{multirow}
\usepackage{makecell}
\usepackage{tikz}
\usepackage{comment}
\usepackage[rightcaption]{sidecap}
\sidecaptionvpos{figure}{t}
\usepackage{enumerate}
\usepackage{booktabs}

\renewcommand{\arraystretch}{2}
\newcolumntype{P}[1]{>{\centering\arraybackslash}p{#1}}
\newcolumntype{M}[1]{>{\centering\arraybackslash}m{#1}}

\usepackage{csquotes}
\usepackage{xcolor} 

\usepackage[colorlinks=true, linkcolor=blue, citecolor=blue, urlcolor=blue]{hyperref}

\usepackage{orcidlink}

\usepackage[capitalise]{cleveref}

\newcommand{\dd}{\mathrm{d}}

\begin{document}

\title{Do equation of state parametrizations of dark energy faithfully capture the dynamics of the late universe?}

\author{\"{O}zg\"{u}r Akarsu\,\orcidlink{0000-0001-6917-6176}}
\email{akarsuo@itu.edu.tr}
\affiliation{Department of Physics, Istanbul Technical University, Maslak 34469 Istanbul, T\"{u}rkiye}

\author{Maria Caruana\,\orcidlink{0000-0002-8989-0462}}
\email{caruanamaria@itu.edu.tr}
\affiliation{Department of Physics, Istanbul Technical University, Maslak 34469 Istanbul, T\"{u}rkiye}

\author{Konstantinos F. Dialektopoulos\,\orcidlink{0000-0002-0672-1496}}
\email{kdialekt@gmail.com}
\affiliation{Institute of Space Sciences and Astronomy, University of Malta, Malta, MSD 2080}

\author{Luis A. Escamilla\,\orcidlink{0000-0003-4334-5140}}
\email{torresl@itu.edu.tr}
\affiliation{Department of Physics, Istanbul Technical University, Maslak 34469 Istanbul, T\"{u}rkiye}

\author{Emre O. Kahya\,\orcidlink{0000-0003-2760-7091}}
\email{eokahya@itu.edu.tr}
\affiliation{Department of Physics, Istanbul Technical University, Maslak 34469 Istanbul, T\"{u}rkiye}

\author{Jackson Levi Said\,\orcidlink{0000-0002-7835-4365}}
\email{jackson.said@um.edu.mt}
\affiliation{Institute of Space Sciences and Astronomy, University of Malta, Malta, MSD 2080}
\affiliation{Department of Physics, University of Malta, Malta}


\begin{abstract}
We investigate how strongly late-time inferences about dark energy (DE) dynamics depend on the functional prior used to represent the expansion history. Using identical late-time combinations of cosmic chronometers, DESI baryon-acoustic-oscillation measurements, the Pantheon+ Type~Ia supernova sample, and, optionally, the Local Distance Network $H_0$ prior (H0DN), we compare a model-agnostic node-based reconstruction of the reduced Hubble function $E(z)\equiv H(z)/H_0$ with a representative family of smooth low-dimensional DE equation-of-state parametrizations, including $w$CDM, CPL, JBP, Barboza--Alcaniz, exponential, and logarithmic forms.
Over the redshift range directly constrained by the data, both approaches yield consistent $H(z)$ and $H(z)/(1+z)$, and, in the absence of external anchoring, compatible values of $H_0$.
However, a clear method dependence emerges at intermediate redshift, most clearly at $z\sim1.7$: the reconstruction favors stronger deceleration, $q_{\rm Rec}(1.7)\simeq 0.56$--0.61, whereas the smooth parametrizations cluster at $q(1.7)\simeq 0.32$--0.40, implying a persistent $\sim 2$--$3\sigma$ discrepancy across dataset combinations and parametrizations.
For the EoS-based parametrizations, whose effective DE densities remain positive by construction, the preferred $w_{\rm DE}(1.7)<-1$ values correspond to NECB-violating (phantom-like) behaviour, but this is a less robust discriminator because $w_{\rm DE}=p_{\rm DE}/\rho_{\rm DE}$ becomes ill-conditioned as $\rho_{\rm DE}\to0$.
In the effective-fluid mapping, the reconstruction accommodates the same late-time kinematical preference through a rapid descent of $\rho_{\rm DE}(z)$ toward very small values and, in some realizations, toward a sign change, whereas the EoS-based parametrizations absorb it through smoother, and in several cases NECB-violating, evolution over $z\sim1$--2.
Although the reconstruction improves the best-fit likelihood, especially when H0DN is imposed, Bayesian evidence continues to favor the simpler parametric descriptions.
Our results isolate $z\sim1.5$--2 as the key window in which EoS-based DE parametrizations can compress localized kinematic structure and associated features of DE that are still permitted by current late-time data.
\end{abstract}

\maketitle

\section{Introduction}
\label{sec:intro}

Late-time cosmic acceleration remains the central empirical clue that the dominant ingredients of the present-day Universe are still incompletely understood.
Within general relativity (GR), the minimal phenomenological description is the spatially flat $\Lambda$ cold dark matter ($\Lambda$CDM) model~\cite{Sahni:1999gb,Peebles:2002gy,Copeland:2006wr}, in which the recent expansion is driven by a strictly constant vacuum-energy density with equation of state (EoS) parameter $w=-1$.
Yet the success of $\Lambda$CDM has not resolved the underlying physical problem of vacuum energy~\cite{Weinberg:1988cp}, and the precision-cosmology era~\cite{Riess:2021jrx,Breuval:2024lsv,Planck:2018vyg,eBOSS:2020yzd,Scolnic:2021amr,Brout:2022vxf,Rubin:2023jdq,DESI:2024mwx,DESI:2025zgx,AtacamaCosmologyTelescope:2025blo,SPT-3G:2025bzu,KiDS:2020suj,DiValentino:2020vvd,DES:2021vln,Wright:2025xka,DES:2025sig} has transformed a qualitative puzzle into a quantitative one: does the late-time expansion history contain structure beyond a featureless cosmological constant?
This question is sharpened not only by the persistent mismatch between local and early-Universe determinations of the Hubble constant, together with milder discrepancies in clustering- and lensing-derived quantities, but also by the emergence of late-time distance datasets with sufficient precision and redshift leverage to probe nontrivial dynamics directly~\cite{Verde:2019ivm,DiValentino:2020zio,DiValentino:2021izs,Perivolaropoulos:2021jda,Schoneberg:2021qvd,Abdalla:2022yfr,DiValentino:2022fjm,Vagnozzi:2023nrq,Khalife:2023qbu,Akarsu:2024qiq,CosmoVerseNetwork:2025alb}.

The present situation is especially timely in the DESI era.
Baryon acoustic oscillation (BAO)  measurements from DESI now trace the late-time expansion history over a broad redshift interval, extending to $z\gtrsim2$, and do so with a precision that makes the question of evolving dark energy (DE) more than merely speculative~\cite{DESI:2024mwx,DESI:2025zgx}.
Because BAO separately probe transverse and line-of-sight distances through $D_M/r_{\rm d}$ and $D_H/r_{\rm d}$, they provide exactly the kind of geometric leverage needed to expose localized structure in the expansion history at intermediate redshift.
The official DESI cosmological analyses show that, although the BAO measurements themselves remain broadly compatible with flat $\Lambda$CDM, combinations with CMB and supernova information can favor a time-dependent DE EoS  parameter, with standard CPL-based $w_0w_a$CDM analyses yielding departures from a cosmological constant at more than $3\sigma$ significance in some cases~\cite{DESI:2024mwx,DESI:2025zgx,DES:2025sig,Giare:2024gpk,Giare:2024oil,Giare:2025pzu,Fazzari:2025lzd,Xu:2026sbw,Pantos:2026rpe}.
\textit{This development raises a methodological question that is as important as the phenomenological one: when late-time data appear to favor dynamics beyond $\Lambda$CDM, are those dynamics properties of the data, or of the functional manifold used to represent them?} 

A conventional way to address this question is to parametrize the DE EoS parameter $w_{\rm DE}(z)$ and confront it with observations.
The Chevallier--Polarski--Linder (CPL) form~\cite{Chevallier:2000qy,Linder:2002et}, ${w(a)=w_0+w_a(1-a)}$, has become the canonical benchmark because it is low-dimensional, smooth, and recovers $\Lambda$CDM in the limit ${(w_0,w_a)=(-1,0)}$.
More generally, however, CPL belongs to a broader class of smooth low-dimensional DE EoS parametrizations that compress possible late-time dynamics into one or two additional degrees of freedom.
Their practical value is obvious, but so is the prior structure they impose.
In the usual spatially flat FLRW framework with separately conserved cosmic components, specifying $w_{\rm DE}(z)$ determines $\rho_{\rm DE}(z)$ through an exponential integral and therefore builds in a sign-preserving, globally smooth effective DE density.
Such parametrizations are naturally suited to describing controlled deformations around the standard $\Lambda$CDM scenario, but not necessarily to charting the full space of late-time effective dynamics.
If the actual expansion history contains localized intermediate-redshift structure, or if the effective DE density becomes very small and tends toward a sign change, then a smooth $w_{\rm DE}(z)$-based description may still fit the data successfully while redistributing that structure into a smoother and more diffuse form.
\textit{The essential issue, therefore, is not simply whether smooth low-dimensional EoS parametrizations fit the data, or even improve the fit relative to the usual cosmological-constant description, but whether they faithfully represent the kinematics of the Universe and the dynamics of DE that the data permit.} 

This limitation is not merely formal.
Explicit counterexamples are provided, for instance, by DE models with sign-switching energy density, such as $\Lambda_{\rm s}$CDM and related AdS-to-dS transition scenarios, as well as by effective sectors arising in modified-gravity constructions, in which the late-time background evolution can remain perfectly regular at the level of $H(z)$, $\rho_{\rm DE}(z)$, and $p_{\rm DE}(z)$ while the ratio $w_{\rm DE}(z)=p_{\rm DE}(z)/\rho_{\rm DE}(z)$ becomes singular when $\rho_{\rm DE}$ crosses zero~\cite{Akarsu:2019hmw,Akarsu:2021fol,Akarsu:2022typ,Akarsu:2023mfb,Akarsu:2025gwi,Akarsu:2026anp,Gokcen:2026pkq}.
These models have also emerged as economical late-time extensions of $\Lambda$CDM proposed to address multiple cosmological tensions, including those in $H_0$, $M_B$, $S_8$, and growth-related observables.
Their physical logic is transparent: a negative effective DE density contribution at intermediate redshift can suppress $H(z)$ relative to $\Lambda$CDM, so consistency with the CMB distance to last scattering requires compensating enhancement at lower redshift, naturally shifting the inferred $H_0$ upward.
In DE scenarios in which a $\Lambda$-like component at low redshift either flips sign and becomes negative, instantaneously or rapidly (e.g., $\Lambda_{\rm s}$CDM~\cite{Akarsu:2019hmw,Akarsu:2021fol,Akarsu:2022typ,Akarsu:2023mfb}), or vanishes (e.g., emergent DE~\cite{Li:2019yem,DeFelice:2020cpt,Yang:2021eud}) at or around a critical redshift $z=z_\dagger$, the EoS parameter can remain approximately $w_{\rm DE}\approx -1$ on both sides of the transition, even though $\rho_{\rm DE}$ undergoes its essential dynamical event, namely sign reversal or emergence.
Thus, the relevant dynamics resides in the localized evolution of the density itself, rather than in any extended evolution of the ratio variable. This is the case, for example, in abrupt $\Lambda_{\rm s}$CDM~\cite{Akarsu:2019hmw,Akarsu:2021fol,Akarsu:2022typ,Akarsu:2023mfb}, in the string-inspired $\Lambda_{\rm s}$CDM$^+$ construction~\cite{Anchordoqui:2023woo,Anchordoqui:2024gfa,Anchordoqui:2024dqc,Soriano:2025gxd}, in the type-II minimally modified-gravity realization $\Lambda_{\rm s}$VCDM~\cite{Akarsu:2024qsi,Akarsu:2024eoo}, in the overall-metric-sign-change construction of Ref.~\cite{Alexandre:2023nmh}, and in the composite-DE extension $w$XCDM~\cite{Gomez-Valent:2024tdb,Gomez-Valent:2024ejh}, which generalizes the abrupt $\Lambda_{\rm s}$CDM by allowing different constant EoS parameter values for DE before and after the transition in $\rho_{\rm DE}$.
In smooth realizations, by contrast, $\rho_{\rm DE}$ can cross zero continuously, so that the associated pole in $w_{\rm DE}$ is purely kinematic, arising because $p_{\rm DE}$ remains finite while $\rho_{\rm DE}$ passes through zero, rather than from any singularity in the underlying background expansion~\cite{Akarsu:2025gwi,Akarsu:2026anp,Gokcen:2026pkq}.
Examples include $\Lambda_{\rm s}$VCDM~\cite{Akarsu:2024qsi,Akarsu:2024eoo}, the phantom-scalar realization Ph-$\Lambda_{\rm s}$CDM~\cite{Akarsu:2025gwi,Akarsu:2025dmj,Adil:2026kfn}, the teleparallel realization $f(T)$-$\Lambda_{\rm s}$CDM~\cite{Akarsu:2024nas,Souza:2024qwd}, and other phenomenological smooth-transition models~\cite{Bouhmadi-Lopez:2025ggl,Bouhmadi-Lopez:2025spo,Ibarra-Uriondo:2026zbp}.
If the transition is sufficiently fast, this nontrivial behaviour is compressed into a narrow redshift layer around $z_\dagger$, outside of which $w_{\rm DE}$ can remain observationally almost indistinguishable from $-1$ even though the background differs nontrivially from $\Lambda$CDM.
Sign-switching scenarios are therefore of particular interest not only because they furnish explicit counterexamples to globally regular $w_{\rm DE}(z)$-based descriptions, but also because phenomenological analyses have repeatedly identified a transition regime around $z_\dagger\sim1.5\text{--}2.5$~\cite{Akarsu:2019hmw,Akarsu:2021fol,Akarsu:2022typ,Akarsu:2023mfb,Anchordoqui:2024gfa,Soriano:2025gxd,Akarsu:2024eoo,Gomez-Valent:2024tdb,Gomez-Valent:2024ejh,Akarsu:2025gwi,Ibarra-Uriondo:2026zbp}.

It was also shown recently in Ref.~\cite{Gokcen:2026pkq} that extending CPL so as to admit negative DE density values at higher redshift tends to reduce the apparent low-redshift departure from a positive cosmological constant.
Moreover, it was shown in~\cite{Akarsu:2022lhx,Escamilla:2024ahl} that, for a given Hubble function---for instance that of $\Lambda$CDM---and still in agreement with CMB measurements of the angular scale of the sound horizon at last scattering, any deviation in the Hubble radius that preserves this agreement must take the form of an admissible wavelet, unless the value of $H_0$ and/or the comoving sound horizon is also modified.
Thus, if these deviations are confined to the late universe, where DE is significant, one generically expects oscillatory behaviour in the DE density, which can also lead to singularities in the EoS parameter if the oscillations temporarily drive the DE density below zero.
Indeed, Ref.~\cite{Escamilla:2024ahl} showed, using datasets including CMB, DESI BAO, and SNIa data, that wavelet-like deviations in the late universe are favored by the BAO measurements, with a significant improvement in the fit at more than $3\sigma$ once the DESI BAO data are included.
Such examples make plain that the standard smooth low-dimensional EoS-based DE parametrizations, precisely because they are constructed as controlled and globally regular deformations of $\Lambda$CDM, are not naturally suited to encoding this class of late-time dynamics as a primary feature.

A complementary strategy is therefore to develop model-agnostic methods for studying the dynamics of DE.
For a broad, though by no means exhaustive, sample of works employing model-agnostic methods to investigate DE and/or the late-time expansion dynamics of the Universe, see Refs.~\cite{Holsclaw:2011wi,AlbertoVazquez:2012ofj,Seikel:2012uu,Keeley:2020aym,Bonilla:2020wbn,Jesus:2021bxq,Bernardo:2021cxi,Escamilla:2021uoj,Elizalde:2022rss,Escamilla:2023shf,Dinda:2024ktd,Gadbail:2024lek,Yang:2025kgc,DESI:2025fii,You:2025uon,Mukherjee:2025ytj,Ye:2024ywg,Wang:2025vfb,Ormondroyd:2025exu,Berti:2025phi,Gonzalez-Fuentes:2025lei,Shafieloo:2007cs,Wang:2018fng,Mitra:2024ahj,Tan:2025xas,Huterer:2002hy,Crittenden:2005wj,Albrecht:2006mqt,Zhao:2012aw,Liu:2015mkm,Zhao:2017cud,Raveri:2017qvt,Dai:2018zwv,Gomez-Valent:2021cbe,Liu:2026rot,Sousa-Neto:2025gpj,DESI:2024aqx,DESI:2025wyn}. In addition to its investigations of the standard EoS-based DE parametrizations, namely the $w$CDM and $w_0w_a$CDM models~\cite{DESI:2024mwx}, the DESI collaboration has also considered a more flexible Chebyshev expansion of the DE EoS parameter up to fourth order~\cite{DESI:2024aqx}.
Using the DESI BAO data together with different compilations of Type~Ia supernova (SNIa) data, that analysis found that the effective DE density can become negligible for $z\gtrsim1$ and, for some dataset combinations, even negative by $z\gtrsim1.5$--2~\cite{DESI:2024aqx}, a trend already indicated by pre-DESI BAO analyses, including those based on SDSS BAO data~\cite{Escamilla:2024ahl,Escamilla:2023shf,Sabogal:2024qxs}.
One way to investigate such behaviour with minimal prior structure is to reconstruct the expansion history itself.
Rather than imposing a specific form for $w_{\rm DE}(z)$, one can reconstruct the reduced Hubble function $E(z)\equiv H(z)/H_0$ directly from late-time observations and only afterwards map the inferred kinematics to effective-fluid quantities under GR.
This separates, as cleanly as possible, the information supplied by the data from the structure imposed by a chosen parametric manifold.
In our companion reconstruction study~\cite{Akarsu:2026anp}, we implemented a node-based inference of $E(z)$ in which a Gaussian-process kernel acts as a smooth interpolant between fixed redshift nodes whose amplitudes are sampled in a Bayesian analysis.
With present late-time data extending to $z\simeq2$--2.3, that reconstruction showed that directly constrained quantities such as $H(z)$ can remain relatively stable while derivative- and mapping-sensitive quantities exhibit richer intermediate-redshift behaviour.
In particular, the GR-mapped effective DE density was found to descend rapidly toward very small values and, in some dataset combinations, toward a sign change, while the same reconstruction also admitted hints of a transient extra-acceleration episode around $z\sim2$~\cite{Akarsu:2026anp}.
When such a sign change was present, the associated transition redshift $z_\dagger$ was found to be anticorrelated with $H_0$: a lower-$z_\dagger$, higher-$H_0$ branch emerged most clearly in several combinations without SN but with an external $H_0$ prior, with $H_0$ approaching $\sim73\ {\rm km\,s^{-1}\,Mpc^{-1}}$ for $z_\dagger\sim1.7$, whereas adding SN generally weakened this degeneracy and shifted the inferred transition to somewhat higher redshift~\cite{Akarsu:2026anp}.
Taken together, these results suggest that the real methodological issue may not be whether evolving DE exists in some smooth $w_{\rm DE}(z)$ form, but whether standard EoS-based parametrizations of DE project a richer late-time kinematics onto a restricted and globally regular language, thereby obscuring genuine dynamical features of DE.

The present paper is a focused companion to that reconstruction analysis carried out in Ref.~\cite{Akarsu:2026anp}. Rather than re-establishing the reconstruction itself, our aim is to determine what becomes of the same late-time kinematics when they are projected onto standard smooth low-dimensional parametrizations of the DE EoS parameter.
To this end, we perform a controlled head-to-head comparison between the node-based reconstruction of $E(z)$ and several widely used EoS-based DE parametrizations, namely the constant-$w_{\rm DE}$ case ($w$CDM), CPL~\cite{Chevallier:2000qy,Linder:2002et}, JBP~\cite{Jassal:2004ej,Jassal:2005qc}, Barboza--Alcaniz~\cite{Barboza:2008rh}, exponential~\cite{Pan:2019brc,Najafi:2024qzm}, and logarithmic forms~\cite{Tripathi:2016slv}.
In all cases, we use identical late-time dataset combinations built from cosmic chronometers (CC)~\cite{Jimenez:2003iv,Simon:2004tf,Moresco:2012by,Zhang:2012mp,Moresco:2015cya,Moresco:2016mzx,Ratsimbazafy:2017vga}, DESI BAO measurements~\cite{DESI:2025zgx,DESI:2025qqy,DESI:2025fii}, and the Pantheon+ SNIa sample~\cite{Scolnic:2021amr,Brout:2022vxf}, with and without an external $H_0$ prior from the Local Distance Network (H0DN)~\cite{H0DN:2025lyy}.
We deliberately restrict the analysis to the late-time background, adopt the DESI convention of calibrating the BAO ruler through BBN~\cite{DESI:2025zgx}, and do not impose a CMB likelihood, in order to isolate as cleanly as possible the role of late-time functional priors.

The aim of the present work is therefore as much methodological as it is phenomenological.
We ask where smooth low-dimensional EoS-based DE parametrizations and the reconstruction agree, where they diverge, and which derived quantities provide the cleanest diagnostics of that divergence.
In practice, this means distinguishing directly constrained kinematical quantities, such as $H(z)$, from derivative-sensitive ones, especially the deceleration history $q(z)$, and then examining how any differences propagate into the GR-mapped effective-fluid variables $\rho_{\rm DE}(z)$, $p_{\rm DE}(z)$, and $w_{\rm DE}(z)$.
A central complication is that $w_{\rm DE}(z)$ becomes ill-conditioned whenever $\rho_{\rm DE}(z)$ approaches zero, so large excursions in $w_{\rm DE}(z)$ need not correspond to equally dramatic structure in the underlying expansion history~\cite{Akarsu:2025gwi,Akarsu:2026anp,Gokcen:2026pkq}.
This is not merely a formal caveat, but a behaviour explicitly realized in $\Lambda_{\rm s}$CDM and similar models.
More broadly, one of the key questions behind the present work is whether the phantom-like behaviour frequently inferred, within a positive-density framework, from smooth low-dimensional parametrizations of the DE EoS parameter at relatively high redshift within the data range should be interpreted as a genuine feature of DE and evidence for exotic DE microphysics, or instead as a compensating projection of more localized late-time kinematics onto a restricted functional family.
Finally, by reporting both $\Delta\chi^2_{\min}$ and Bayesian-evidence ratios, we distinguish improvements in best fit from dynamics that are genuinely required by the data once model complexity is taken into account.

The paper is organized as follows.
In~\cref{sec:methodology} we summarize the datasets and inference pipeline, and define both the reconstruction setup and the parametric DE models used throughout.
In~\cref{sec:discussion} we present the constraints on $H(z)$, $H(z)/(1+z)$, and $q(z)$, together with the corresponding GR-mapped effective-fluid quantities $\rho_{\rm DE}(z)$, $p_{\rm DE}(z)$, and $w_{\rm DE}(z)$, as well as the associated scalar-sector diagnostics $\Delta\mathcal{X}(z)$ and $V(z)$.
We then interpret these results in terms of the information content of current late-time data and the role of functional priors, with particular emphasis on the intermediate-redshift regime where smooth low-dimensional EoS parametrizations can act as compressive projections of more localized kinematic structure.
We conclude in~\cref{sec:conc} with the implications for using smooth EoS models as effective descriptions in the era of precision BAO and deep Type~Ia supernova samples.

\section{Methodology}\label{sec:methodology}

Our analysis compares two classes of late-time background description under identical likelihoods: a non-parametric node-based reconstruction of the reduced Hubble function $E(z)\equiv H(z)/H_0$ and a family of smooth low-dimensional DE EoS parametrizations. From each inferred expansion history, we derive the kinematical diagnostics the Hubble parameter $H(z)$, the conformal Hubble parameter $H(z)/(1+z)=aH=\dot a$, and the deceleration parameter $q(z)$, shown in~\cref{fig:H_and_q}. Assuming GR at the background level, we then map the same histories to effective-fluid variables, namely the DE energy density, pressure, and EoS parameter, $\rho_{\rm DE}(z)$, $p_{\rm DE}(z)$, and $w_{\rm DE}(z)$, as well as the associated scalar-sector diagnostics, namely the effective kinetic contribution $\Delta\mathcal{X}(z)$ and the potential $V(z)$, shown in~\cref{fig:rhode_pde_wde,fig:Keff_and_V}. The scalar-sector quantities are derived \emph{a posteriori} from the inferred background and do not enter the likelihood directly. Their effective two-field interpretation is summarized in~\cref{app:two_scalar}.

Following the companion analysis of Ref.~\cite{Akarsu:2026anp}, the reconstruction is performed at the level of $E(z)$ rather than $w_{\rm DE}(z)$. The new element of the present work is not the reconstruction pipeline itself, but the parallel Bayesian inference for the flat $\Lambda$CDM baseline and for a family of smooth low-dimensional EoS parametrizations of DE that extend the standard $\Lambda$CDM dark-energy sector, in which DE is represented by a positive-density cosmological constant with $w=-1$. These include the constant-$w_{\rm DE}$ case ($w$CDM) and the CPL, JBP, Barboza--Alcaniz, exponential, and logarithmic forms. This setup enables a controlled comparison between a flexible non-parametric description and a broad class of low-dimensional EoS models, with $w$CDM providing the one-parameter extension of the $\Lambda$CDM DE sector and the remaining cases providing standard two-parameter extensions. We summarize below the reconstruction setup, the parametric models, and the datasets and priors adopted throughout.

\subsection{Model-agnostic reconstruction}
\label{subsec:methodology}

Our reconstruction targets the dimensionless expansion history $E(z)\equiv H(z)/H_0$, thereby providing a data-driven description of the late-time expansion that can reveal dynamics preferred by the observations.

The reconstruction makes use of Gaussian Processes (GP), but not in the usual regression sense~\cite{Velazquez:2024aya}. Rather, the GP kernel is employed as a smooth interpolant between a set of fixed redshift nodes whose amplitudes are treated as free parameters in a Bayesian inference. Node-based GP interpolation has been used in the literature to explore departures from the $\Lambda$CDM paradigm; for example, it has been applied to reconstructions of the dark energy equation of state and of dark-sector interactions~\cite{Gerardi:2019obr,Escamilla:2023shf}. We adopt the squared-exponential (RBF) kernel
\begin{equation}
K(z,z')=\exp\!\big[-\theta(z-z')^2\big],
\end{equation}
where the hyperparameter $\theta$ controls the correlation scale. In our implementation we keep $\theta$ fixed, as in Ref.~\cite{Akarsu:2026anp}, so that the reconstruction remains a controlled finite-dimensional inference rather than a mixed node--hyperparameter fit. This avoids additional degeneracies while retaining an infinitely differentiable interpolant, which is especially useful when reconstructing derivative-sensitive quantities such as $H'(z)$ and $q(z)$.

The free parameters of the reconstruction are the expansion amplitudes $E_i\equiv E(z_i)$ at fixed node positions $z_i$, together with the present-day Hubble constant $H_0$. We adopt five nodes located at $z_i=\{0.6,\,1.2,\,1.8,\,2.4,\,3.0\}$, while the normalization at $z=0$ is provided by $H_0$. Thus, the reconstruction is described by the six-dimensional parameter set $\{H_0,E_1,\dots,E_5\}$. Because the available data do not populate the interval $2.4<z<3.0$, the node at $z=3.0$ acts primarily as an extrapolation anchor. Consequently, reconstructed behaviour in this high-redshift interval, particularly for derivative-based quantities, should be interpreted with caution.

A practical advantage of this setup, compared with applying GP regression directly to the observables, is that derived quantities involving derivatives of the expansion history can be obtained straightforwardly by differentiating the interpolant for each posterior sample. In this way, uncertainties in $E'(z)$, $H'(z)$, and $q(z)$ are propagated directly through the sampling procedure rather than reconstructed in a separate post-processing step.

For parameter estimation and evidence evaluation, we use the \texttt{SimpleMC} code~\cite{simplemc}, which employs the nested-sampling package \texttt{dynesty}~\cite{speagle2020dynesty} to implement nested sampling~\cite{skilling}. Each run is performed with 500 live points and a convergence tolerance of 0.01. In the reconstruction framework we adopt broad flat priors $H_0\in[40,90]\ {\rm km\,s^{-1}\,Mpc^{-1}}$, $E_i\in[0.5,4.0]$. These node locations and priors are chosen to cover the redshift range directly probed by the late-time data while remaining agnostic about the detailed form of the expansion history.

\subsection{Dark energy equation of state parametrizations}

For the dark energy sector, we consider the standard low-dimensional EoS parametrizations and assume that DE and the dust component (cold dark matter plus baryons) are minimally interacting, i.e., coupled only through gravity and separately conserved. We neglect the radiation contribution, which remains negligible over the redshift range relevant for the present analysis, $z\lesssim3$. Assuming that gravity is governed by general relativity and that the late-time Universe is described by a spatially flat Friedmann--Lema\^itre--Robertson--Walker (FLRW) spacetime, the background expansion is written as
\begin{equation}
E^2(z)=\Omega_{\rm m0}(1+z)^3
+(1-\Omega_{\rm m0})
\exp\!\left[
3\int_0^z \frac{1+w_{\rm DE}(\tilde z)}{1+\tilde z}\,\dd \tilde z
\right],
\label{eq:Ez_general_wz}
\end{equation}
so that each model is fully determined once a functional form for $w_{\rm DE}(z)$ is specified. Under this construction, the DE density contribution defined by~\cref{eq:Ez_general_wz} is sign-preserving by design: once spatial flatness is imposed and $\Omega_{\rm m0}<1$, the prefactor $(1-\Omega_{\rm m0})$ is positive, while the exponential factor never changes sign. These parametrizations are therefore used here as phenomenological compressions of possible late-time dynamics rather than as literal microphysical models of DE. Their role in the present work is to assess how strongly the inferred kinematics depend on the assumed functional prior.

We consider the following benchmark models:
\begin{itemize}

\item \textbf{$\Lambda$CDM:}
\begin{equation}
w_{\rm DE}(z)=-1.
\end{equation}
This corresponds to a cosmological constant $\Lambda$ (or, equivalently at the background level, to a vacuum-energy component with $w_{\rm DE}=-1$) and provides the minimal late-time baseline of the $\Lambda$CDM model. The $\Lambda$CDM limit is recovered by all of the two-parameter ans\"atze for $(w_0,w_a)=(-1,0)$.

\item \textbf{$w$CDM:}
\begin{equation}
w_{\rm DE}(z)=w_0=\mathrm{const.}
\end{equation}
This one-parameter extension tests whether the data prefer a constant departure of the DE EoS parameter from the cosmological-constant value $-1$, without introducing explicit redshift dependence.

\item \textbf{Chevallier--Polarski--Linder (CPL)}~\cite{Chevallier:2000qy,Linder:2002et}:
\begin{equation}
w_{\rm DE}(z)=w_0+w_a\frac{z}{1+z}.
\end{equation}
This is the standard two-parameter benchmark. It may be regarded as a first-order expansion in $1-a$, is regular on $z\in[-1,\infty)$, and approaches $w_0+w_a$ at high redshift.

\item \textbf{Jassal--Bagla--Padmanabhan (JBP)}~\cite{Jassal:2004ej,Jassal:2005qc}:
\begin{equation}
w_{\rm DE}(z)=w_0+w_a\frac{z}{(1+z)^2}.
\end{equation}
In this case, the departure from $w_0$ is concentrated mainly at intermediate redshift, while $w_{\rm DE}(0)=w_0$ and $w_{\rm DE}(z\rightarrow\infty)\rightarrow w_0$.

\item \textbf{Barboza--Alcaniz}~\cite{Barboza:2008rh}:
\begin{equation}
w_{\rm DE}(z)=w_0+w_a\frac{z(1+z)}{1+z^2}.
\end{equation}
This form remains finite on $z\in[-1,\infty)$ and, like CPL, tends to $w_0+w_a$ at high redshift, though with a different interpolation through the intermediate-redshift regime.

\item \textbf{Exponential}~\cite{Pan:2019brc,Najafi:2024qzm}:
\begin{equation}
w_{\rm DE}(z)=w_0+w_a\left[\exp\!\left(\frac{z}{1+z}\right)-1\right].
\end{equation}
This provides a nonlinear but still smooth generalization of CPL. Its high-redshift limit is finite, $w_{\rm DE}(z\rightarrow\infty)\rightarrow w_0+w_a(e-1)$,
while near $z=0$ it behaves similarly to other smooth two-parameter expansions.

\item \textbf{Logarithmic}~\cite{Tripathi:2016slv}:
\begin{equation}
w_{\rm DE}(z)=w_0+w_a\ln(1+z).
\end{equation}
Over the redshift range relevant for our late-time data, this ansatz yields a comparatively slow and gradual evolution of the equation of state, although, unlike the previous bounded forms, it does not asymptote to a finite constant as $z\to\infty$.

\end{itemize}

\subsection{Datasets and priors}

Given that the late-time expansion history is the central focus of this work, we employ several complementary low-redshift probes to constrain the dark energy dynamics. In all cases, we use the same late-time dataset combinations and the same likelihood structure for the reconstruction and for the smooth EoS parametrizations, so that any differences can be attributed to the assumed functional description rather than to changes in the data.

\begin{itemize}
    \item \textbf{Cosmic Chronometers (CC):} We use 31 model-independent measurements of $H(z)$ spanning the redshift range $0.07<z<1.96$~\cite{Zhang:2012mp,Jimenez:2003iv,Simon:2004tf,Moresco:2012by,Moresco:2016mzx,Ratsimbazafy:2017vga,Moresco:2015cya}. These provide direct constraints on the expansion rate without assuming a specific cosmological model.
    
    \item \textbf{Type Ia Supernovae (SN):} We employ the Pantheon+ compilation~\cite{Scolnic:2021amr,Brout:2022vxf}, which consists of 1701 light curves from 1550 distinct SNe~Ia. Spanning $0.01<z<2.26$, this sample provides luminosity-distance information to constrain the distance--redshift relation.
    
    \item \textbf{Baryon Acoustic Oscillations (DESI):} We include measurements from the recent DESI DR2 release~\cite{DESI:2025zgx,DESI:2025fii,DESI:2025qqy}. These constrain combinations of the expansion rate and angular-diameter distance normalized by the sound horizon scale $r_{\rm d}$. Following the DESI analysis, we calibrate $r_{\rm d}$ using Big Bang Nucleosynthesis (BBN) information~\cite{DESI:2025zgx}, rather than imposing a full CMB likelihood, in order to preserve the late-time focus of the present analysis.
    
    \item \textbf{Hubble Constant Prior (H0DN):} To anchor the absolute distance scale, we adopt an external Gaussian prior on $H_0$ (in km\,s$^{-1}$\,Mpc$^{-1}$) from the Local Distance Network~\cite{H0DN:2025lyy}. This acts effectively as a data point at $z=0$ and is denoted as H0DN in our results.
\end{itemize}

For parameter estimation and evidence evaluation, we use the same \texttt{SimpleMC}/\texttt{dynesty} setup described above. For the smooth dark energy parametrizations we assume spatial flatness and neglect the radiation contribution, given its negligible effect on late-time dynamics. The corresponding parameter sets are $\{H_0,\Omega_{\rm m0}\}$ for flat $\Lambda$CDM, $\{H_0,\Omega_{\rm m0},w_0\}$ for $w$CDM, and $\{H_0,\Omega_{\rm m0},w_0,w_a\}$ for the remaining two-parameter models. We adopt broad flat priors $H_0 \in [40,90]~{\rm km\,s^{-1}\,Mpc^{-1}}$, $\Omega_{\rm m0} \in [0.1,0.9]$,
$w_0 \in [-2.0,0.0]$, and 
$w_a \in [-3.5,2.0]$. The equation of state parameters $w_0$ and $w_a$ are varied only in the models where they are active, and remain fixed in the $\Lambda$CDM and $w$CDM limits where appropriate.

\section{Results and Discussion}
\label{sec:discussion}

\begin{figure*}[ht!]
    \centering
    \includegraphics[width=0.32\linewidth]{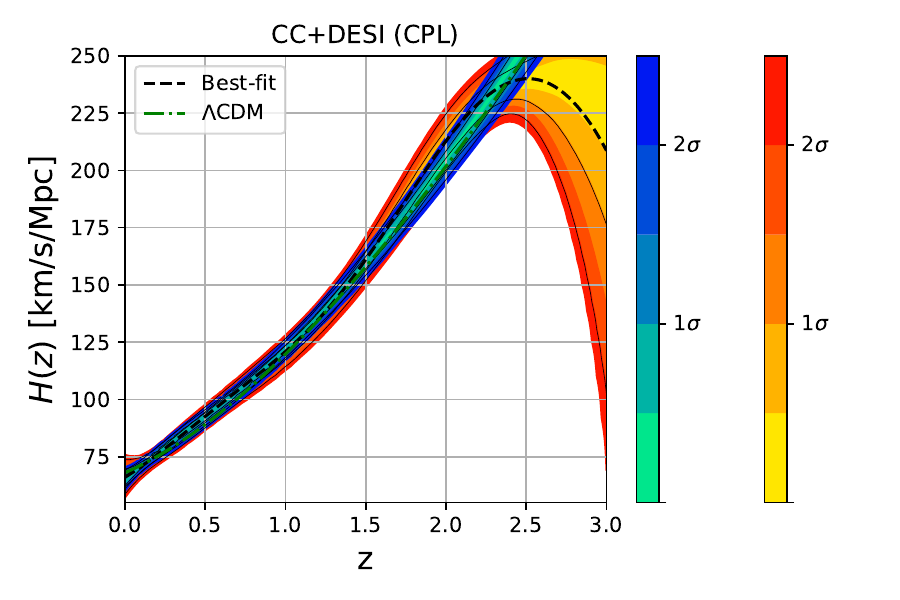}
    \includegraphics[width=0.32\linewidth]{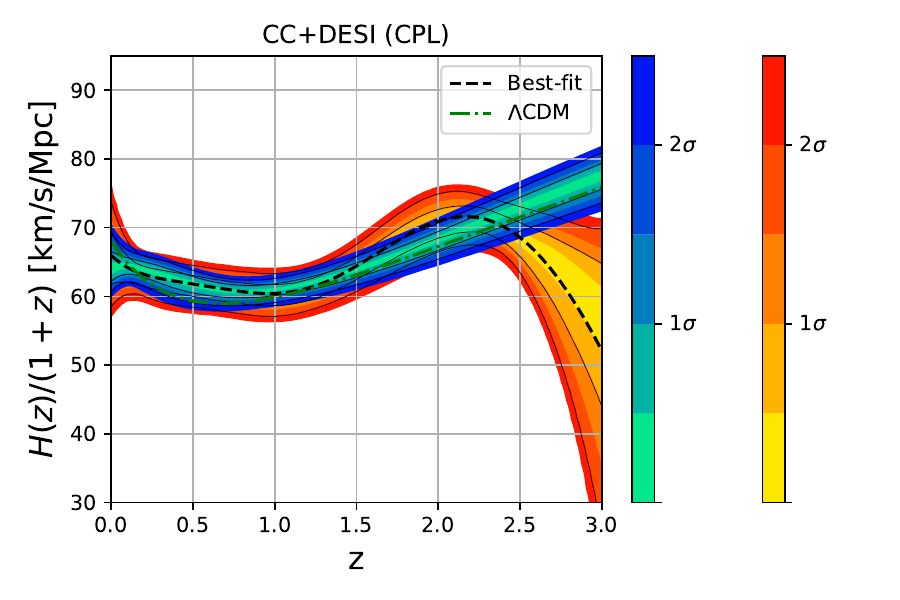}
    \includegraphics[width=0.32\linewidth]{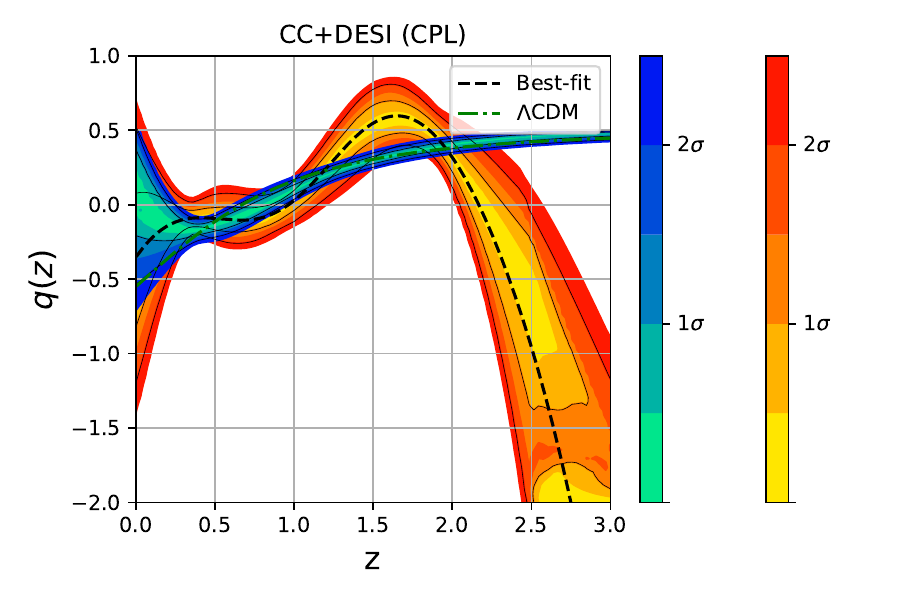}
    
    \includegraphics[width=0.32\linewidth]{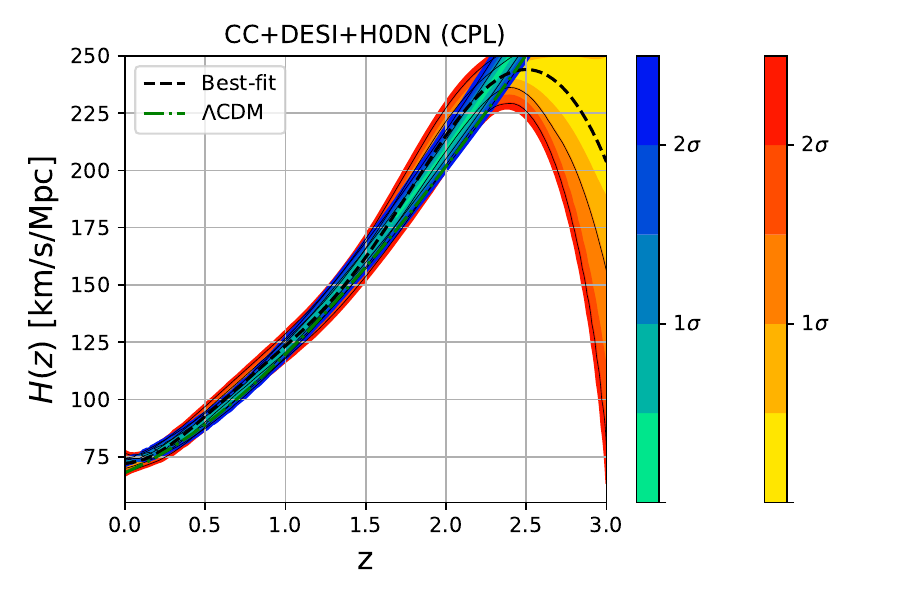}
    \includegraphics[width=0.32\linewidth]{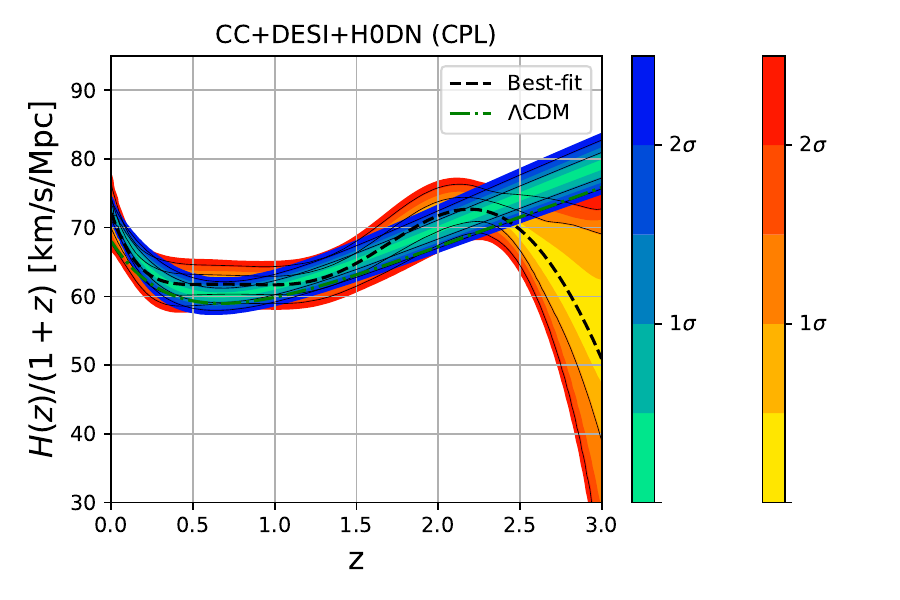}
    \includegraphics[width=0.32\linewidth]{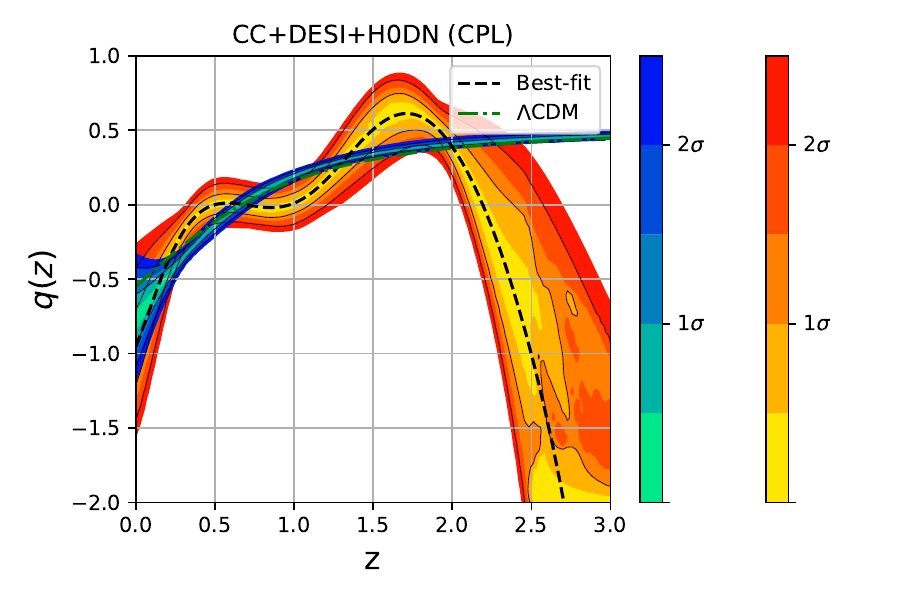}
    
    \includegraphics[width=0.32\linewidth]{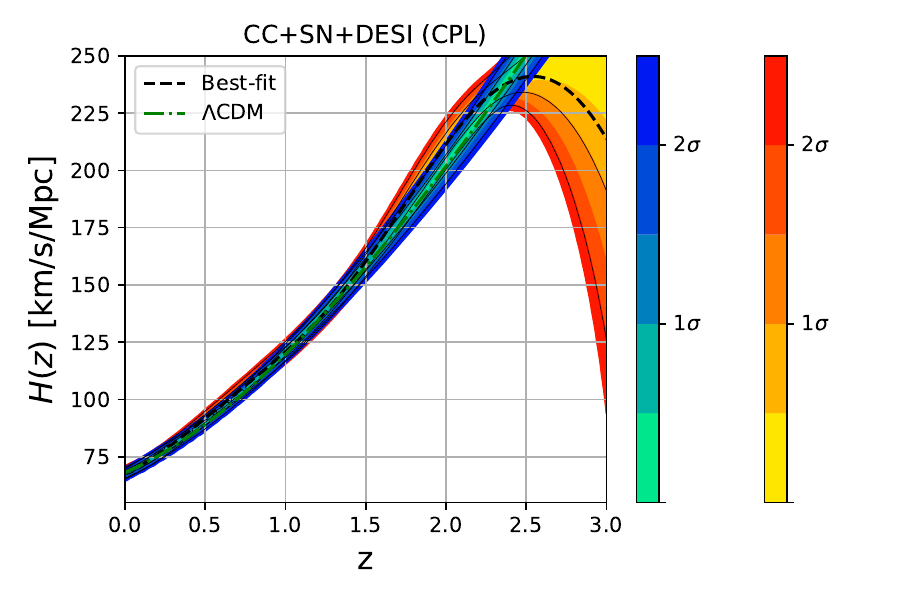}
    \includegraphics[width=0.32\linewidth]{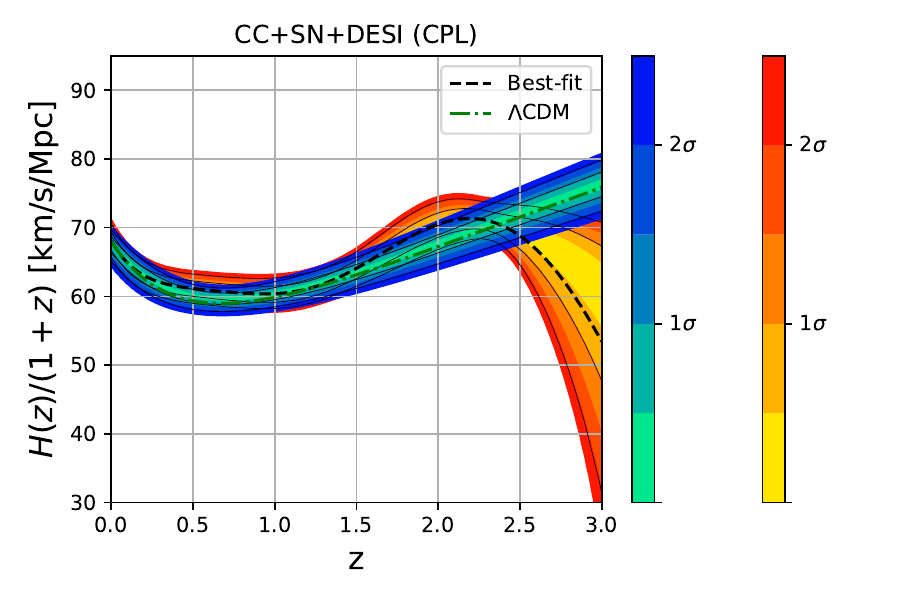}
    \includegraphics[width=0.32\linewidth]{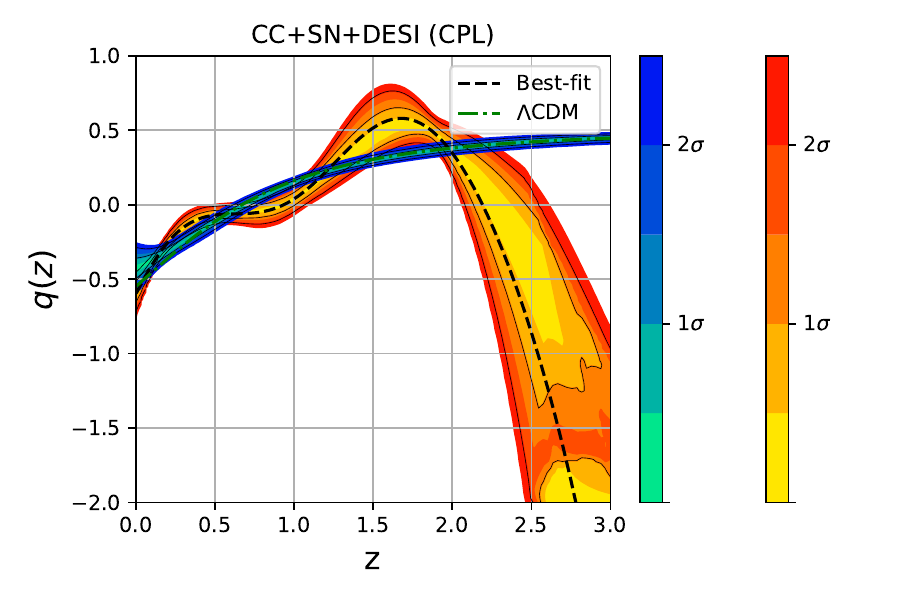}

       \includegraphics[width=0.32\linewidth]{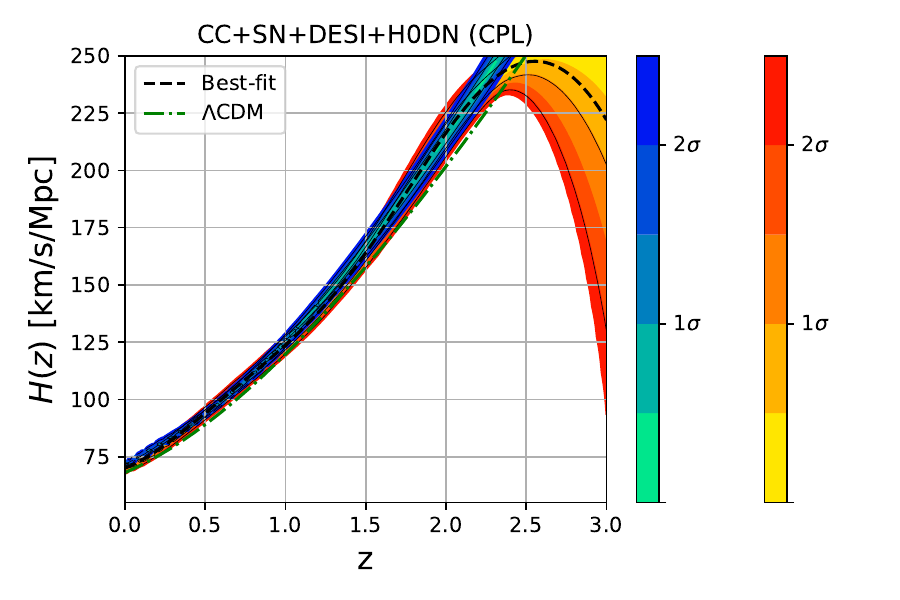}
    \includegraphics[width=0.32\linewidth]{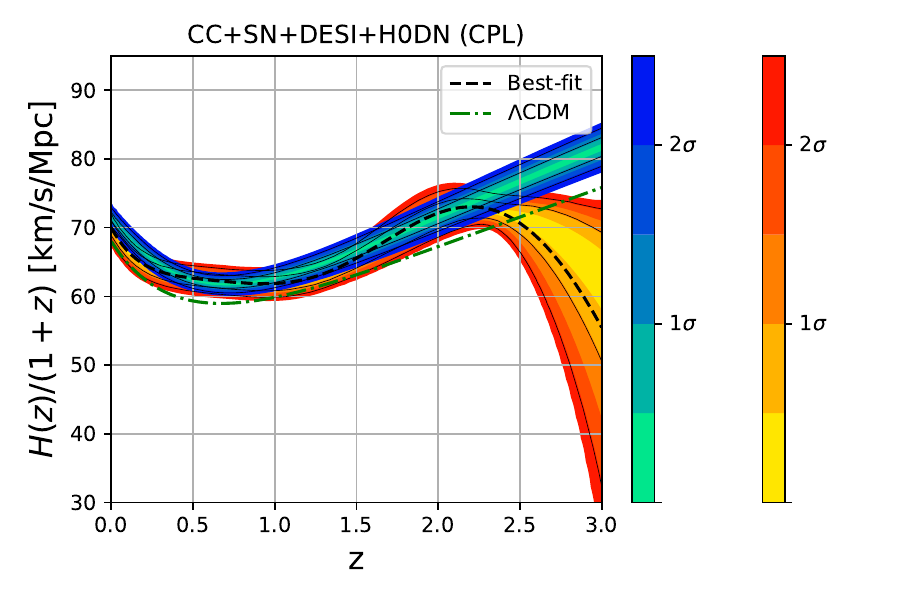}
    \includegraphics[width=0.32\linewidth]{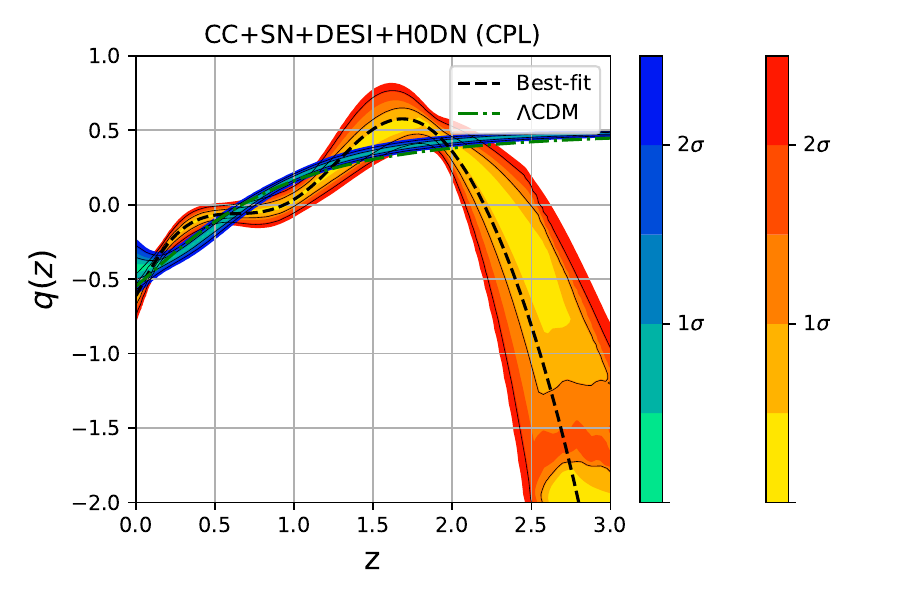}
\caption{
Posterior predictive regions for the kinematical diagnostics, namely the Hubble parameter $H(z)$, the conformal Hubble parameter $H(z)/(1+z)$, and the deceleration parameter $q(z)$, comparing the non-parametric node-based reconstruction (Rec.) with the CPL parametrization, shown here as a representative EoS-based DE benchmark. The four rows correspond, from top to bottom, to the dataset combinations CC+DESI, CC+DESI+H0DN, CC+SN+DESI, and CC+SN+DESI+H0DN, with SN denoting Pantheon+. In each panel, the cool (blue--green) and warm (yellow--red) shaded bands show, respectively, the CPL and reconstruction constraints; the accompanying color strips indicate the $\sigma$-equivalent credible levels, with inner and outer shading corresponding approximately to $1\sigma$ and $2\sigma$ (i.e. $\sim68\%$ and $\sim95\%$ for a Gaussian posterior). The black dashed curve shows the best-fit reconstruction, while the green dash-dotted curve shows the best-fit $\Lambda$CDM baseline for the same dataset combination. Because the highest-redshift node is fixed at $z=3$ and no data lie in the interval $2.4<z<3.0$, the behaviour approaching this boundary should be interpreted cautiously, particularly for the derivative-based diagnostic $q(z)$. Corresponding comparisons with the JBP, Barboza--Alcaniz, exponential, and logarithmic parametrizations are shown in~\cref{app:reconstructions}.
}
    \label{fig:H_and_q}
\end{figure*}

The kinematical constraints are summarized in~\cref{fig:H_and_q}, with the corresponding parameter estimates collected in~\cref{tab:results2}. This analysis is intentionally minimalist in its focus on late-time data, yet broader in scope than a typical CPL-versus-reconstruction comparison. By expanding the parametric side beyond the standard CPL benchmark to include the $\Lambda$CDM baseline, $w$CDM, JBP, Barboza--Alcaniz, exponential, and logarithmic forms, we find that the choice of a specific smooth EoS-based parametrization does not materially alter the background picture. Across the redshift range populated by the data, these diverse ans\"atze yield mutually consistent results that remain close to the flat $\Lambda$CDM baseline. Consequently, the primary qualitative contrast is not between individual models, but between the entire class of smooth low-dimensional EoS-based DE parametrizations and the more flexible, non-parametric reconstruction. This reinforces the central lesson of the study: \textit{the choice of description class matters far more than the particular functional form assumed for the redshift dependence of the DE EoS parameter.}

\begin{table*}[t!]
\centering
\caption{Marginalized parameter constraints and model-comparison diagnostics for the non-parametric node-based reconstruction (Rec.) and the smooth low-dimensional EoS-based DE parametrizations considered in this work, for each dataset combination. Quoted uncertainties correspond to $68\%$ credible intervals. For the reconstruction, $w_0$ denotes the present-day DE EoS parameter inferred from the reconstructed background under the GR mapping, while $w_a$ is not defined. For each model column $i$, the last two rows report the improvement in the best-fit likelihood and the Bayesian evidence relative to the $\Lambda$CDM baseline, defined as $\Delta\chi^2_{\min,i}\equiv \chi^2_{\min}(i)-\chi^2_{\min}(\Lambda\mathrm{CDM})$ and $\ln B_{\Lambda\mathrm{CDM},i}\equiv \ln(Z_{\Lambda\mathrm{CDM}}/Z_i)$. Thus, $\Delta\chi^2_{\min,i}<0$ indicates a better best fit than $\Lambda$CDM, while $\ln B_{\Lambda\mathrm{CDM},i}>0$ indicates that the Bayesian evidence favors the simpler $\Lambda$CDM baseline once the Occam penalty is taken into account. By construction, the $\Lambda$CDM entries in these two rows are shown as dashes.}
\label{tab:results2}

\footnotesize
\setlength{\tabcolsep}{3.0pt}
\renewcommand{\arraystretch}{1.08}

\resizebox{\textwidth}{!}{%
\begin{tabular}{@{}llcccccccc@{}}
\toprule
\textbf{Dataset} & \textbf{Parameter}
& \textbf{Rec.} & \textbf{$\Lambda$CDM} & \textbf{$w$CDM} & \textbf{CPL} & \textbf{JBP} & \textbf{Barb.\ Alc.} & \textbf{Exp.} & \textbf{Log.} \\
\midrule

\multirow{6}{*}{\makecell[l]{\textbf{CC+DESI}}}
& $H_0$
& $65.96 \pm 3.80$
& $68.50 \pm 0.53$ & $67.0\pm 2.0$ & $64.69 \pm 2.12$
& $64.6^{+2.2}_{-2.6}$ & $64.8^{+1.8}_{-2.2}$ & $64.9^{+1.6}_{-2.4}$
& $65.3^{+1.7}_{-2.1}$ \\

& $q_0$
& $-0.35^{+0.27}_{-0.31}$
& $-0.55\pm0.01$ & $-0.486\pm 0.077$ & $0.02^{+0.34}_{-0.22}$
& $-0.05^{+0.39}_{-0.20}$ & $-0.03^{+0.28}_{-0.24}$ & $0.03^{+0.37}_{-0.22}$
& $-0.02\pm 0.26$ \\

& $w_0$
& $-0.81^{+0.26}_{-0.30}$
& $-1$ & $-0.939\pm 0.071$ & $-0.48\pm0.26$
& $-0.54^{+0.37}_{-0.20}$ & $-0.54\pm 0.24$ & $-0.47^{+0.35}_{-0.23}$
& $-0.52^{+0.24}_{-0.28}$ \\

& $w_a$
& $-$
& $-$ & $-$ & $-1.66^{+0.87}_{-1.0}$
& $-2.03^{+0.95}_{-1.8}$ & $-0.85\pm 0.48$ & $-1.99^{+0.93}_{-1.5}$
& $-1.29^{+0.79}_{-0.72}$ \\

\addlinespace[1pt]
\cmidrule(lr){2-10}

& $\Delta\chi^2_{\min,i}$
& $-3.47$
& $-$ & $-0.91$ & $-3.42$
& $-3.21$ & $-3.49$ & $-3.41$
& $-3.34$ \\

& $\ln B_{\Lambda\mathrm{CDM},i}$
& $10.76$
& $-$ & $2.01$ & $1.27$
& $1.26$ & $1.89$ & $1.28$
& $1.38$ \\
\midrule

\multirow{6}{*}{\makecell[l]{\textbf{CC+DESI}\\\textbf{+H0DN}}}
& $H_0$
& $72.41 \pm 1.71$
& $68.92 \pm 0.42$ & $72.14\pm 0.85$ & $70.10\pm1.09$
& $72.38\pm 0.77$ & $72.36\pm 0.78$ & $72.76\pm 0.79$
& $72.60\pm 0.79$ \\

& $q_0$
& $-0.61\pm 0.20$
& $-0.55 \pm 0.01$ & $-0.709\pm 0.045$ & $-0.46\pm0.23$
& $-0.918^{+0.093}_{-0.17}$ & $-0.74\pm 0.15$ & $-0.81\pm 0.18$
& $-0.67\pm 0.14$ \\

& $w_0$
& $-1.02\pm 0.19$
& $-1$ & $-1.143\pm 0.037$ & $-0.92\pm0.19$
& $-1.314^{+0.074}_{-0.14}$ & $-1.17^{+0.10}_{-0.12}$ & $-1.22^{+0.13}_{-0.16}$
& $-1.11^{+0.10}_{-0.11}$ \\

& $w_a$
& $-$
& $-$ & $-$ & $-0.45^{+0.76}_{-0.59}$
& $1.07^{+0.87}_{-0.41}$ & $0.03^{+0.31}_{-0.25}$ & $0.41^{+0.84}_{-0.63}$
& $-0.14^{+0.43}_{-0.34}$ \\

\addlinespace[1pt]
\cmidrule(lr){2-10}

& $\Delta\chi^2_{\min,i}$
& $-6.88$
& $-$ & $-4.03$ & $-4.87$
& $-5.15$ & $-4.33$ & $-4.82$
& $-4.81$ \\

& $\ln B_{\Lambda\mathrm{CDM},i}$
& $10.59$
& $-$ & $2.14$ & $1.54$
& $1.22$ & $1.78$ & $1.55$
& $1.55$ \\
\midrule

\multirow{6}{*}{\makecell[l]{\textbf{CC+SN+DESI}}}
& $H_0$
& $68.05 \pm 1.63$
& $68.54 \pm 0.52$ & $66.7\pm 1.2$ & $67.42\pm1.16$
& $66.8^{+1.3}_{-1.2}$ & $67.2^{+1.6}_{-1.3}$ & $67.2^{+1.5}_{-1.3}$
& $67.5^{+1.7}_{-1.3}$ \\

& $q_0$
& $-0.56^{+0.16}_{-0.22}$
& $-0.54\pm 0.01$ & $-0.471\pm 0.037$ & $-0.43\pm0.07$
& $-0.435\pm 0.092$ & $-0.434\pm 0.059$ & $-0.428\pm 0.078$
& $-0.428\pm 0.064$ \\

& $w_0$
& $-1.01^{+0.15}_{-0.21}$
& $-1$ & $-0.926\pm 0.038$ & $-0.89\pm0.06$
& $-0.894\pm 0.080$ & $-0.899\pm 0.048$ & $-0.891\pm 0.065$
& $-0.895^{+0.048}_{-0.055}$ \\

& $w_a$
& $-$
& $-$ & $-$ & $-0.30^{+0.34}_{-0.29}$
& $-0.30\pm 0.62$ & $-0.15^{+0.19}_{-0.16}$ & $-0.29^{+0.47}_{-0.41}$
& $-0.23^{+0.30}_{-0.25}$ \\

\addlinespace[1pt]
\cmidrule(lr){2-10}

& $\Delta\chi^2_{\min,i}$
& $-6.61$
& $-$ & $-3.81$ & $-4.11$
& $-3.86$ & $-4.22$ & $-4.32$
& $-4.23$ \\

& $\ln B_{\Lambda\mathrm{CDM},i}$
& $12.23$
& $-$ & $1.08$ & $2.11$
& $1.87$ & $2.27$ & $2.08$
& $2.07$ \\
\midrule

\multirow{6}{*}{\makecell[l]{\textbf{CC+SN+DESI}\\\textbf{+H0DN}}}
& $H_0$
& $70.12 \pm 1.31$
& $68.92 \pm 0.42$ & $71.24\pm 0.63$ & $71.24 \pm 0.85$
& $71.02\pm 0.69$ & $71.61\pm 0.65$ & $71.57\pm 0.65$
& $71.69\pm 0.65$ \\

& $q_0$
& $-0.57^{+0.11}_{-0.15}$
& $-0.548\pm 0.011$ & $-0.598\pm 0.029$ & $-0.43^{+0.06}_{-0.07}$
& $-0.427\pm 0.095$ & $-0.456\pm 0.063$ & $-0.417\pm 0.078$
& $-0.433\pm 0.064$ \\

& $w_0$
& $-1.01^{+0.08}_{-0.20}$
& $-1$ & $-1.060\pm 0.027$ & $-0.91\pm0.06$
& $-0.906\pm 0.087$ & $-0.942\pm 0.055$ & $-0.902\pm 0.070$
& $-0.921\pm 0.057$ \\

& $w_a$
& $-$
& $-$ & $-$ & $-0.46^{+0.36}_{-0.27}$
& $-1.23^{+0.68}_{-0.62}$ & $-0.46^{+0.19}_{-0.16}$ & $-1.04^{+0.46}_{-0.40}$
& $-0.71^{+0.28}_{-0.23}$ \\

\addlinespace[1pt]
\cmidrule(lr){2-10}

& $\Delta\chi^2_{\min,i}$
& $-7.24$
& $-$ & $-1.77$ & $-3.11$
& $-2.66$ & $-3.31$ & $-3.13$
& $-3.23$ \\

& $\ln B_{\Lambda\mathrm{CDM},i}$
& $12.12$
& $-$ & $2.92$ & $2.11$
& $2.85$ & $2.16$ & $2.09$
& $2.14$ \\
\bottomrule
\end{tabular}%
}
\end{table*}

\begin{figure*}[ht!]
    \centering
    \includegraphics[width=0.32\linewidth]{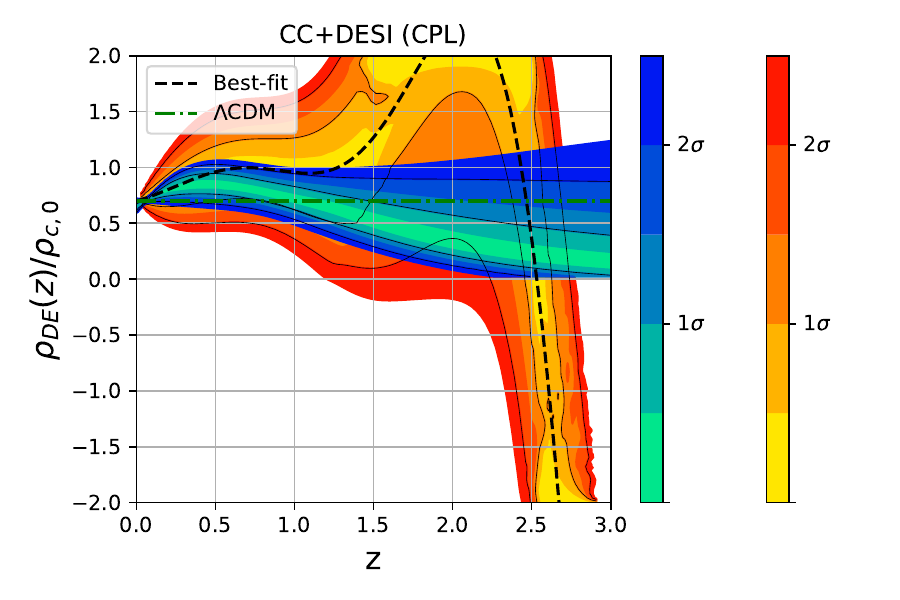}
    \includegraphics[width=0.32\linewidth]{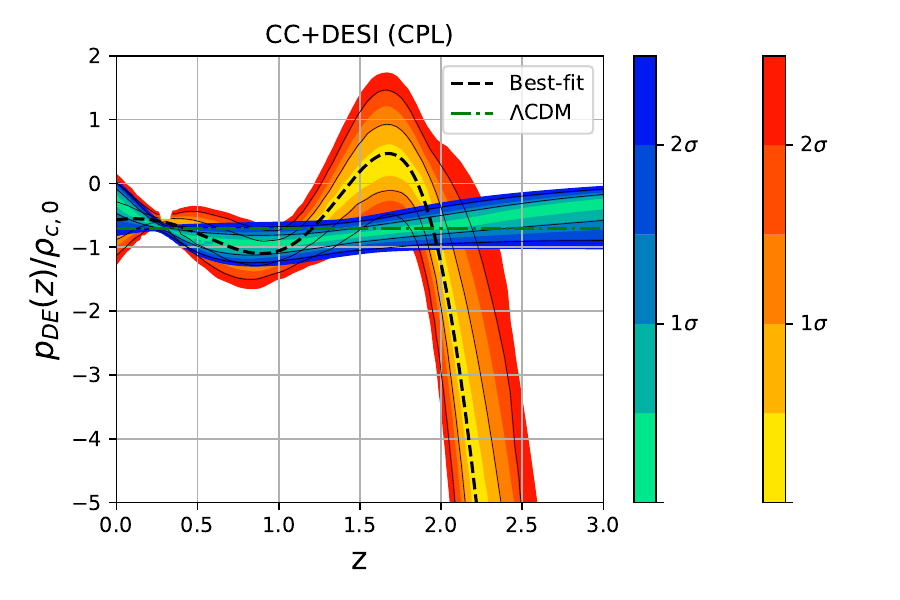}
    \includegraphics[width=0.32\linewidth]{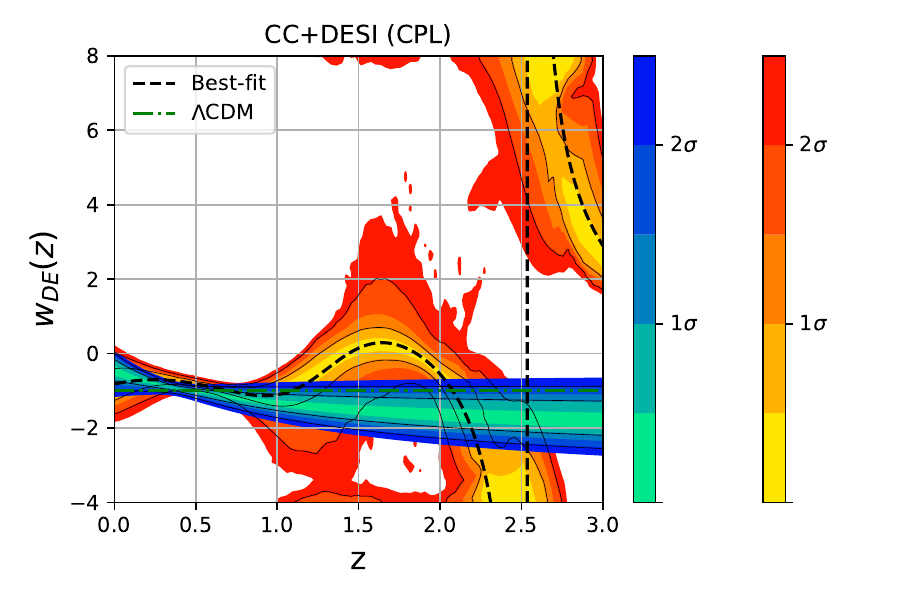}
    
    \includegraphics[width=0.32\linewidth]{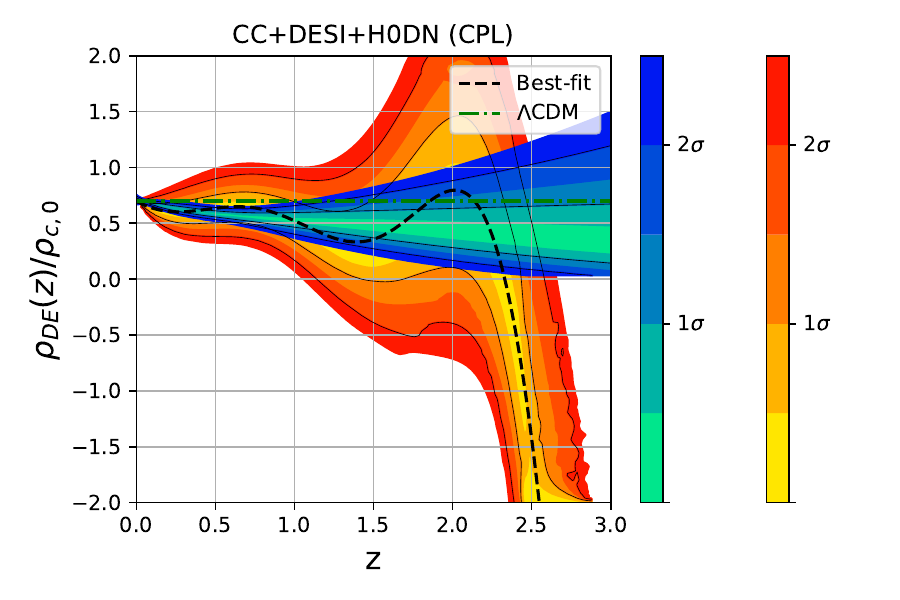}
    \includegraphics[width=0.32\linewidth]{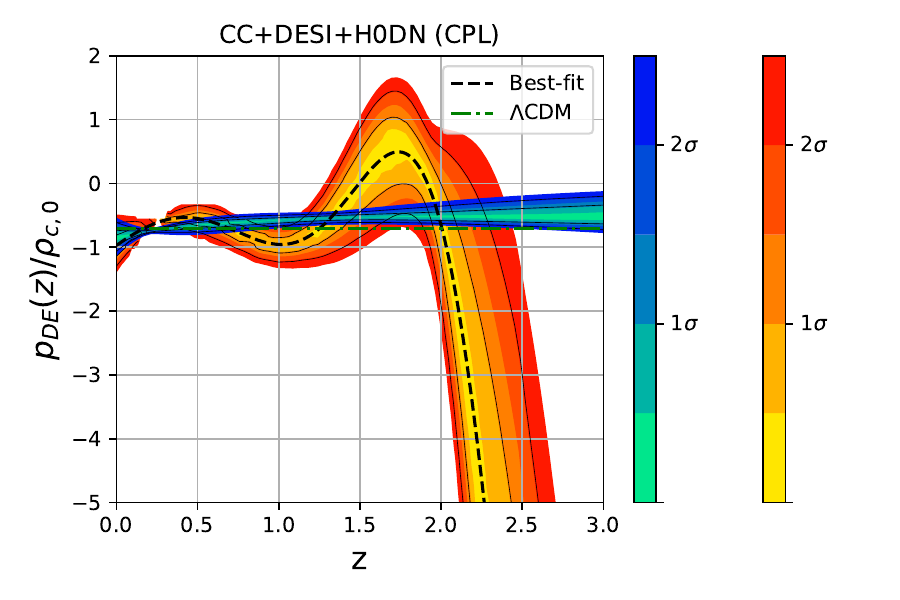}
    \includegraphics[width=0.32\linewidth]{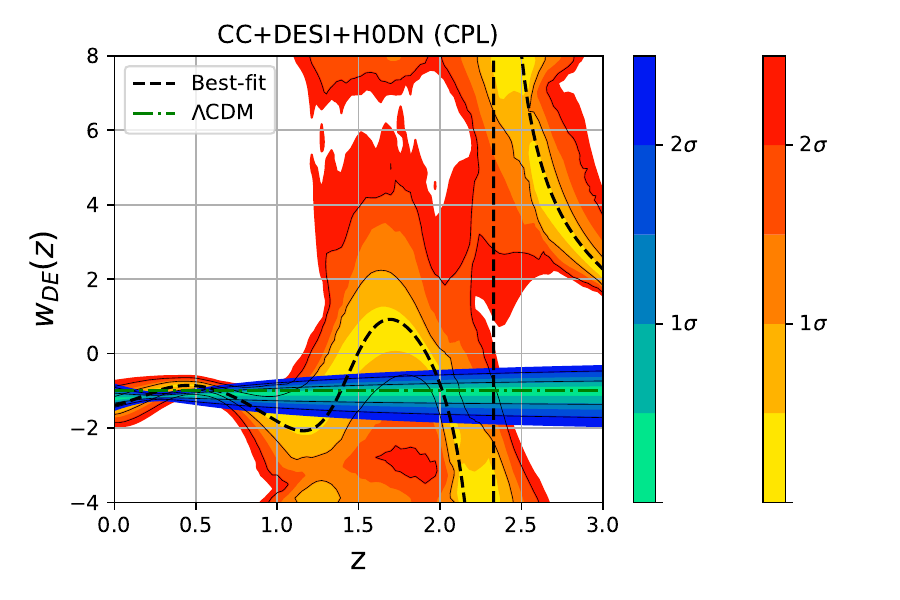}
    
    \includegraphics[width=0.32\linewidth]{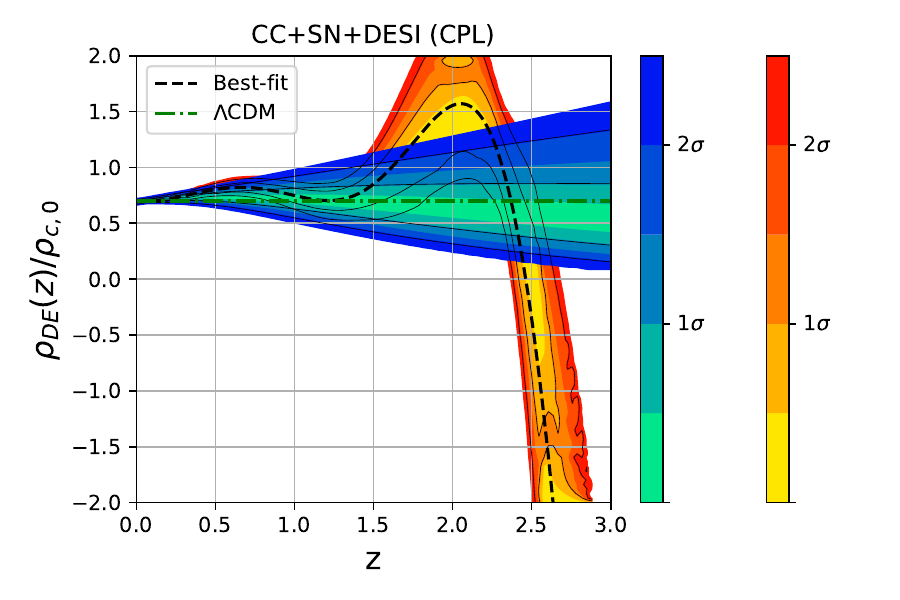}
    \includegraphics[width=0.32\linewidth]{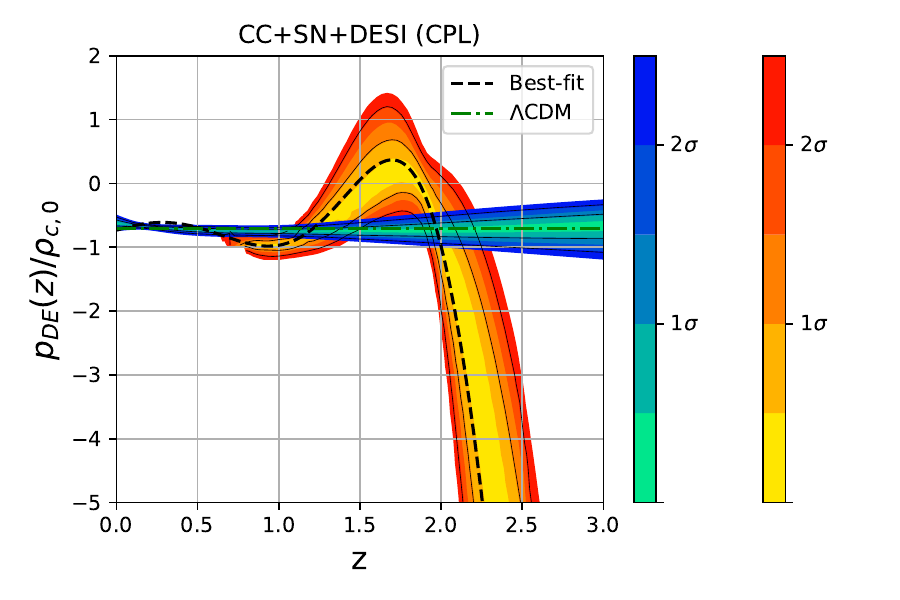}
    \includegraphics[width=0.32\linewidth]{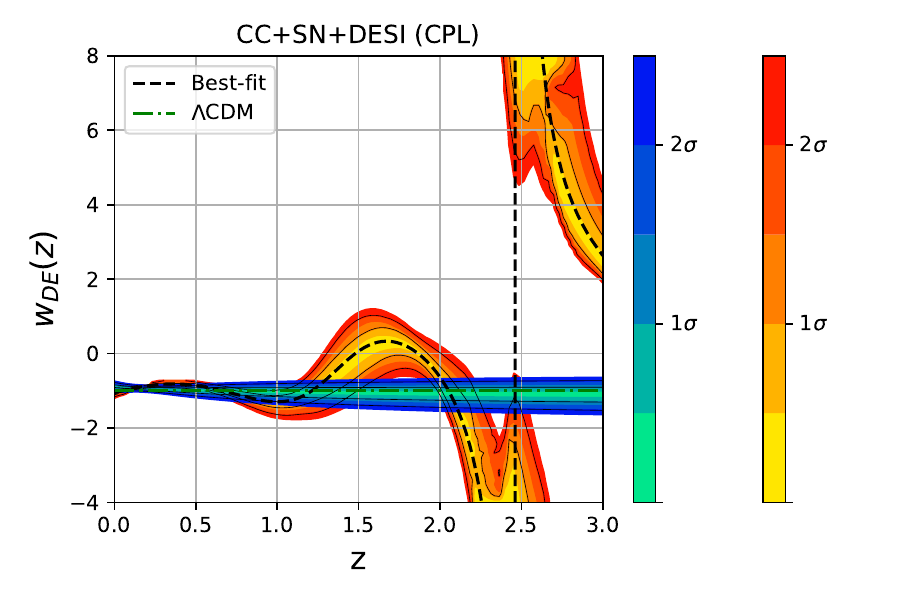}

      \includegraphics[width=0.32\linewidth]{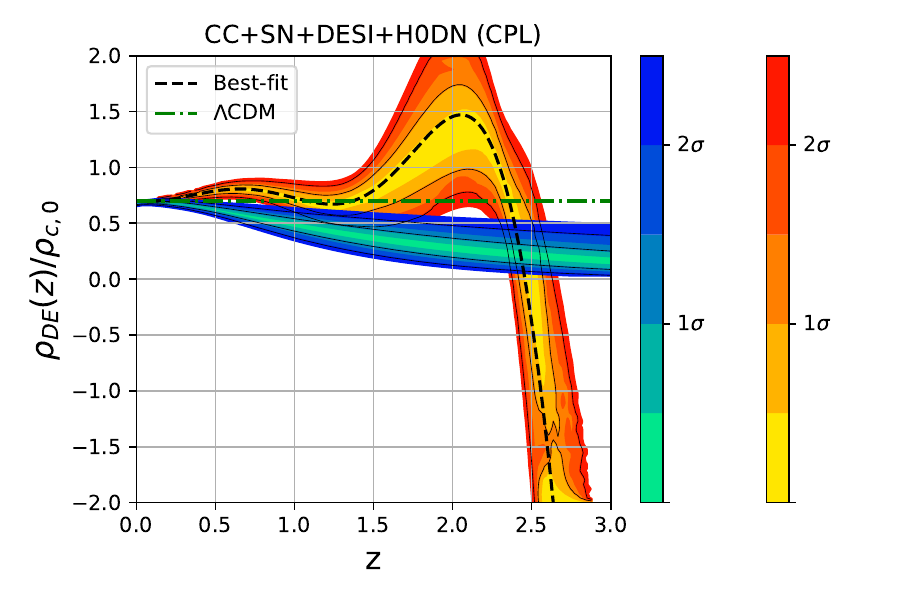}
    \includegraphics[width=0.32\linewidth]{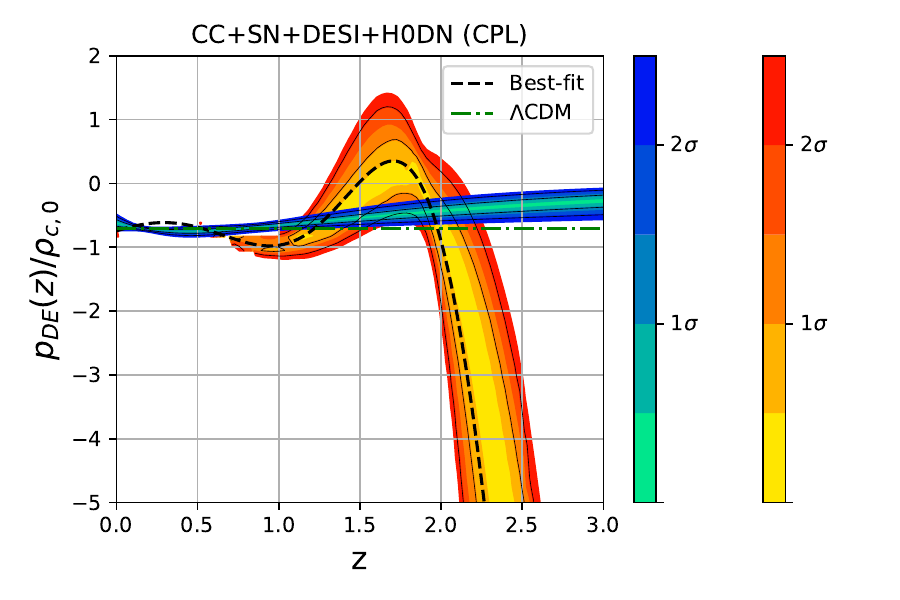}
    \includegraphics[width=0.32\linewidth]{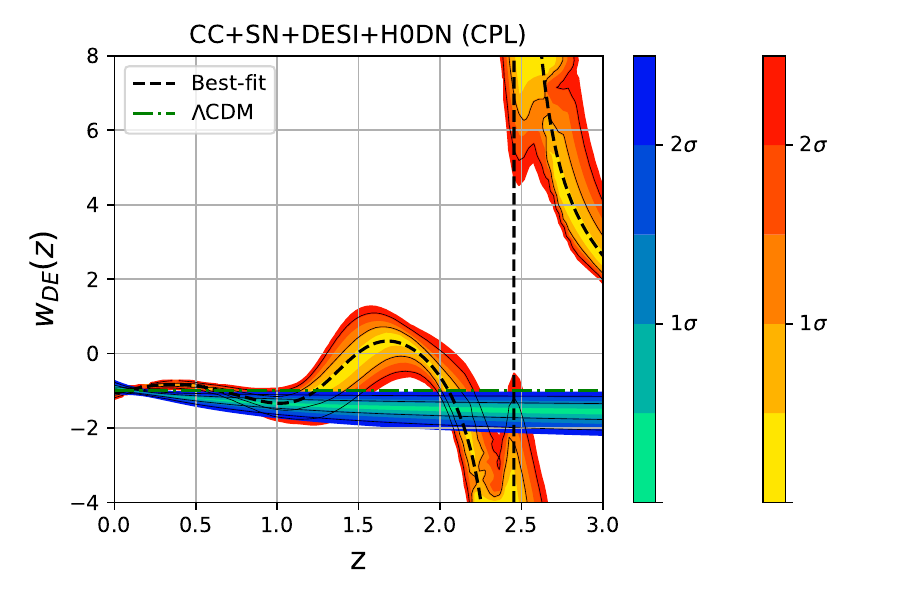}
\caption{
Effective dark energy fluid variables obtained by mapping the inferred expansion histories to an effective GR dark energy sector: the density $\rho_{\rm DE}(z)$, pressure $p_{\rm DE}(z)$ (both normalized to $\rho_{c,0}$), and the EoS parameter $w_{\rm DE}(z)=p_{\rm DE}/\rho_{\rm DE}$. The layout, dataset ordering (top to bottom: CC+DESI, CC+DESI+H0DN, CC+SN+DESI, CC+SN+DESI+H0DN), and visual conventions are the same as in~\cref{fig:H_and_q}. In particular, the cool (blue--green) and warm (yellow--red) shaded bands correspond to CPL and the reconstruction, respectively, with the $\sigma$-equivalent credible levels indicated by the adjacent color strips. The black dashed curve denotes the best-fit reconstruction and the green dash-dotted curve the best-fit $\Lambda$CDM baseline. Apparent divergences or fragmentation in $w_{\rm DE}(z)$ should be interpreted as ratio effects: they arise when $\rho_{\rm DE}(z)$ becomes very small or crosses zero, and do not correspond to singular behaviour in the underlying expansion history. As in~\cref{fig:H_and_q}, the interval $2.4<z<3.0$ is extrapolation-dominated in our node setup and should be treated conservatively.
}
    \label{fig:rhode_pde_wde}
\end{figure*}

\begin{figure*}[ht!]
    \centering
    \includegraphics[width=0.32\linewidth]{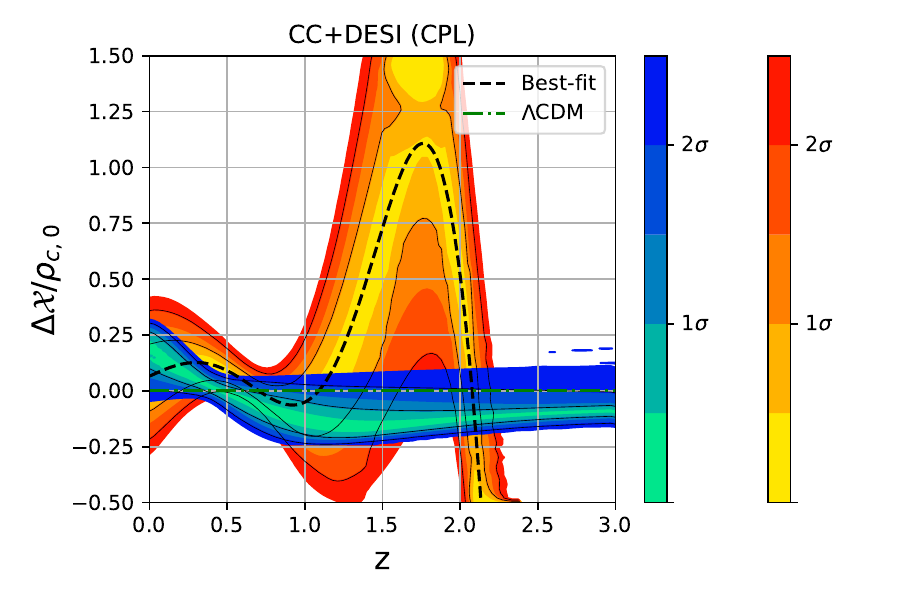}
    \includegraphics[width=0.32\linewidth]{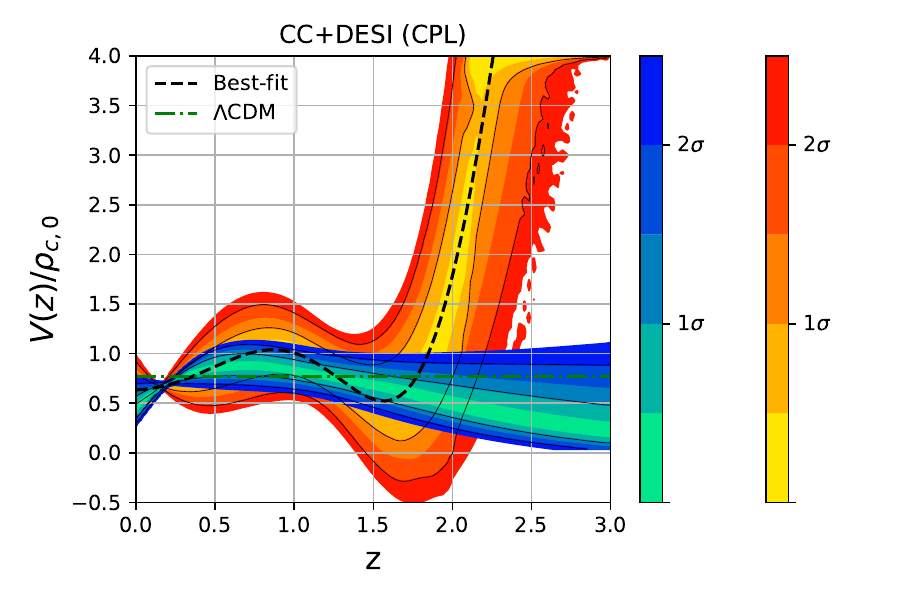}
    
    \includegraphics[width=0.32\linewidth]{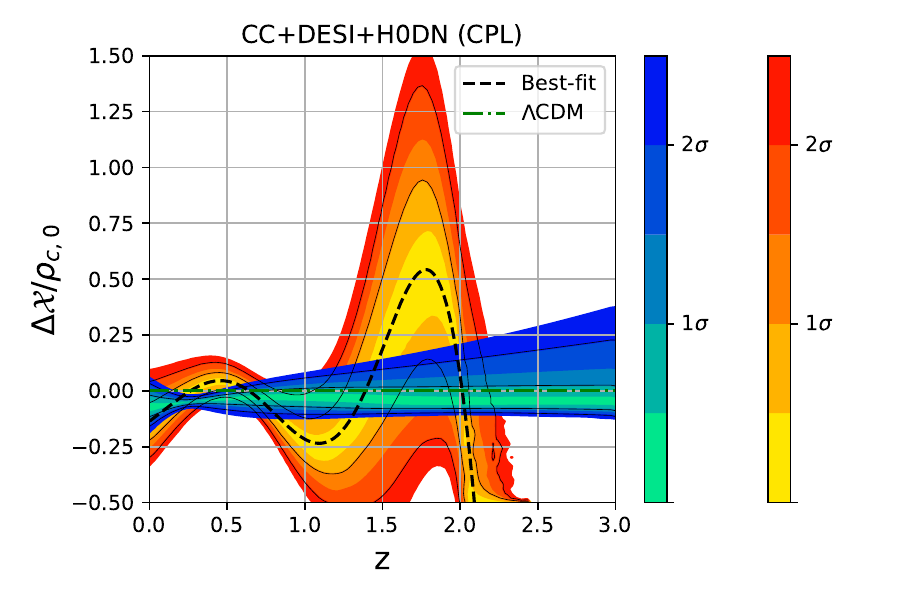}
    \includegraphics[width=0.32\linewidth]{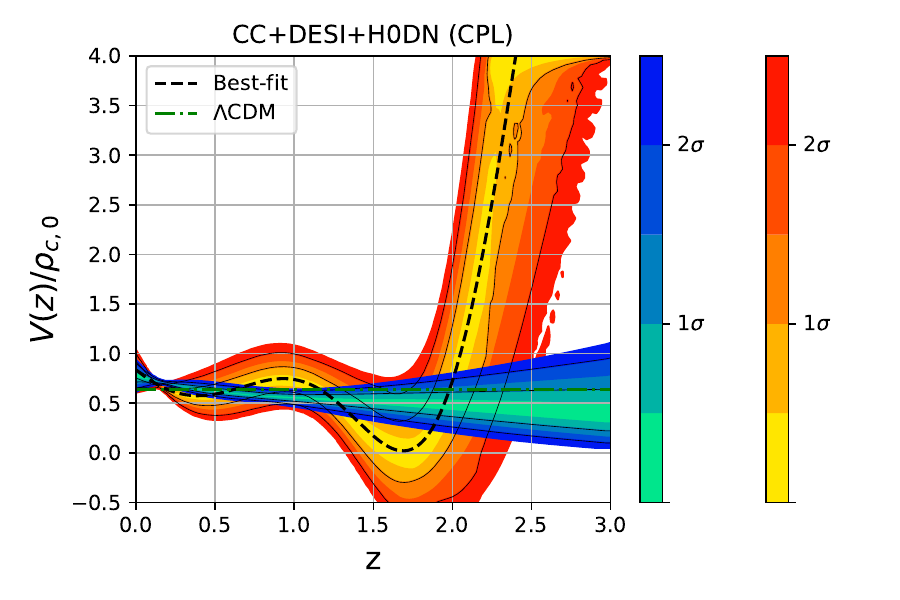}
    
    \includegraphics[width=0.32\linewidth]{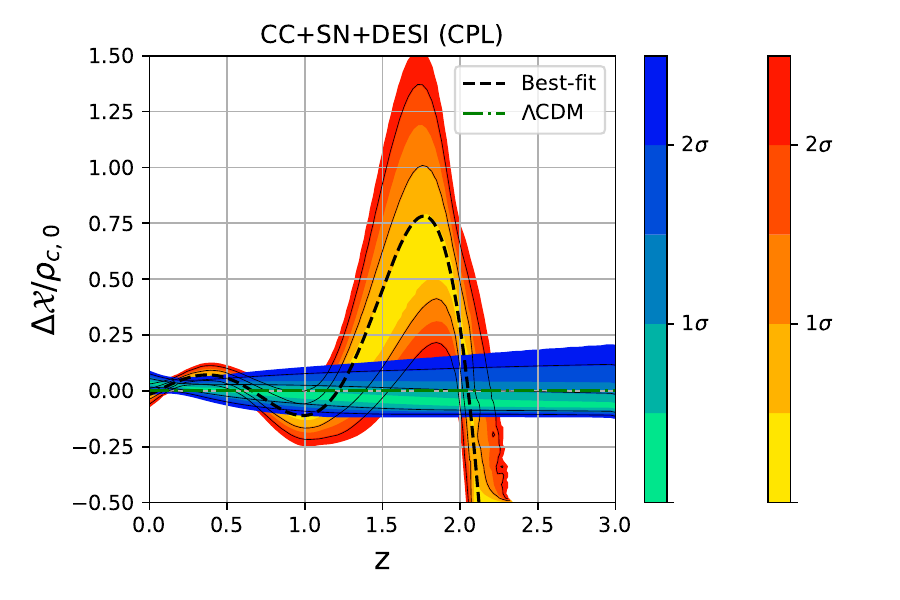}
    \includegraphics[width=0.32\linewidth]{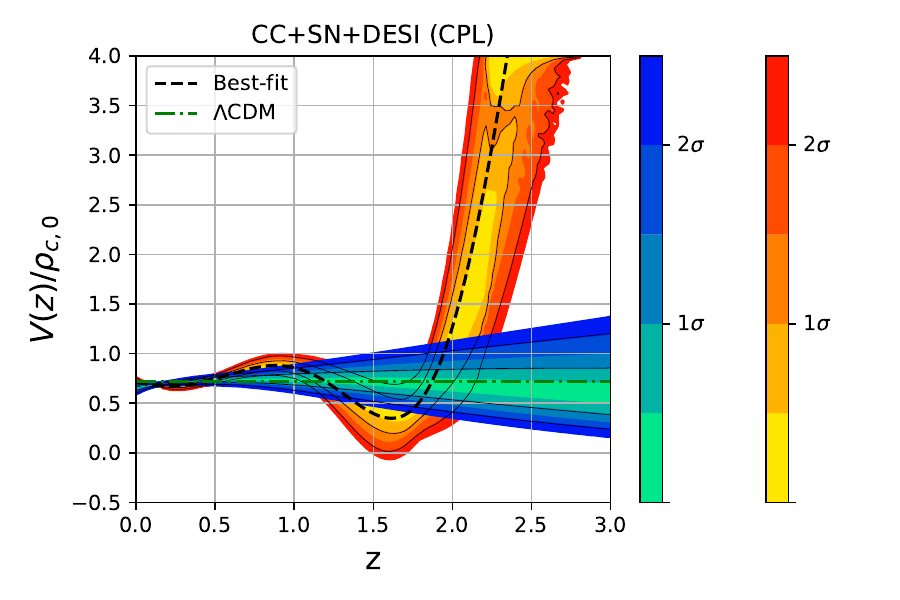}

    \includegraphics[width=0.32\linewidth]{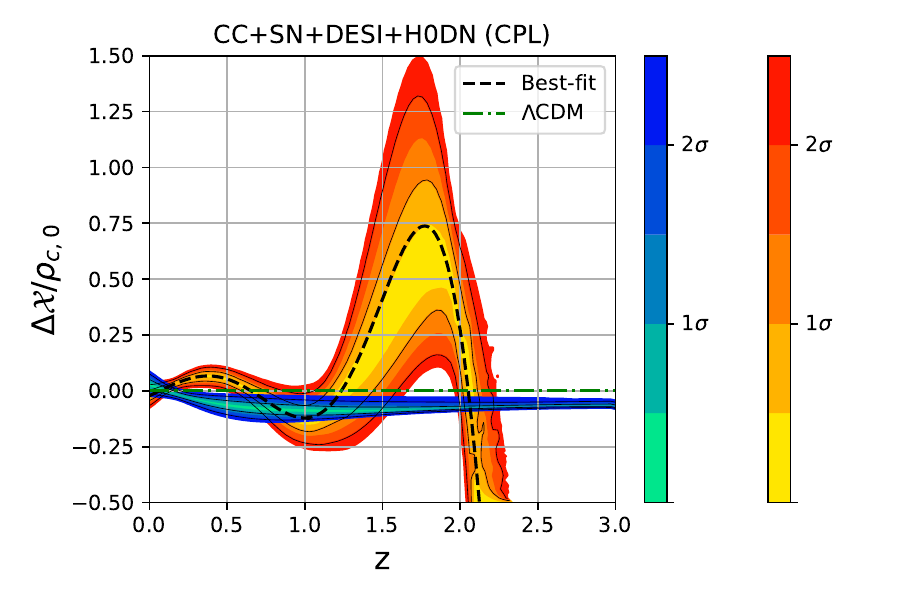}
    \includegraphics[width=0.32\linewidth]{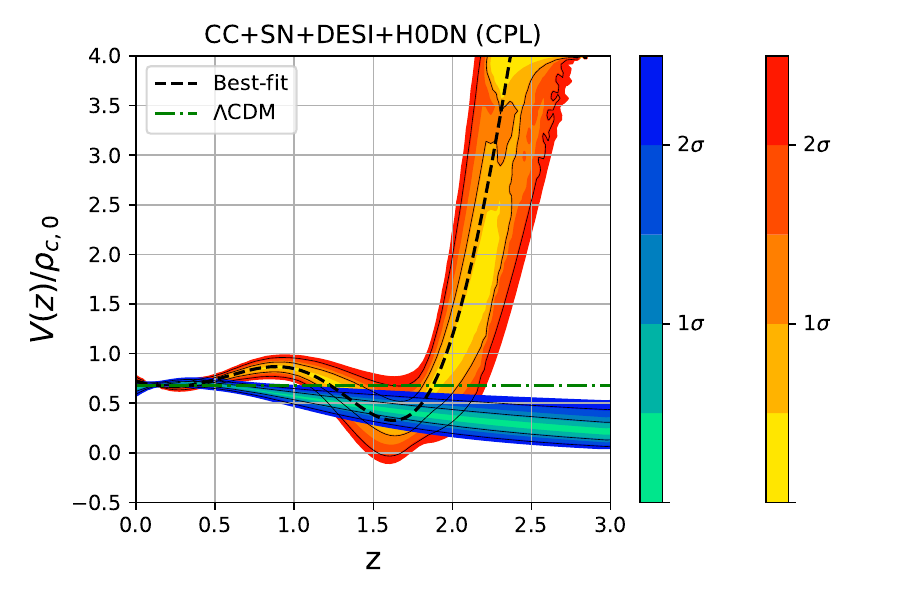}
\caption{
Scalar-sector diagnostics inferred from the same background relations by interpreting the effective fluid in terms of a scalar sector: the effective kinetic contribution $\Delta\mathcal{X}(z)$ and the total effective potential $V(z)$ (both normalized to $\rho_{c,0}$). The four rows correspond, from top to bottom, to the dataset combinations CC+DESI, CC+DESI+H0DN, CC+SN+DESI, and CC+SN+DESI+H0DN. As in~\cref{fig:H_and_q,fig:rhode_pde_wde}, the cool (blue--green) and warm (yellow--red) shaded bands correspond to CPL and the reconstruction, respectively, with inner and outer regions indicating approximately $1\sigma$ and $2\sigma$ credible levels. The black dashed curve shows the best-fit reconstruction, while the green dash-dotted curve shows the best-fit $\Lambda$CDM baseline. Because $\Delta\mathcal{X}(z)$ and $V(z)$ depend on derivatives of the inferred expansion history, their detailed behaviour is especially sensitive to the reconstruction near the high-redshift boundary; consequently, features developing primarily in the extrapolation interval $2.4<z<3.0$ should be interpreted with caution. A change of sign in $\Delta\mathcal{X}$ may be viewed either as an effective phantom-like contribution in a single-field reading, or as the net kinetic balance in a two-field (quintom) realization; the reconstruction itself remains agnostic, since it is performed at the level of background kinematics.
}
    \label{fig:Keff_and_V}
\end{figure*}

This statement is already visible at the level of the directly constrained kinematics shown in~\cref{fig:H_and_q}. In all four dataset combinations, the posterior bands for $H(z)$ and $H(z)/(1+z)$ from the reconstruction and from CPL overlap substantially throughout the redshift range directly populated by the data, with the largest differences appearing only as one approaches the extrapolation-dominated boundary $2.4<z<3.0$. The appendix comparisons show that this behaviour is not specific to CPL: for JBP, Barboza--Alcaniz, exponential, and logarithmic forms, the same broad overlap persists for the directly constrained quantities. Thus, for the observables most directly tied to the geometric information content of the data, the late-time expansion history remains reasonably stable under changes in the smooth low-dimensional functional ansatz assumed for the redshift dependence of the DE EoS parameter.

This stability is already evident in the inferred Hubble constant. For the CC+SN+DESI combination, the EoS-based DE parametrizations span $H_0 \simeq 66.7$--$68.5\,\mathrm{km\,s^{-1}\,Mpc^{-1}}$, while the reconstruction yields $H_0 = 68.05 \pm 1.63\,\mathrm{km\,s^{-1}\,Mpc^{-1}}$. When the local anchor is included (CC+DESI+H0DN), the EoS-based DE models cluster around $H_0 \simeq 70.1$--$72.8\,\mathrm{km\,s^{-1}\,Mpc^{-1}}$, compared with $72.41 \pm 1.71\,\mathrm{km\,s^{-1}\,Mpc^{-1}}$ for the reconstruction. These results show that late-time data define a reasonably stable geometric backbone for the directly constrained background quantities. Importantly, the inferred expansion history does not vary dramatically under changes in the assumed smooth low-dimensional EoS-based DE ansatz; the differences among CPL, JBP, Barboza--Alcaniz, exponential, and logarithmic forms are secondary compared with the common information imposed by the data. In this regime, the prior on the specific functional form is least consequential.

A similar pattern of convergence holds for present-day kinematics. Without supernovae, the CC+DESI combination still allows appreciable method dependence in $q_0$: the reconstruction prefers $q_0<0$, albeit with broad uncertainty, whereas some smooth low-dimensional EoS-based DE parametrizations---most notably the CPL, JBP, Barboza--Alcaniz, exponential, and logarithmic forms---remain compatible with a nearly coasting present Universe. Once supernovae are included, however, the picture tightens substantially, and all models converge on a robustly accelerating Universe. For the CC+SN+DESI combination, the EoS-based DE parametrizations cluster around $q_0 \simeq -0.43$ to $-0.54$, in close agreement with the reconstruction value $q_0=-0.56^{+0.16}_{-0.22}$. This indicates that the low-redshift expansion history is already strongly anchored, and that the more consequential model dependence is pushed away from $z=0$ and into the intermediate-redshift regime.

That intermediate-redshift regime isolates a well-defined window in which the functional prior exerts a quantitatively nontrivial influence, as highlighted in~\cref{tab:q_w_transposed}. At $z=1.7$---which lies safely within the data-supported range and is therefore not controlled by the $z=3$ boundary node---the reconstruction consistently prefers stronger deceleration, with $q(1.7)\simeq 0.56$--0.61. In contrast, all of the smooth low-dimensional EoS-based DE parametrizations cluster at markedly smaller values, $q(1.7)\simeq 0.32$--0.40. Importantly, this is not a CPL-specific effect. The JBP, Barboza--Alcaniz, exponential, and logarithmic forms all reinforce the same qualitative behaviour, and even the $\Lambda$CDM and $w$CDM baselines lie in the same lower-$q(1.7)$ band. The primary contrast, therefore, is not between the reconstruction and one particular parametrization, but between a flexible, model-agnostic reconstruction and the entire class of smooth low-dimensional EoS-based DE parametrizations.

The focus on $z=1.7$ is not merely empirical. In our companion reconstruction analysis, the GR-mapped transition redshift $z_\dagger$ was found to be anticorrelated with $H_0$, with a lower-$z_\dagger$, higher-$H_0$ branch emerging most clearly in several combinations without SN but with an external $H_0$ prior~\cite{Akarsu:2026anp}. In particular, the cases with the strongest high-$H_0$ preference tended to reconstruct transitions around $z_\dagger\sim1.7$, whereas adding SN generally weakened this degeneracy and shifted the inferred transition to somewhat higher redshift~\cite{Akarsu:2026anp}. This is also close to the transition range in which $\Lambda_{\rm s}$CDM-like sign-switching scenarios have been found to be most effective in mitigating multiple late-time tensions~\cite{Akarsu:2019hmw,Akarsu:2021fol,Akarsu:2022typ,Akarsu:2023mfb} (see also Refs. ~\cite{Anchordoqui:2024gfa,Soriano:2025gxd,Akarsu:2024eoo,Gomez-Valent:2024tdb,Gomez-Valent:2024ejh,Akarsu:2025gwi,Ibarra-Uriondo:2026zbp}). The present comparison therefore isolates, at $z=1.7$, not an arbitrary reference point, but a physically suggestive redshift at which the class-level kinematical separation between the smooth low-dimensional EoS-based DE parametrizations and the reconstruction becomes especially clear. At the same time, this should not be read as implying that all dataset combinations reconstruct $z_\dagger\simeq1.7$; once SN are included, the inferred sign change in the companion analysis typically shifts to somewhat higher redshift, even though $z=1.7$ remains a particularly discriminating point for the deceleration history.

\cref{tab:tension_qw17_merged} quantifies this discrepancy. The Gaussianized tensions in $q(1.7)$ persist at the $\sim1.8$--$2.9\sigma$ level across all model and dataset combinations. This shows that the mismatch at intermediate redshift is a robust, class-wide feature of the EoS-based DE parametrizations considered here. The fact that the tension persists not only for the two-parameter EoS-based forms but also for the $\Lambda$CDM and $w$CDM baselines further strengthens the interpretation that the relevant distinction is between functional classes of description rather than between individual parametrizations. In this sense, the present analysis sharpens the message beyond a purely CPL-versus-reconstruction comparison.

By contrast, the corresponding tensions in $w_{\rm DE}(1.7)$, though often nominally larger, must be interpreted with considerably greater caution. \cref{tab:q_w_transposed} shows that the EoS-based DE parametrizations cluster around negative and, in several cases, NECB-violating (phantom-like) values of $w_{\rm DE}(1.7)$, whereas the reconstruction yields a broad and strongly non-Gaussian posterior. This is especially striking for the CC+DESI+H0DN combination, in which the reconstructed $w_{\rm DE}(1.7)$ posterior broadens to $-0.81^{+2.61}_{-2.38}$, causing the nominal tensions reported in~\cref{tab:tension_qw17_merged} to collapse to $\lesssim 0.2\sigma$ despite the continued mismatch in $q(1.7)$. The correct inference is therefore not that the discrepancy in $w_{\rm DE}(1.7)$ disappears, but that $w_{\rm DE}(1.7)$ is intrinsically a less stable discriminator once the reconstructed effective DE density becomes small. Accordingly, $q(1.7)$ provides the cleaner and more robust measure of the class-level separation identified in this work.

It is important to distinguish the most robust quantitative result of the present paper from the broader intermediate-redshift phenomenology suggested by the reconstruction. Within the redshift range directly supported by the data, the clearest class-level discriminator is the persistent separation at $z=1.7$, where the reconstruction prefers systematically stronger deceleration than the smooth low-dimensional EoS-based DE parametrizations. At the same time, the reconstructed $q(z)$ profiles also admit, in some cases, hints of a further localized intermediate-redshift feature at $z\gtrsim2$, including a possible additional accelerated-expansion interval with $q(z)<0$, although this lies closer to the boundary of the directly constrained region and must therefore be interpreted more cautiously. This is fully consistent with the companion reconstruction analysis, which showed that some dataset combinations can accommodate a transient additional accelerated phase around $z\sim1.7$--2.3 only at the level of a hint and with non-negligible sensitivity to both the dataset combination and the treatment of the high-redshift boundary~\cite{Akarsu:2026anp}. The broader physical relevance of this possibility is underscored by smooth $\Lambda_{\rm s}$VCDM realizations, in which a finite-width AdS-to-dS transition can generate an additional accelerated-expansion interval around the transition epoch, whereas a genuine super-acceleration phase with $\dot H>0$ requires a stronger condition and is not generic~\cite{Akarsu:2024qsi,Akarsu:2024eoo,Akarsu:2025gwi}. The point of the present paper is therefore not to claim a detection of an intermediate accelerated phase, but to show that the kind of localized intermediate-redshift kinematics from which such a phase could emerge is systematically compressed when projected onto smooth low-dimensional EoS-based DE parametrizations.

\begin{table*}[t!]
\centering
\caption{Marginalized constraints on $q(1.7)$ and $w_{\rm DE}(1.7)$ for the non-parametric node-based reconstruction (Rec.) and the smooth low-dimensional EoS-based DE parametrizations considered in this work, for each dataset combination. Quoted uncertainties correspond to $68\%$ credible intervals. For the reconstruction, $w_{\rm DE}(1.7)$ denotes the effective DE EoS parameter inferred from the reconstructed expansion history under the GR background mapping.}
\label{tab:q_w_transposed}

\footnotesize
\setlength{\tabcolsep}{3.0pt}
\renewcommand{\arraystretch}{1.08}

\resizebox{\textwidth}{!}{%
\begin{tabular}{@{}llcccccccc@{}}
\toprule
\textbf{Dataset} & \textbf{Parameter}
& \textbf{Rec.} & \textbf{$\Lambda$CDM} & \textbf{$w$CDM} & \textbf{CPL} & \textbf{JBP} & \textbf{Barb.\ Alc.} & \textbf{Exp.} & \textbf{Log.} \\
\midrule

\multirow{2}{*}{\makecell[l]{\textbf{CC+DESI}}}
& $q(1.7)$
& $0.589^{+0.090}_{-0.11}$
& $0.340^{+0.0056}_{-0.0056}$
& $0.324^{+0.020}_{-0.020}$
& $0.350^{+0.024}_{-0.024}$
& $0.335^{+0.020}_{-0.020}$
& $0.360^{+0.029}_{-0.029}$
& $0.347^{+0.026}_{-0.022}$
& $0.359^{+0.030}_{-0.030}$ \\
& $w_{\rm DE}(1.7)$
& $0.18^{+0.27}_{-0.34}$
& $-1$
& $-0.939^{+0.071}_{-0.071}$
& $-1.53^{+0.32}_{-0.32}$
& $-1.01^{+0.09}_{-0.09}$
& $-1.53^{+0.34}_{-0.34}$
& $-1.40^{+0.24}_{-0.33}$
& $-1.8^{+0.49}_{-0.49}$ \\
\midrule

\multirow{2}{*}{\makecell[l]{\textbf{CC+DESI}\\\textbf{+H0DN}}}
& $q(1.7)$
& $0.61^{+0.13}_{-0.11}$
& $0.3422^{+0.0055}_{-0.0055}$
& $0.374^{+0.009}_{-0.009}$
& $0.375^{+0.011}_{-0.016}$
& $0.359^{+0.012}_{-0.016}$
& $0.376^{+0.011}_{-0.018}$
& $0.370^{+0.010}_{-0.016}$
& $0.381^{+0.010}_{-0.016}$ \\
& $w_{\rm DE}(1.7)$
& $-0.81^{+2.61}_{-2.38}$
& $-1$
& $-1.143^{+0.037}_{-0.037}$
& $-1.10^{+0.26}_{-0.21}$
& $-1.064^{+0.08}_{-0.05}$
& $-1.13^{+0.26}_{-0.19}$
& $-1.03^{+0.23}_{-0.18}$
& $-1.25^{+0.32}_{-0.25}$ \\
\midrule

\multirow{2}{*}{\makecell[l]{\textbf{CC+SN+DESI}}}
& $q(1.7)$
& $0.565^{+0.078}_{-0.10}$
& $0.3440^{+0.0049}_{-0.0049}$
& $0.321^{+0.014}_{-0.013}$
& $0.334^{+0.018}_{-0.019}$
& $0.329^{+0.021}_{-0.021}$
& $0.337^{+0.021}_{-0.024}$
& $0.332^{+0.020}_{-0.020}$
& $0.336^{+0.018}_{-0.023}$ \\
& $w_{\rm DE}(1.7)$
& $0.08^{+0.36}_{-0.27}$
& $-1$
& $-0.926^{+0.038}_{-0.038}$
& $-1.06^{+0.17}_{-0.16}$
& $-0.963^{+0.083}_{-0.083}$
& $-1.08^{+0.19}_{-0.17}$
& $-1.02^{+0.16}_{-0.16}$
& $-1.13^{+0.26}_{-0.23}$ \\
\midrule

\multirow{2}{*}{\makecell[l]{\textbf{CC+SN+DESI}\\\textbf{+H0DN}}}
& $q(1.7)$
& $0.560^{+0.080}_{-0.10}$
& $0.3457^{+0.0048}_{-0.0048}$
& $0.362^{+0.008}_{-0.008}$
& $0.391^{+0.015}_{-0.015}$
& $0.384^{+0.013}_{-0.013}$
& $0.395^{+0.015}_{-0.015}$
& $0.387^{+0.014}_{-0.014}$
& $0.391^{+0.015}_{-0.015}$ \\
& $w_{\rm DE}(1.7)$
& $0.07^{+0.37}_{-0.31}$
& $-1$
& $-1.060^{+0.027}_{-0.027}$
& $-1.48^{+0.18}_{-0.15}$
& $-1.19^{+0.075}_{-0.075}$
& $-1.49^{+0.19}_{-0.15}$
& $-1.39^{+0.15}_{-0.13}$
& $-1.62^{+0.23}_{-0.19}$ \\
\bottomrule
\end{tabular}%
}
\end{table*}

The origin of this behaviour becomes more transparent when the inferred kinematics are mapped onto an effective DE sector, as shown in~\cref{fig:rhode_pde_wde}. The resulting pattern closely mirrors the phenomenology of sign-switching and transition-like DE models, such as $\Lambda_{\rm s}$CDM and its smooth realizations, in which a negative intermediate-redshift effective DE density contribution lowers $H(z)$ relative to $\Lambda$CDM and, once the early-Universe distance anchor is maintained, requires a compensating enhancement of the expansion rate at lower redshift, thereby tending to raise the inferred $H_0$~\cite{Akarsu:2019hmw,Akarsu:2021fol,Akarsu:2022typ,Akarsu:2023mfb,Akarsu:2024qsi,Akarsu:2024eoo,Akarsu:2025gwi}. The present analysis does not identify the reconstruction uniquely with that class, but it does show that the kind of localized kinematics embodied by such models is among the structures that EoS-based DE parametrizations are predisposed to smear out. The reconstruction generically prefers structured intermediate-redshift evolution, characterized by a rapid descent of $\rho_{\rm DE}(z)$ toward very small values and, in some realizations, toward a sign change at $z_\dagger$. By contrast, the EoS-based DE parametrizations---of which CPL serves as a representative benchmark in the main figures---project these same late-time kinematics onto sign-preserving, globally smooth effective densities over most of the constrained interval. Within that positive-density framework, they compensate for the lack of localized structure by shifting the EoS parameter to more negative values, often into the NECB-violating (phantom-like) regime at intermediate redshift. In this sense, the apparent phantom-like behaviour returned by the EoS-based parametrizations is best interpreted not as direct evidence for exotic DE microphysics, but as a compensating projection of a more localized kinematical feature onto a globally smooth low-dimensional manifold.


This interpretation also clarifies how the EoS parameter of the DE, $w_{\rm DE}(z)=p_{\rm DE}(z)/\rho_{\rm DE}(z)$, should be read in~\cref{fig:rhode_pde_wde}. In the reconstruction, the large excursions and fragmented morphology of $w_{\rm DE}(z)$ near the transition are primarily a ratio effect: as $\rho_{\rm DE}(z)\to0$, the ratio becomes ill-conditioned even when the underlying pressure and kinematics remain smooth. The physically robust information is therefore carried by $\rho_{\rm DE}(z)$, $p_{\rm DE}(z)$, and the overall location of the rapid intermediate-redshift descent, rather than by the detailed morphology of $w_{\rm DE}(z)$ in the immediate vicinity of a near-crossing or crossing. Seen in this light, the additional EoS-based DE parametrizations strengthen rather than weaken the main interpretation: they show that a broad class of smooth models responds to the same underlying late-time kinematical preference in essentially the same way, namely by distributing the required flexibility into a smoother evolution, often extending into NECB-violating (phantom-like) regimes within their positive-density framework. More generally, the difficulty is not only that $w_{\rm DE}(z)$ can become singular near $\rho_{\rm DE}=0$, but also that, for sufficiently fast transition-like histories, it can remain observationally close to $-1$ over most of the redshift range while the actual dynamics is carried by the localized evolution of $\rho_{\rm DE}(z)$ itself~\cite{Akarsu:2025gwi,Akarsu:2026anp,Gokcen:2026pkq,Akarsu:2024qsi,Akarsu:2024eoo}.

The scalar-field diagnostics shown in~\cref{fig:Keff_and_V} provide an effective-field interpretation of the same background relations. A central point is that the reconstruction does not merely shift the inferred DE EoS parameter; it produces a localized enhancement of the effective kinetic contribution $\Delta\mathcal{X}(z)$ around $z\simeq 1.7$--2, followed by a rapid turnover as the solution approaches the regime in which $\rho_{\rm DE}$ becomes very small. This localized feature is largely absent from the EoS-based DE parametrizations, which remain comparatively featureless in $\Delta\mathcal{X}(z)$ and instead accommodate the same late-time kinematics by shifting $w_{\rm DE}(z)$ toward more negative values. In the effective scalar-field mapping, this distinction is physically meaningful:~\cref{app:two_scalar} shows that $\rho_{\rm DE}+p_{\rm DE}=2\Delta\mathcal{X}$ and $\dot{\rho}_{\rm DE}=-6H\Delta\mathcal{X}$, so the sign and magnitude of $\Delta\mathcal{X}$ directly control both the NEC character of the effective DE sector and the rate at which $\rho_{\rm DE}(z)$ evolves~\cite{Akarsu:2025gwi,Akarsu:2025dmj,Akarsu:2026anp,Adil:2026kfn}.

\begin{table*}[t!]
\centering
\caption{Gaussianized tensions, expressed in units of $\sigma$, between each parametric model and the non-parametric node-based reconstruction (Rec.) at $z=1.7$, for each dataset combination. For each model column $i$, the two rows in each dataset block report $T_{q,i}$ and $T_{w_{\rm DE},i}$, defined by
$T_{x,i}\equiv |x_i-x_{\rm Rec.}|/\sqrt{\sigma_i^2+\sigma_{\rm Rec.}^2}$ for $x\in\{q(1.7),w_{\rm DE}(1.7)\}$, where the uncertainties are symmetrized $68\%$ credible intervals, $\sigma=(\sigma_+ + \sigma_-)/2$.}
\label{tab:tension_qw17_merged}

\footnotesize
\setlength{\tabcolsep}{3.0pt}
\renewcommand{\arraystretch}{1.08}

\resizebox{0.7\textwidth}{!}{%
\begin{tabular}{@{}llccccccc@{}}
\toprule
\textbf{Dataset} & \textbf{Parameter}
& \textbf{$\Lambda$CDM} & \textbf{$w$CDM} & \textbf{CPL} & \textbf{JBP} & \textbf{Barb.\ Alc.} & \textbf{Exp.} & \textbf{Log.} \\
\midrule

\multirow{2}{*}{\makecell[l]{\textbf{CC+DESI}}}
& $T_q$
& $2.49$ & $2.60$ & $2.32$ & $2.49$ & $2.20$ & $2.35$ & $2.20$ \\
& $T_{w_{\rm DE}}$
& $3.87$ & $3.55$ & $3.87$ & $3.74$ & $3.74$ & $3.79$ & $3.43$ \\
\midrule

\multirow{2}{*}{\makecell[l]{\textbf{CC+DESI}\\\textbf{+H0DN}}}
& $T_q$
& $2.10$ & $1.79$ & $1.95$ & $2.08$ & $1.94$ & $1.99$ & $1.90$ \\
& $T_{w_{\rm DE}}$
& $0.08$ & $0.14$ & $0.12$ & $0.10$ & $0.13$ & $0.09$ & $0.18$ \\
\midrule

\multirow{2}{*}{\makecell[l]{\textbf{CC+SN+DESI}}}
& $T_q$
& $2.67$ & $2.86$ & $2.54$ & $2.58$ & $2.48$ & $2.55$ & $2.51$ \\
& $T_{w_{\rm DE}}$
& $3.20$ & $3.16$ & $3.21$ & $3.20$ & $3.20$ & $3.11$ & $3.03$ \\
\midrule

\multirow{2}{*}{\makecell[l]{\textbf{CC+SN+DESI}\\\textbf{+H0DN}}}
& $T_q$
& $2.38$ & $2.19$ & $1.85$ & $1.94$ & $1.81$ & $1.90$ & $1.85$ \\
& $T_{w_{\rm DE}}$
& $3.15$ & $3.31$ & $4.10$ & $3.62$ & $4.10$ & $3.97$ & $4.23$ \\
\bottomrule
\end{tabular}%
}
\end{table*}

In a two-field reading, the evolution---and, where present, the sign change---of $\Delta\mathcal{X}$ may be interpreted as a shift in the balance between canonical and phantom sectors, whereas in an effective single-field reading it simply signals a transient episode in which the kinetic sector becomes dynamically non-negligible; the reconstruction itself remains agnostic between these interpretations. The localized positive excursion in $\Delta\mathcal{X}$ around $z\simeq1.7$--2 therefore provides the scalar-sector counterpart of the enhanced deceleration seen in $q(z)$ and of the rapid intermediate-redshift descent of $\rho_{\rm DE}(z)$. This point is further corroborated by the appendix comparisons: the JBP, Barboza--Alcaniz, exponential, and logarithmic forms all remain much smoother in $\Delta\mathcal{X}(z)$ and do not reproduce the same localized kinetic activation preferred by the reconstruction.

The reconstructed effective potential $V(z)$ likewise develops more structure than do the EoS-based DE parametrizations. Most clearly in the SN-including cases, one observes a mild but recurrent non-monotonic evolution at $z\lesssim2$, followed by a steep rise toward higher redshift as the reconstruction approaches the near-crossing regime. The appendix comparisons confirm that this contrast is not specific to CPL: across the EoS-based family, the scalar-sector evolution remains much smoother and does not reproduce the localized kinetic transition preferred by the reconstruction. At the same time, because both $\Delta\mathcal{X}(z)$ and $V(z)$ are reconstructed from derivatives of the inferred expansion history, their detailed amplitudes and fine structure are necessarily more sensitive than $H(z)$, $q(z)$, or $\rho_{\rm DE}(z)$ to the treatment of the high-redshift boundary. Accordingly, the most robust scalar-field statement is not the detailed shape of the reconstructed effective potential, but the existence and approximate redshift location of a localized intermediate-redshift kinetic transition that is present in the reconstruction and systematically absent from the EoS-based DE parametrizations.

The joint behaviour of the best-fit likelihoods and Bayesian evidences follows a familiar Occam pattern. The additional flexibility of the reconstruction is demonstrably useful: it improves the best fit in all four dataset combinations, with $\Delta\chi^2_{\min,\mathrm{Rec.}}\simeq -3.47$, $-6.88$, $-6.61$, and $-7.24$ for CC+DESI, CC+DESI+H0DN, CC+SN+DESI, and CC+SN+DESI+H0DN, respectively. The improvement is especially pronounced once the external $H_0$ anchor is imposed, precisely where the intermediate-redshift kinematical differences become most visible. Yet this extra freedom is not currently demanded by the data in an evidence-based sense: the corresponding $\ln B_{\Lambda\mathrm{CDM},\mathrm{Rec.}}$ values remain large and positive, $\simeq 10.6$--12.2, showing that the Bayesian evidence continues to favor the simpler $\Lambda$CDM baseline once parameter-volume effects are taken into account. While this global parameter-volume penalty strongly favors $\Lambda$CDM, it does not invalidate the localized kinematic preference for stronger deceleration around $z\sim1.7$ exposed by the reconstruction. The correct interpretation is therefore not that the reconstruction has already established a specific new physical mechanism, but that it isolates a well-localized regime in which EoS-based DE parametrizations act as compressive projections of the kinematics still permitted by current late-time data.

The expanded parametric comparison thus strengthens the central message of this study. The discrepancy near $z\simeq1.7$ is not an isolated feature of CPL, but a robust, class-wide property of the smooth low-dimensional EoS-based DE parametrizations considered here. This redshift interval therefore emerges as the critical arena in which future observations will be most decisive in determining whether the localized structure preferred by the reconstruction is physical, prior-driven, or systematic in origin. If it is physical, the enhanced deceleration and the associated rapid evolution in the effective dark energy sector should remain stable under controlled changes of the reconstruction setup and sharpen as high-redshift BAO and SN constraints improve. If, instead, the discrepancy is mainly induced by prior choices or residual systematics, it should weaken as those tests are carried out. Either outcome is informative, because it will determine whether smooth low-dimensional EoS-based DE parametrizations remain adequate effective surrogates for the late-time Universe, or whether forthcoming data begin to require explicitly localized phenomenology in the expansion history.

\section{Conclusions}
\label{sec:conc}

Read together with the companion reconstruction analysis of Ref.~\cite{Akarsu:2026anp}, the present work sharpens a basic but often under-emphasized point in late-time cosmology: the background expansion history can be relatively well constrained even when the inferred DE dynamics remain sensitive to the functional assumptions used to represent them. When late-time data are used to infer $H(z)$ with minimal structural input, the resulting reconstruction of $E(z)=H(z)/H_0$ is broadly consistent, over the redshift range directly populated by CC, SN, and BAO measurements, with the class of smooth low-dimensional EoS-based DE parametrizations considered here. This includes not only the CPL benchmark, but also $w$CDM, JBP, Barboza--Alcaniz, exponential, and logarithmic forms. In this sense, current late-time observations already define a fairly stable geometric backbone, and much of the apparent diversity in inferred DE behaviour is still driven by the adopted functional prior rather than by a uniquely selected physical history.

The value of the present comparison, however, lies not simply in re-fitting $H(z)$, but in isolating where the choice of functional description becomes dynamically consequential. The key discriminants are derivative- and mapping-sensitive quantities, especially the deceleration history $q(z)$ and the effective-fluid reconstruction of $\rho_{\rm DE}(z)$ and $p_{\rm DE}(z)$. In the intermediate-redshift window where the data still have appreciable leverage, all of the smooth low-dimensional EoS-based DE parametrizations considered here yield qualitatively similar predictions, while the reconstruction prefers systematically stronger deceleration. Around $z\simeq1.7$, the reconstruction gives $q(1.7)\simeq0.56$--0.61, whereas the smooth EoS-based models cluster at $q(1.7)\simeq0.32$--0.40. This separation persists across all dataset combinations and translates into a moderate but robust Gaussianized tension, typically at the $\sim2$--$3\sigma$ level. The main lesson is therefore sharper than in a purely CPL-based comparison: a broad family of globally smooth low-dimensional parametrizations of the DE EoS parameter can reproduce the directly constrained background expansion while still systematically compressing localized intermediate-redshift kinematical structure permitted by the data. In particular, abrupt or sufficiently fast sign-switching realizations show that the essential dynamics may reside in the density sign reversal itself, with $w_{\rm DE}(z)$ remaining identically or observationally close to $-1$ away from a narrow transition layer; this is precisely the kind of behaviour that standard smooth low-dimensional EoS-based DE parametrizations are not designed to capture. Seen in this light, the present results also strengthen the methodological motivation for exploring transition-like and sign-switching DE sectors, such as $\Lambda_{\rm s}$CDM and related realizations, not because the reconstruction has uniquely selected them, but because they explicitly embody the kind of localized late-time kinematics that smooth low-dimensional EoS-based DE parametrizations tend to compress~\cite{Akarsu:2019hmw,Akarsu:2021fol,Akarsu:2022typ,Akarsu:2023mfb,Akarsu:2024qsi,Akarsu:2024eoo,Akarsu:2025gwi,Akarsu:2026anp}.

The enlarged comparison also clarifies how phantom-like behaviour should be interpreted. For the EoS-based DE parametrizations studied here, the effective DE density is sign-preserving by construction, so the condition $w_{\rm DE}(1.7)<-1$ is equivalent to $\rho_{\rm DE}+p_{\rm DE}<0$ and therefore to phantom-like behaviour in the NECB sense~\cite{Akarsu:2025gwi,Akarsu:2026anp,Gokcen:2026pkq}. The fact that this tendency appears across several distinct ans\"atze should not, however, be over-read as direct evidence for a specific DE microphysics. Rather, it is more naturally understood as a compensating projection onto a smooth low-dimensional functional manifold when the underlying kinematics favour a more localized feature. By contrast, the reconstruction accommodates the same late-time kinematical preference through a rapid intermediate-redshift descent of $\rho_{\rm DE}(z)$ toward very small values and, in some realizations, toward a sign change, without requiring a globally smooth $w_{\rm DE}<-1$ history. Once $\rho_{\rm DE}(z)$ is allowed to become small or change sign, the physically meaningful discriminator is the NECB, $\rho_{\rm DE}+p_{\rm DE}=0$, rather than the line $w_{\rm DE}=-1$ taken in isolation. Near such a regime, the ratio $w_{\rm DE}=p_{\rm DE}/\rho_{\rm DE}$ becomes ill-conditioned, and large excursions in $w_{\rm DE}(z)$ may be largely algebraic rather than physically revealing. Accordingly, the more robust quantities for inter-model comparison are $\rho_{\rm DE}(z)$, $p_{\rm DE}(z)$, $\rho_{\rm DE}(z)+p_{\rm DE}(z)$, and the deceleration history $q(z)$, rather than the detailed morphology of $w_{\rm DE}(z)$ in the immediate vicinity of a near-crossing or crossing. In particular, the present analysis suggests that $q(1.7)$ is a cleaner discriminator than $w_{\rm DE}(1.7)$, whose reconstructed posterior can remain broad and strongly non-Gaussian.

Finally, the joint behaviour of $\Delta\chi^2_{\min}$ and the Bayesian evidence places these results in their proper statistical context. Additional functional freedom is demonstrably useful: it improves the best fit and exposes where smooth low-dimensional EoS-based DE parametrizations compress structure that is still permitted by current late-time data. Yet that extra freedom is not, at present, demanded by the data once the associated parameter-volume penalty is taken into account. The practical implication is that the intermediate-redshift window isolated here, $z\sim1.5$--2, should be regarded as the critical arena for future tests. If the deceleration discrepancy identified in this work reflects genuine structure in the expansion history, it should persist under controlled changes of the reconstruction setup and sharpen as BAO and SN constraints improve in the same redshift range. If, instead, it is driven primarily by prior choices or residual systematics, it should weaken under such tests. Either outcome is informative, because it will determine whether smooth low-dimensional EoS-based DE parametrizations remain adequate effective surrogates for late-time data, or whether forthcoming observations begin to require explicitly localized phenomenology in the expansion history. Read together with the companion reconstruction analysis of Ref.~\cite{Akarsu:2026anp}, the present results suggest not that current data have already established a sign-switching DE sector, but rather that derivative-sensitive intermediate-redshift inferences remain substantially more fragile to functional priors than the directly measured expansion history itself.

\newpage

\begin{acknowledgments}
Project BridgingCosmology is financed by Xjenza Malta and the Scientific and Technological Research Council of T\"{U}B\.{I}TAK, through the Xjenza Malta--T\"{U}B\.{I}TAK 2024 Joint Call for R\&I projects.
\"{O}.A.\ acknowledges support from the Turkish Academy of Sciences through the Outstanding Young Scientist Award programme (T\"{U}BA-GEB\.{I}P).
This work was supported by T\"{U}B\.{I}TAK under Grant No.~124N627.
M.C.\ and L.E.\ acknowledge support from T\"{U}B\.{I}TAK through postdoctoral researcher fellowships associated with Grant No.~124N627.
The authors thank T\"{U}B\.{I}TAK for their support.
This article is based upon work from COST Action CA21136 \emph{Addressing observational tensions in cosmology with systematics and fundamental physics} (CosmoVerse), supported by COST (European Cooperation in Science and Technology).
This initiative is part of the PRIMA Programme supported by the European Union.
\end{acknowledgments}

\appendix
\crefalias{section}{appendix}
\section{Two scalar field dark energy}
\label{app:two_scalar}

In this appendix we summarize the effective two-field framework underlying the scalar-sector interpretation used in~\cref{sec:discussion}. We consider a dark energy sector described by two minimally coupled scalar fields: a canonical (quintessence-like) field $Q$ and a phantom field $P$. At the level of the homogeneous background, more general collections of canonical and phantom fields can always be represented by effective single canonical and phantom degrees of freedom along the background trajectory. The $(Q,P)$ system is therefore the minimal setup that can accommodate both NECB-satisfying and NECB-violating dark energy phases within a single effective description. In particular, a single minimally coupled scalar field with a fixed-sign kinetic term cannot change the sign of its kinetic contribution, and hence cannot smoothly cross the dark-sector null-energy-condition boundary (NECB).

For generality the scalar potential may be an arbitrary function of the two fields. For notational simplicity we restrict to a separable form,
\begin{equation}
V(Q,P)=V_1(Q)+V_2(P),
\end{equation}
so that the potential-sector contributions are explicit while the fields remain coupled gravitationally through the Hubble rate $H$. The action is
\begin{equation}
\begin{aligned}
\mathcal{S}=\mathcal{S}_{\rm m}+\int \dd^4x\,\sqrt{-g}\Bigg[
\frac{R}{2\kappa^2}
-\frac{1}{2}(\nabla Q)^2 - V_1(Q)
 \\ 
+\frac{1}{2}(\nabla P)^2 - V_2(P)
\Bigg].
\end{aligned}
\end{equation}
In a spatially flat FLRW background the corresponding field equations are
\begin{equation}
\begin{aligned}
3H^2 = \kappa^2 \Big[
\rho_{\rm m}
+\frac{1}{2}\big(\dot Q^2-\dot P^2\big)
+V_1(Q)+V_2(P)
\Big],
\end{aligned}
\end{equation}
\begin{equation}
\begin{aligned}
3H^2+2\dot H = \kappa^2 \Big[
- p_{\rm m}
-\frac{1}{2}\big(\dot Q^2-\dot P^2\big)
+V_1(Q)+V_2(P)
\Big],
\end{aligned}
\end{equation}
while the Klein--Gordon equations read
\begin{equation}
\ddot Q+3H\dot Q=-V_{1,Q},
\qquad
\ddot P+3H\dot P=V_{2,P}.
\end{equation}

It is convenient to define the total effective potential and the effective kinetic contribution as
\begin{equation}
\label{eq:potential_kinetic_twoscalar}
V \equiv V_1(Q)+V_2(P),
\qquad
\Delta\mathcal{X}\equiv \frac{1}{2}\big(\dot Q^2-\dot P^2\big).
\end{equation}
In terms of these quantities, the dark energy density and pressure are
\begin{equation}
\rho_{\rm DE}=\Delta\mathcal{X}+V,
\qquad
p_{\rm DE}=\Delta\mathcal{X}-V,
\end{equation}
so that
\begin{equation}
\rho_{\rm DE}+p_{\rm DE}=2\Delta\mathcal{X},
\qquad
w_{\rm DE}=\frac{p_{\rm DE}}{\rho_{\rm DE}}
=-1+\frac{2\Delta\mathcal{X}}{\rho_{\rm DE}}.
\end{equation}
These relations make clear that $\Delta\mathcal{X}$ controls the NECB character of the effective dark energy sector: $\Delta\mathcal{X}>0$ corresponds to canonical dominance and NECB satisfaction, whereas $\Delta\mathcal{X}<0$ corresponds to phantom dominance and NECB violation.

Using the Friedmann equations, one obtains the background identities
\begin{gather}
\kappa^2 V
=
3H^2+\dot H-\frac{\kappa^2}{2}\big(\rho_{\rm m}-p_{\rm m}\big),
\\
\kappa^2 \Delta\mathcal{X}
=
-\dot H-\frac{\kappa^2}{2}\big(\rho_{\rm m}+p_{\rm m}\big).
\end{gather}
These are precisely the relations used in the main text to reconstruct the effective potential and kinetic sector from the inferred background expansion under the GR mapping. In particular, $\Delta\mathcal{X}$ is determined directly by $\dot H$ and by the matter-sector contribution.

The dark energy continuity equation takes the form
\begin{equation}
\dot\rho_{\rm DE}
=
-3H\big(\rho_{\rm DE}+p_{\rm DE}\big)
=
-6H\Delta\mathcal{X}.
\end{equation}
Thus, in an expanding universe, the sign of $\Delta\mathcal{X}$ directly controls the direction of evolution of the dark energy density: $\Delta\mathcal{X}>0$ implies $\dot\rho_{\rm DE}<0$, while $\Delta\mathcal{X}<0$ implies $\dot\rho_{\rm DE}>0$.

For the potential one finds
\begin{equation}
\dot V
=
-\dot{\Delta\mathcal{X}}-6H\Delta\mathcal{X}
=
-a^{-6}\frac{\dd}{\dd t}\!\left(a^6\Delta\mathcal{X}\right).
\end{equation}
Hence the evolution of the total potential depends on both the magnitude and time variation of the kinetic difference. In a near-plateau regime with $\dot V\simeq0$, one obtains $\Delta\mathcal{X}\propto a^{-6}\to0$ as the universe expands. The dark energy then approaches
\begin{equation}
\rho_{\rm DE}\simeq V,
\qquad
p_{\rm DE}\simeq -V,
\qquad
w_{\rm DE}\simeq -1,
\end{equation}
independently of whether the earlier evolution was temporarily quintessence-like or phantom-like.

Away from $\rho_{\rm DE}=0$, the dark-sector NECB is given by
\begin{equation}
w_{\rm DE}=-1
\quad\Longleftrightarrow\quad
\Delta\mathcal{X}=0.
\end{equation}
Near such a point, one has $\dot V=-\dot{\Delta\mathcal{X}}$ at the boundary, so the sign of $\dot V$ determines which side of the boundary the system enters: $\dot V<0$ drives the evolution toward $\Delta\mathcal{X}>0$ (quintessence-like), whereas $\dot V>0$ drives it toward $\Delta\mathcal{X}<0$ (phantom-like).

Now consider the neighborhood of a zero crossing of the dark energy density, $\rho_{\rm DE}=0$. Since $\rho_{\rm DE}=\Delta\mathcal{X}+V$, the crossing corresponds to
\begin{equation}
V=-\Delta\mathcal{X}.
\end{equation}
Writing
\begin{equation}
\rho_{\rm DE}=s\,\delta,
\qquad
\Delta\mathcal{X}=\sigma\,\xi,
\end{equation}
with $\delta,\xi\ge0$ and $s,\sigma=\pm1$, one obtains
\begin{equation}
w_{\rm DE}
=
-1+2\,\frac{\sigma}{s}\,\frac{\xi}{\delta}.
\end{equation}
The sign of $w_{\rm DE}+1$ is therefore controlled by the ratio $\Delta\mathcal{X}/\rho_{\rm DE}$, and the two-field system admits four branches corresponding to NEC-satisfying or NEC-violating behaviour on both positive- and negative-density sides of the crossing.

As $\delta\to0$, the behaviour of $w_{\rm DE}$ depends on the limiting ratio $\xi/\delta$: it may diverge, approach a finite constant, or tend to $-1$. Accordingly, a $\rho_{\rm DE}=0$ crossing need not correspond to any singularity in the underlying stress-energy sector; rather, it signals that the ratio $w_{\rm DE}=p_{\rm DE}/\rho_{\rm DE}$ has become an ill-conditioned diagnostic. The physically meaningful quantities remain $\rho_{\rm DE}$, $p_{\rm DE}$, $\Delta\mathcal{X}$, and $V$. This is why, in the main text, large excursions in the reconstructed $w_{\rm DE}(z)$ near the transition should be interpreted as a ratio effect rather than as a singularity in the underlying expansion history.

In the context of the present work, the reconstructed $\Delta\mathcal{X}(z)$ and $V(z)$ shown in~\cref{fig:Keff_and_V} should therefore be read in this effective sense. A localized excursion in $\Delta\mathcal{X}(z)$ around $z\sim1.7$--2 corresponds to a temporary activation of the kinetic sector, while the apparent fragmentation or divergence of $w_{\rm DE}(z)$ as $\rho_{\rm DE}(z)$ becomes small reflects the breakdown of the EoS parameter as a faithful diagnostic, not a pathology of the background dynamics. This illustrates the richer and more flexible behaviour allowed by a two-field effective description compared with a single minimally coupled scalar field.

\newpage

\section{Comparison between EoS-based DE Parametrizations and the Model-Agnostic Reconstruction}
\label{app:reconstructions}

This appendix presents the corresponding comparisons between the non-parametric node-based reconstruction and the JBP, Barboza--Alcaniz, exponential, and logarithmic EoS-based DE parametrizations for the full CC+SN+DESI+H0DN dataset combination. The panel layout follows the same logic as in the main text, showing the kinematical quantities $H(z)$, the conformal Hubble parameter $H(z)/(1+z)$, and $q(z)$, the effective-fluid variables $\rho_{\rm DE}(z)$, $p_{\rm DE}(z)$, and $w_{\rm DE}(z)$, and the scalar-sector diagnostics $\Delta\mathcal{X}(z)$ and $V(z)$. In all panels, the cool (blue--green) and warm (yellow--red) shaded bands denote the EoS-based parametrization and the reconstruction, respectively, while the black dashed and green dash-dotted curves indicate the best-fit reconstruction and the best-fit $\Lambda$CDM baseline.

These figures confirm that the qualitative picture discussed in~\cref{sec:discussion} is not peculiar to CPL. For all four alternative smooth low-dimensional EoS-based DE parametrizations, the directly constrained quantities $H(z)$ and $H(z)/(1+z)$ remain broadly consistent with the reconstruction, whereas the derivative- and mapping-sensitive quantities continue to display the same class-wide pattern: weaker intermediate-redshift deceleration and smoother effective-fluid and scalar-sector evolution than in the reconstruction. To avoid visual redundancy, we show only the full CC+SN+DESI+H0DN combination here; the quantitative constraints for all dataset combinations are reported in~\cref{tab:results2}.

\clearpage

\begin{figure}[t]
    \centering
    \includegraphics[width=0.49\linewidth]{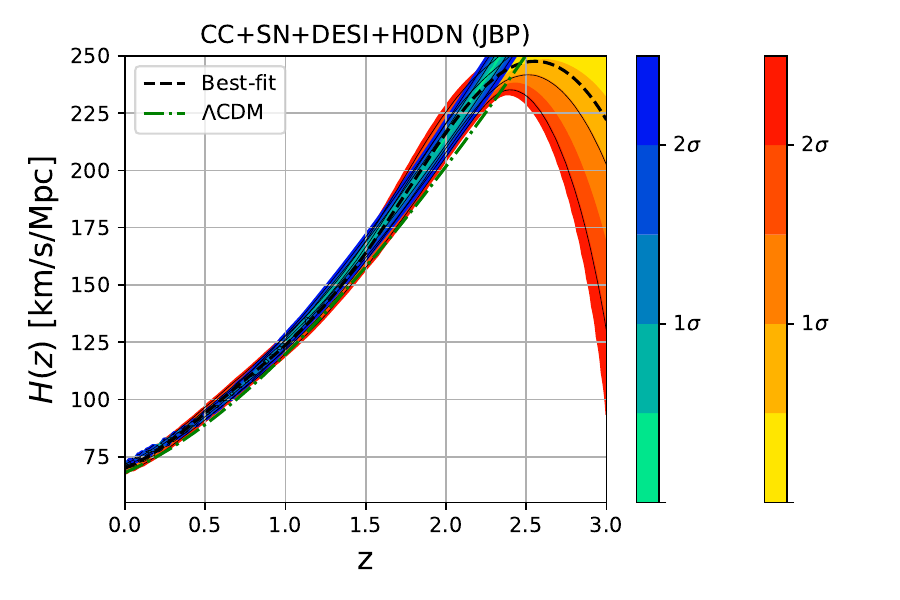}\hfill
    \includegraphics[width=0.49\linewidth]{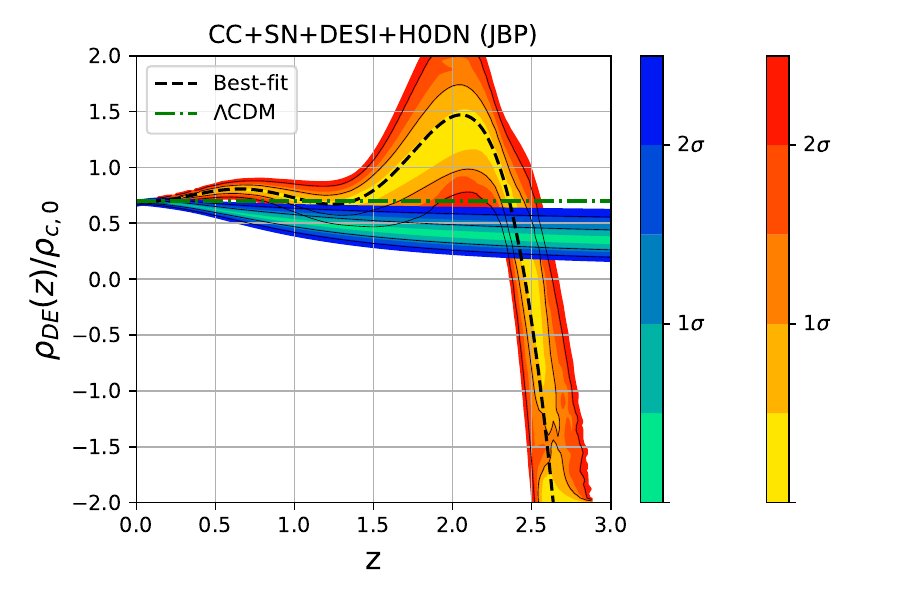}

    \includegraphics[width=0.49\linewidth]{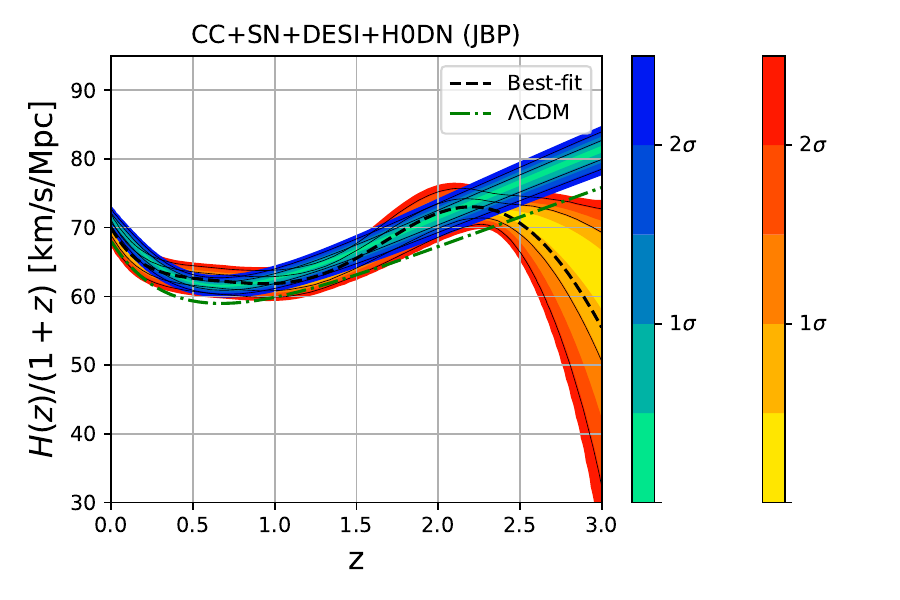}\hfill
    \includegraphics[width=0.49\linewidth]{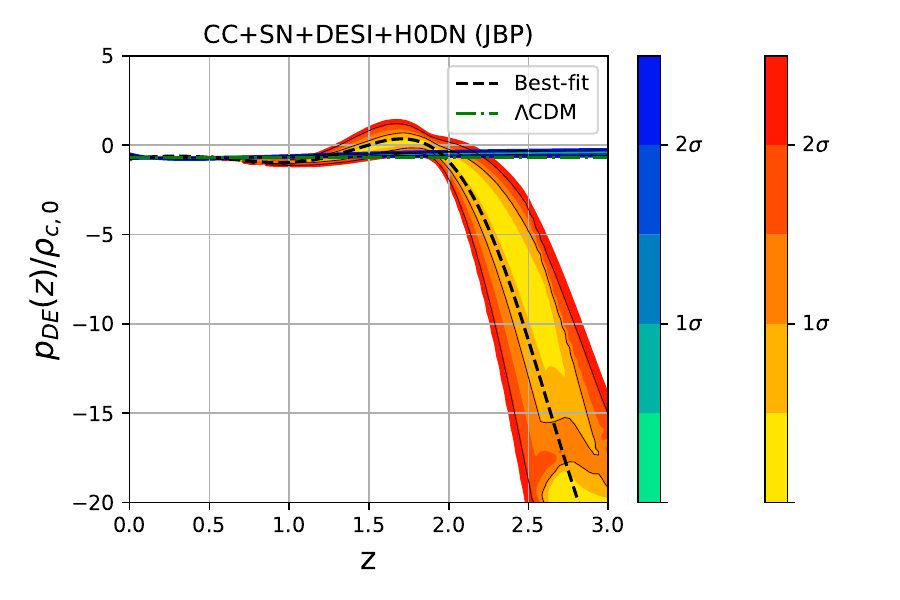}

    \includegraphics[width=0.49\linewidth]{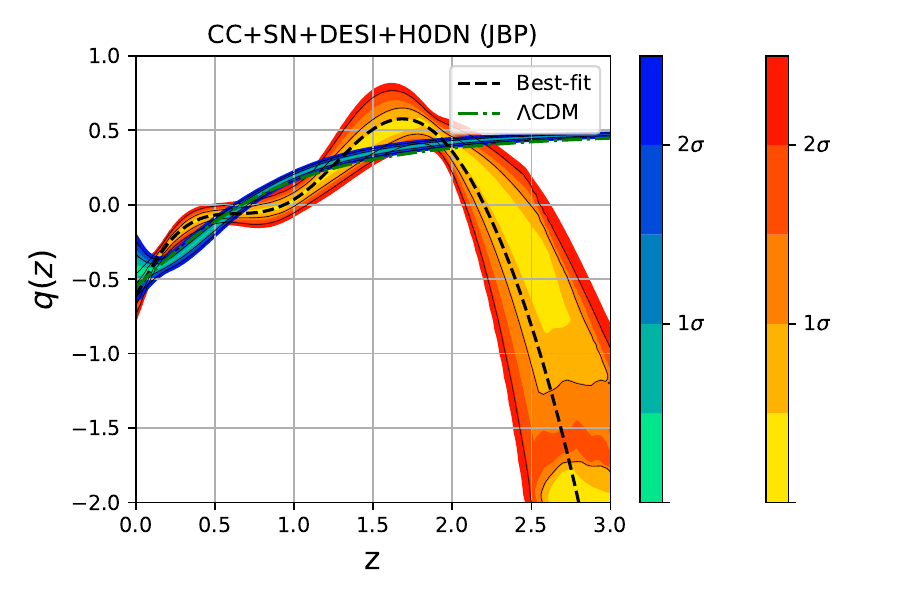}\hfill
    \includegraphics[width=0.49\linewidth]{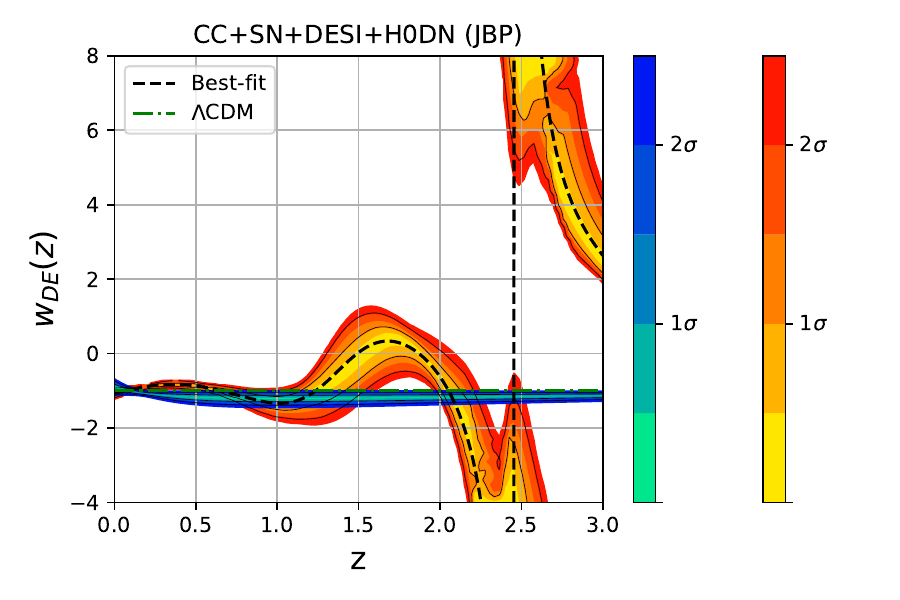}

    \includegraphics[width=0.49\linewidth]{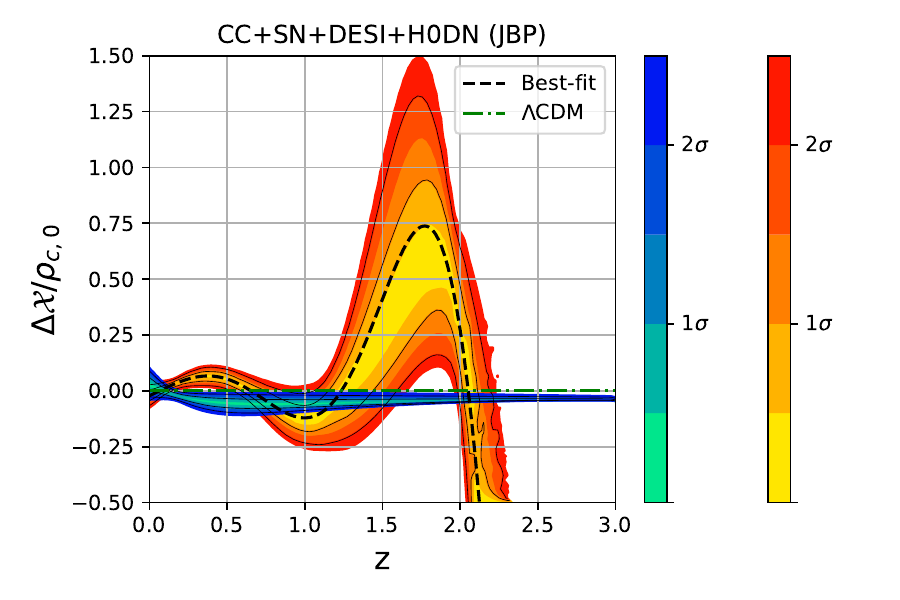}\hfill
    \includegraphics[width=0.49\linewidth]{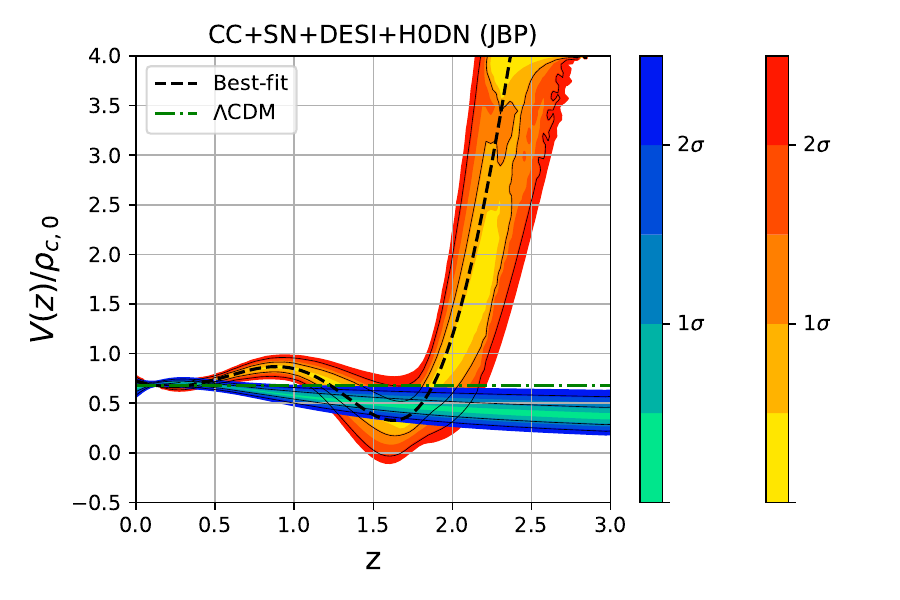}
\caption{
Posterior predictive regions for the CC+SN+DESI+H0DN dataset combination obtained from the non-parametric node-based reconstruction (Rec.) and the JBP parametrization. The left column, from top to bottom, shows the Hubble parameter $H(z)$, the conformal Hubble parameter $H(z)/(1+z)$, the deceleration parameter $q(z)$, and the effective kinetic contribution $\Delta\mathcal{X}/\rho_{c,0}$; the right column shows the effective DE density $\rho_{\rm DE}/\rho_{c,0}$, pressure $p_{\rm DE}/\rho_{c,0}$, EoS parameter $w_{\rm DE}(z)$, and effective potential $V(z)/\rho_{c,0}$. The cool (blue--green) and warm (yellow--red) bands correspond, respectively, to the JBP and reconstruction constraints, with inner and outer regions indicating approximately $1\sigma$ and $2\sigma$ credible levels. The black dashed curve shows the best-fit reconstruction, while the green dash-dotted curve shows the best-fit $\Lambda$CDM baseline. As in the main figures, results near $2.4<z<3.0$ should be interpreted with caution, especially for derivative-based quantities.
}
    \label{fig:app_jbp}
\end{figure}

\begin{figure}[H]
    \centering
    \includegraphics[width=0.49\linewidth]{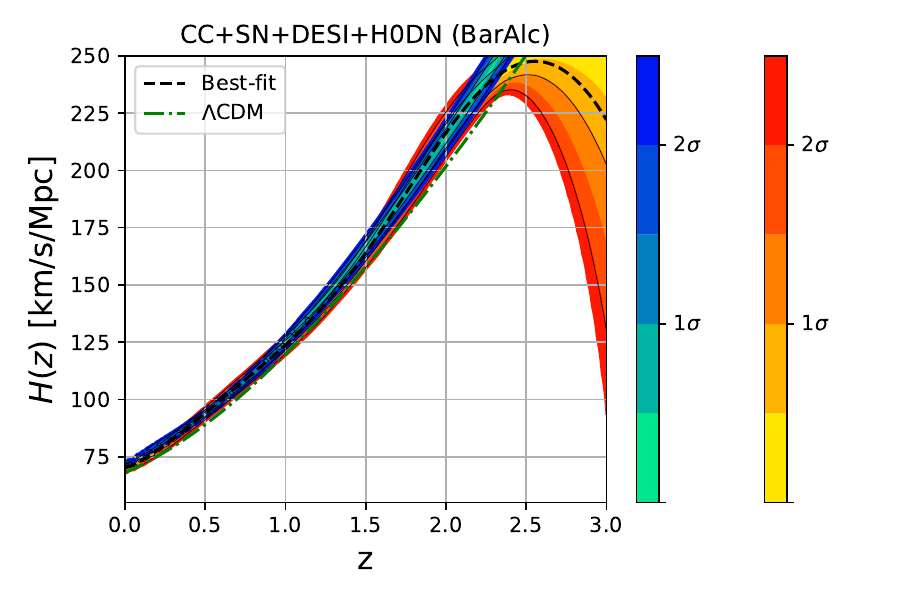}\hfill
    \includegraphics[width=0.49\linewidth]{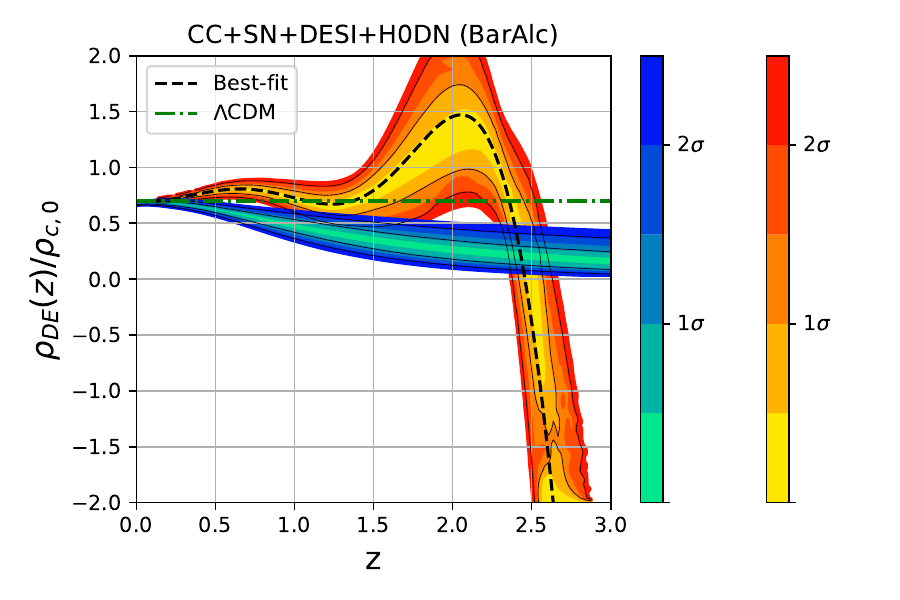}

    \includegraphics[width=0.49\linewidth]{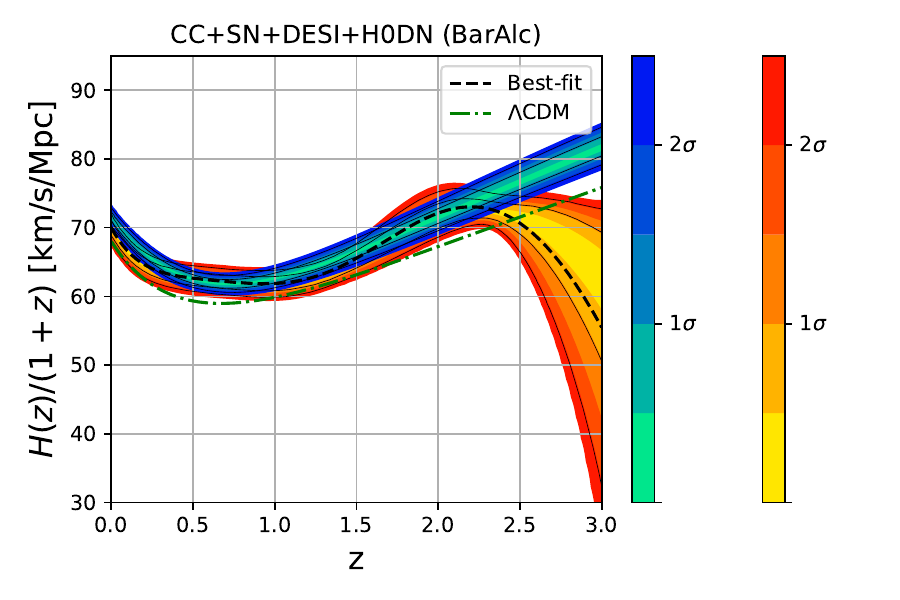}\hfill
    \includegraphics[width=0.49\linewidth]{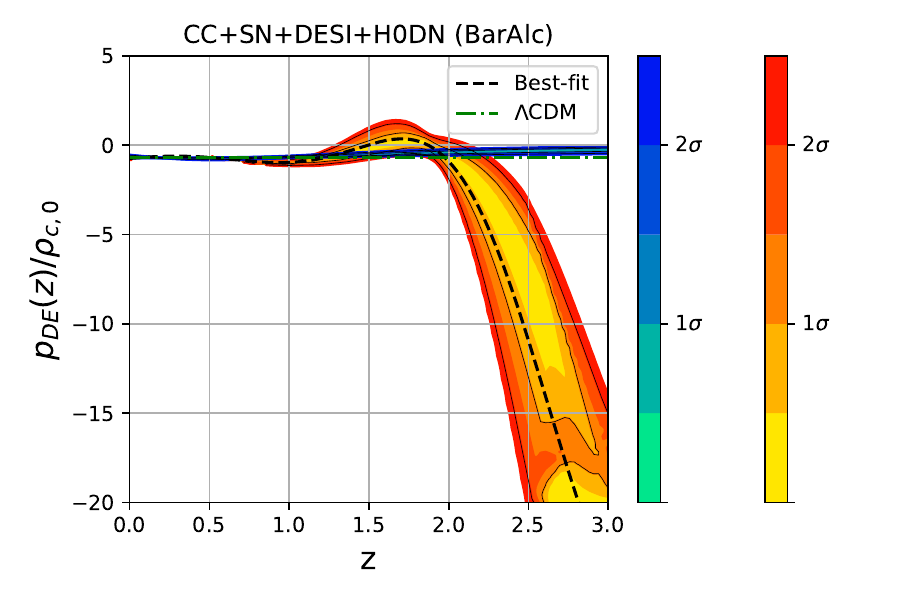}

    \includegraphics[width=0.49\linewidth]{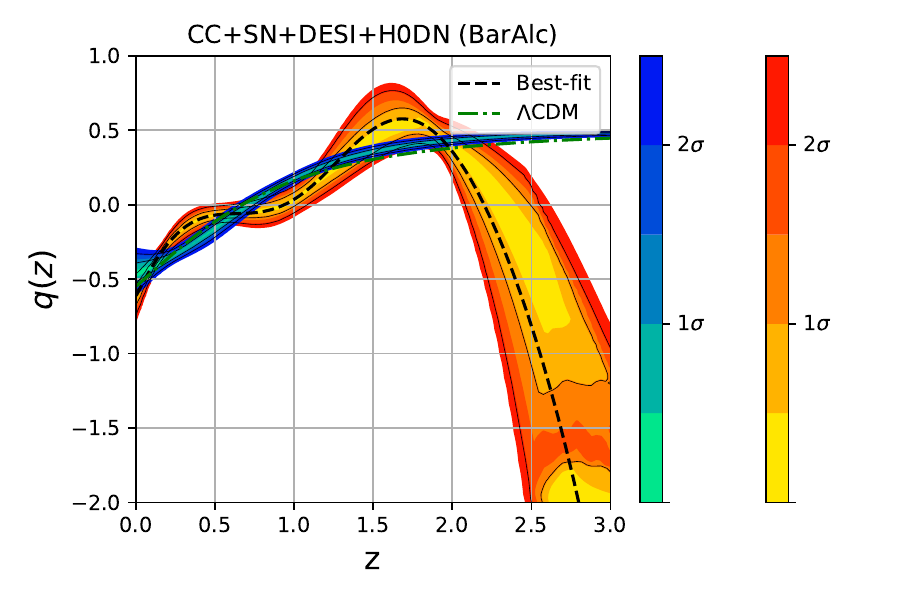}\hfill
    \includegraphics[width=0.49\linewidth]{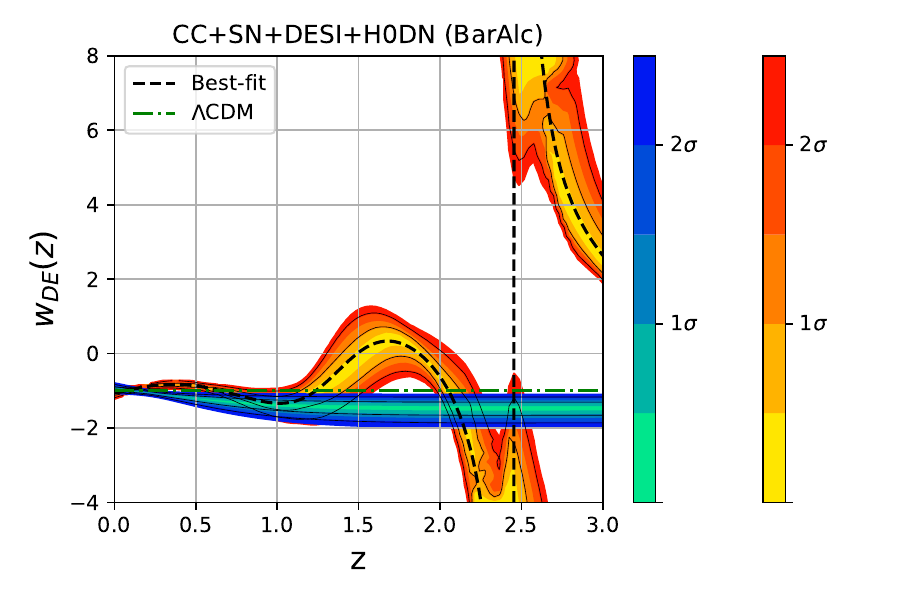}

    \includegraphics[width=0.49\linewidth]{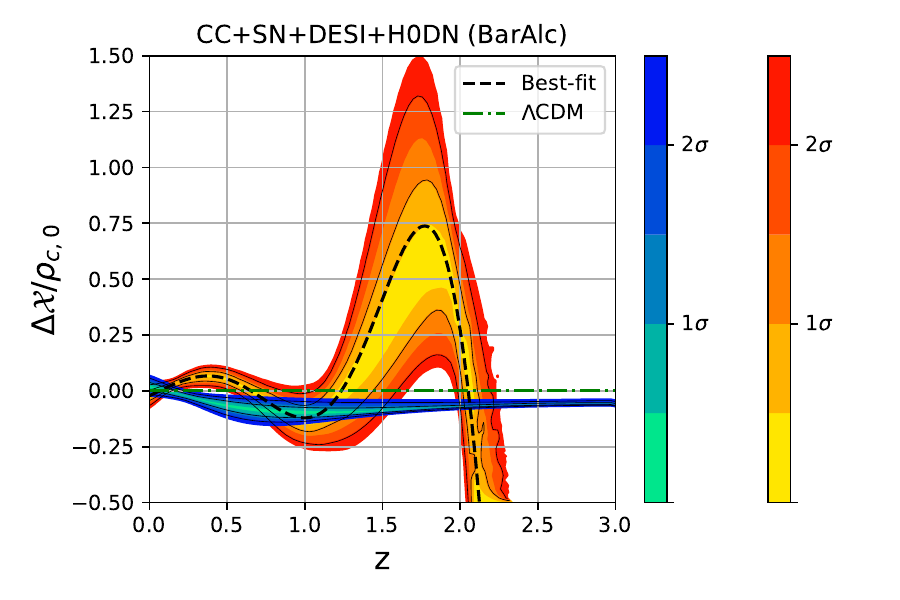}\hfill
    \includegraphics[width=0.49\linewidth]{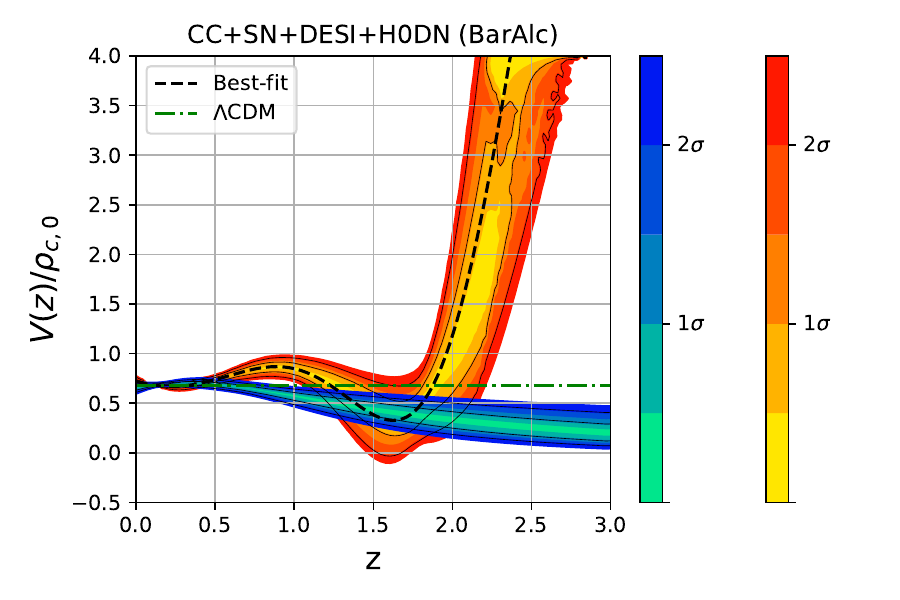}
\caption{
Posterior predictive regions for the CC+SN+DESI+H0DN dataset combination obtained from the non-parametric node-based reconstruction (Rec.) and the Barboza--Alcaniz parametrization. The left column, from top to bottom, shows the Hubble parameter $H(z)$, the conformal Hubble parameter $H(z)/(1+z)$, the deceleration parameter $q(z)$, and the effective kinetic contribution $\Delta\mathcal{X}/\rho_{c,0}$; the right column shows the effective DE density $\rho_{\rm DE}/\rho_{c,0}$, pressure $p_{\rm DE}/\rho_{c,0}$, EoS parameter $w_{\rm DE}(z)$, and effective potential $V(z)/\rho_{c,0}$. The cool (blue--green) and warm (yellow--red) bands correspond, respectively, to the Barboza--Alcaniz and reconstruction constraints, with inner and outer regions indicating approximately $1\sigma$ and $2\sigma$ credible levels. The black dashed curve shows the best-fit reconstruction, while the green dash-dotted curve shows the best-fit $\Lambda$CDM baseline. As in the main figures, results near $2.4<z<3.0$ should be interpreted with caution, especially for derivative-based quantities.
}
    \label{fig:app_baralc}
\end{figure}

\begin{figure}[H]
    \centering
    \includegraphics[width=0.49\linewidth]{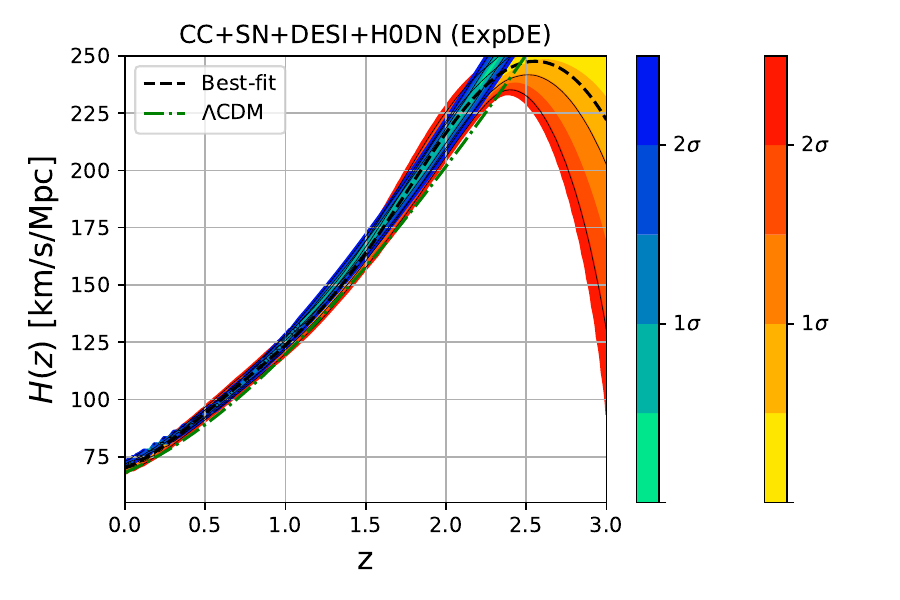}\hfill
    \includegraphics[width=0.49\linewidth]{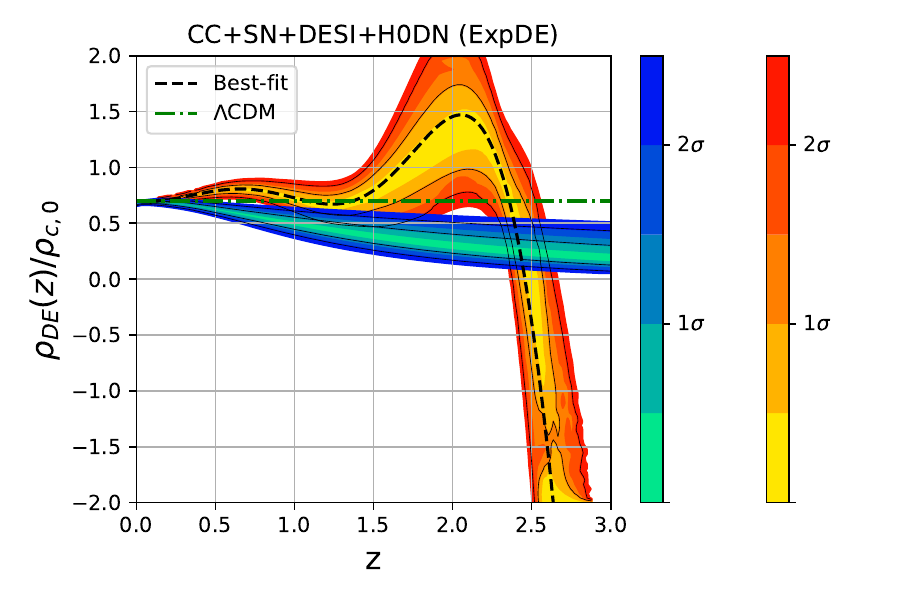}

    \includegraphics[width=0.49\linewidth]{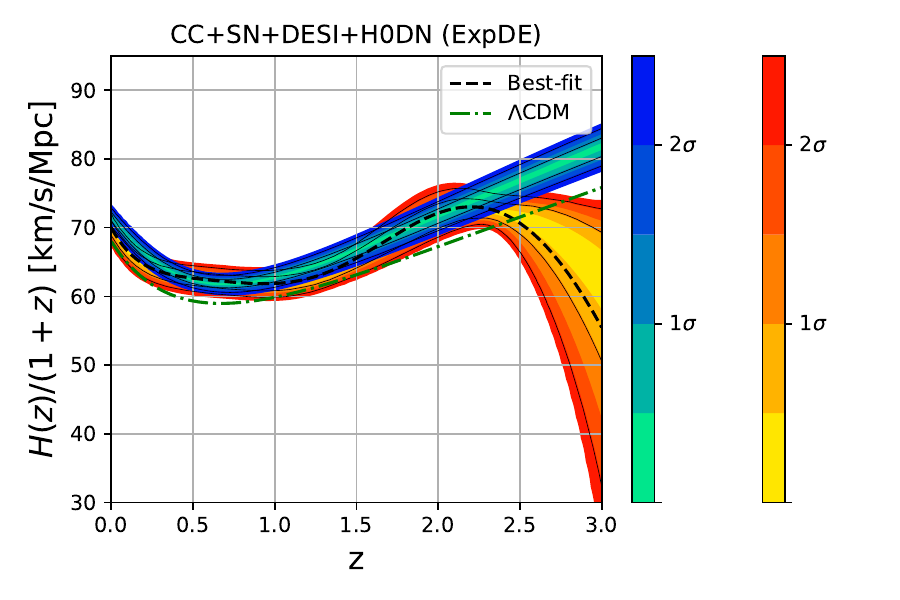}\hfill
    \includegraphics[width=0.49\linewidth]{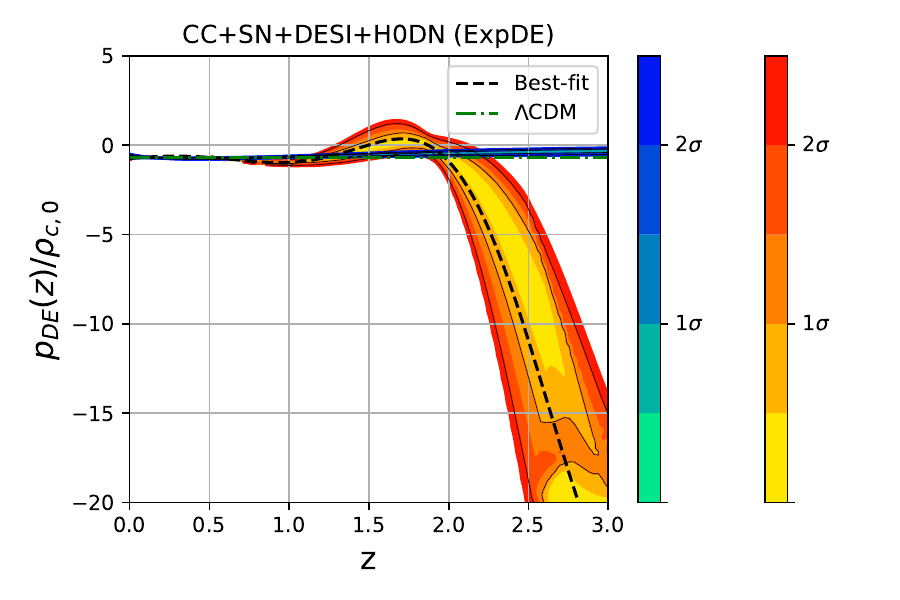}

    \includegraphics[width=0.49\linewidth]{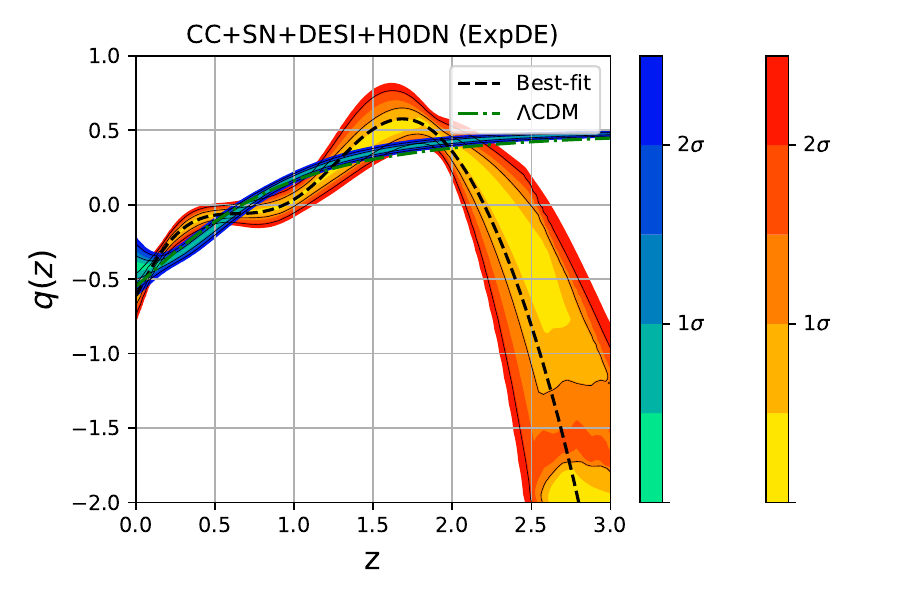}\hfill
    \includegraphics[width=0.49\linewidth]{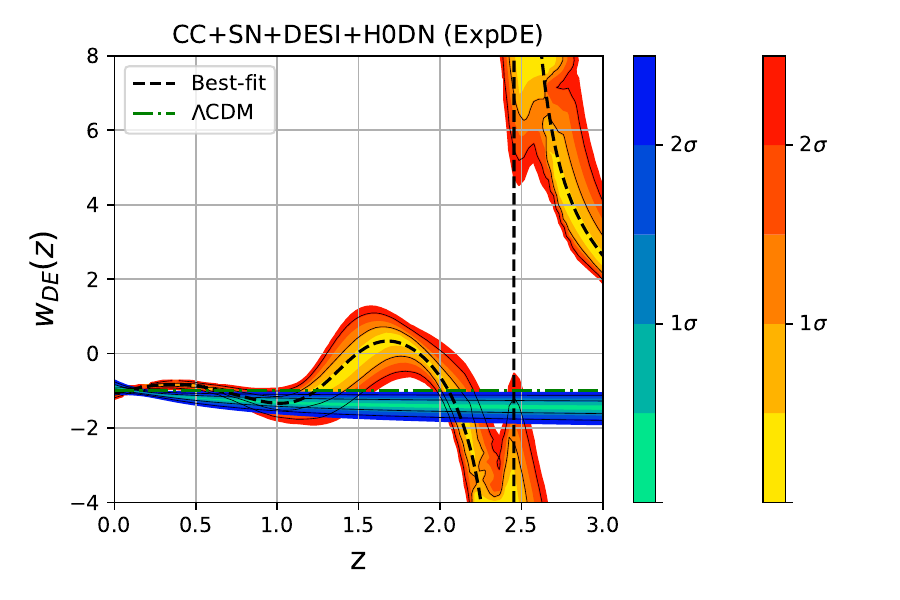}

    \includegraphics[width=0.49\linewidth]{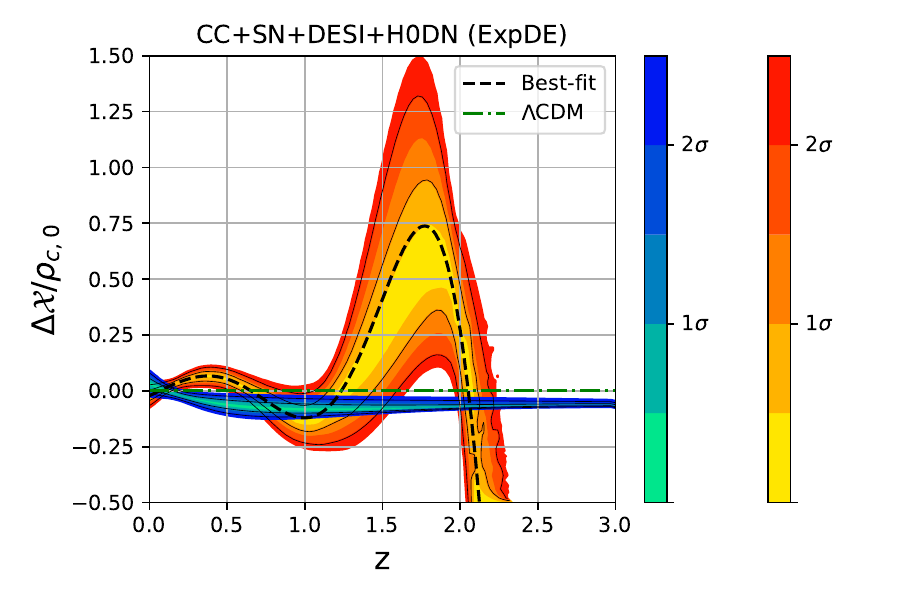}\hfill
    \includegraphics[width=0.49\linewidth]{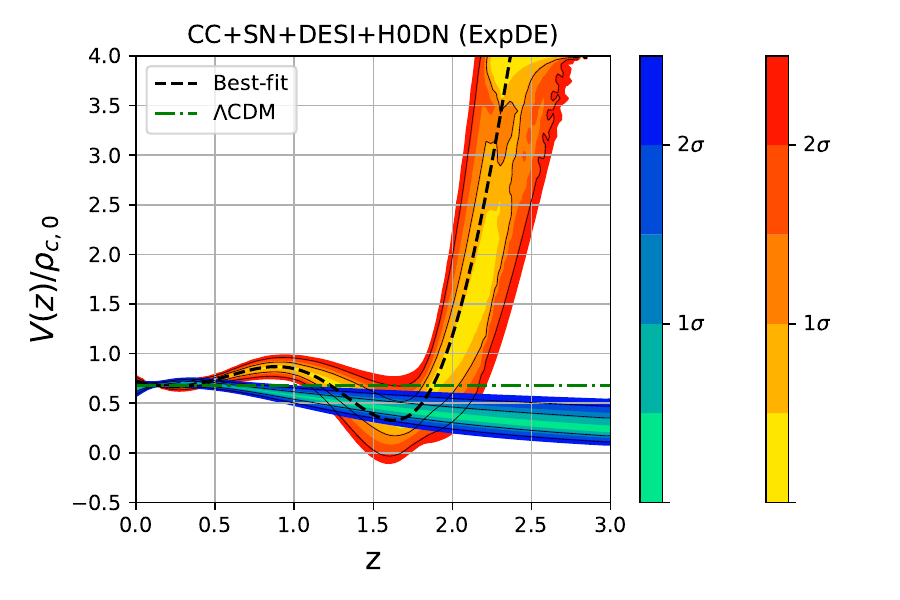}
\caption{
Posterior predictive regions for the CC+SN+DESI+H0DN dataset combination obtained from the non-parametric node-based reconstruction (Rec.) and the exponential parametrization. The left column, from top to bottom, shows the Hubble parameter $H(z)$, the conformal Hubble parameter $H(z)/(1+z)$, the deceleration parameter $q(z)$, and the effective kinetic contribution $\Delta\mathcal{X}/\rho_{c,0}$; the right column shows the effective DE density $\rho_{\rm DE}/\rho_{c,0}$, pressure $p_{\rm DE}/\rho_{c,0}$, EoS parameter $w_{\rm DE}(z)$, and effective potential $V(z)/\rho_{c,0}$. The cool (blue--green) and warm (yellow--red) bands correspond, respectively, to the exponential parametrization and reconstruction constraints, with inner and outer regions indicating approximately $1\sigma$ and $2\sigma$ credible levels. The black dashed curve shows the best-fit reconstruction, while the green dash-dotted curve shows the best-fit $\Lambda$CDM baseline. As in the main figures, results near $2.4<z<3.0$ should be interpreted with caution, especially for derivative-based quantities.
}
    \label{fig:app_exp}
\end{figure}

\begin{figure}[H]
    \centering
    \includegraphics[width=0.49\linewidth]{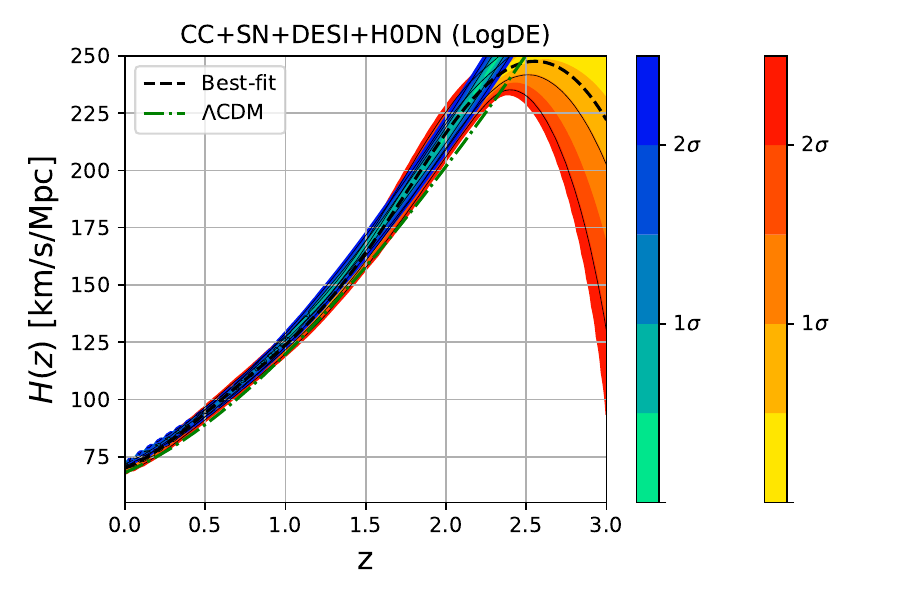}\hfill
    \includegraphics[width=0.49\linewidth]{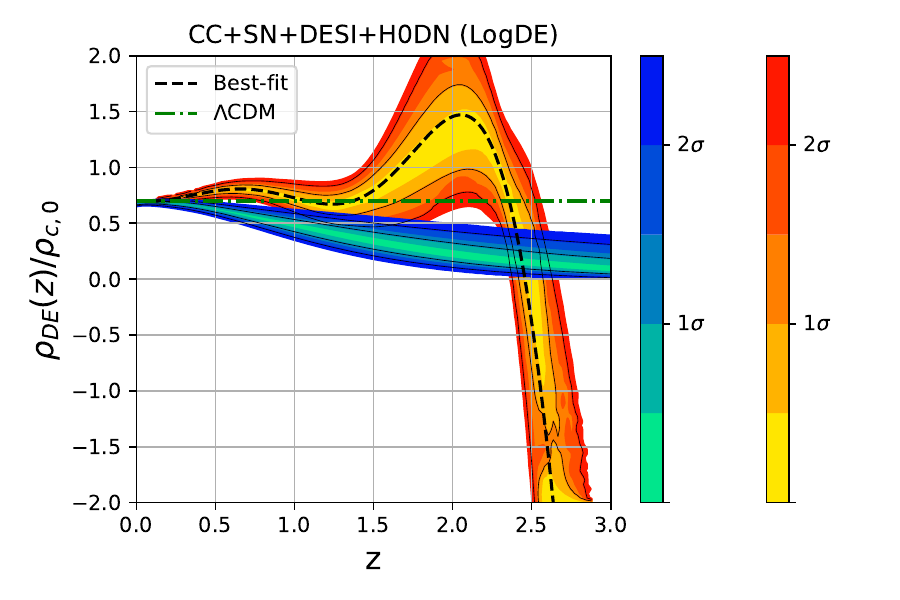}

    \includegraphics[width=0.49\linewidth]{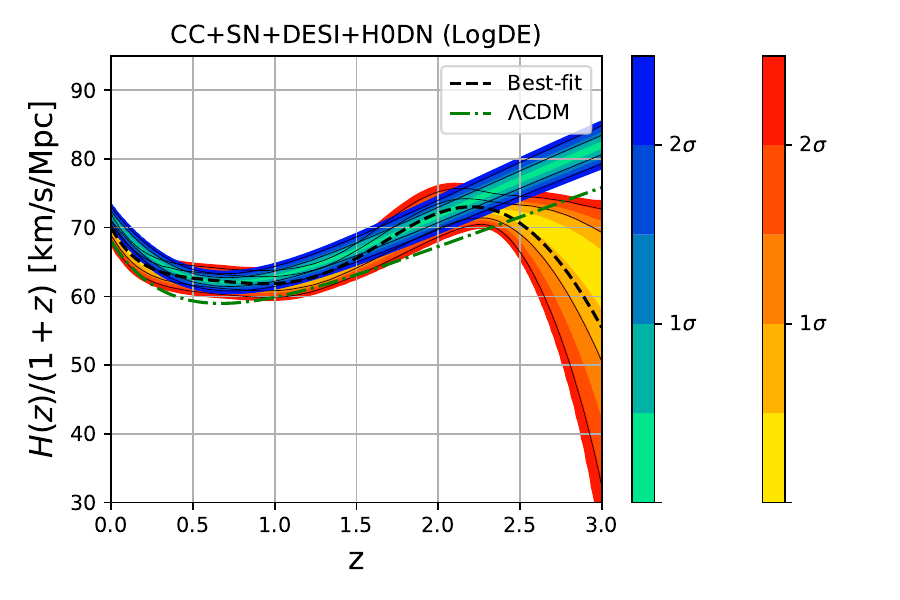}\hfill
    \includegraphics[width=0.49\linewidth]{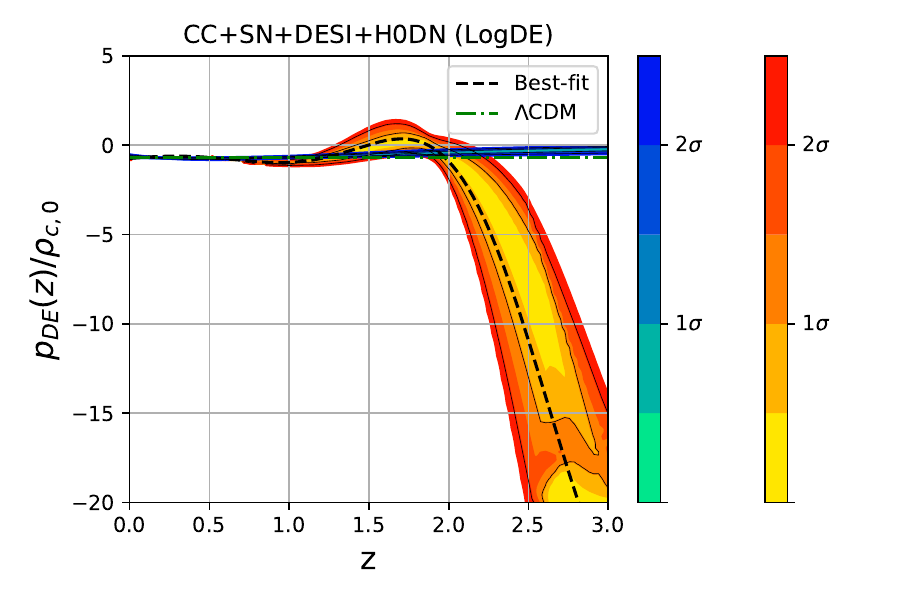}

    \includegraphics[width=0.49\linewidth]{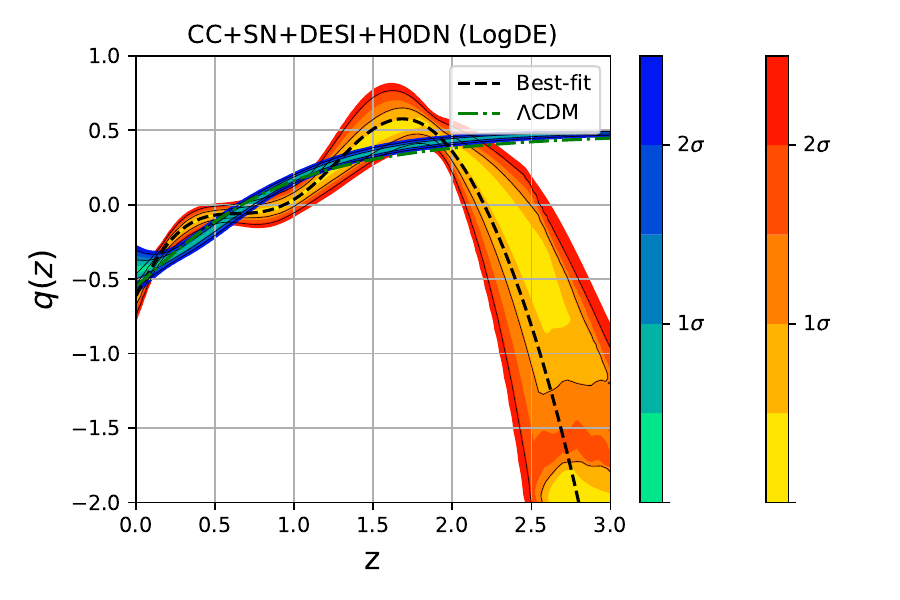}\hfill
    \includegraphics[width=0.49\linewidth]{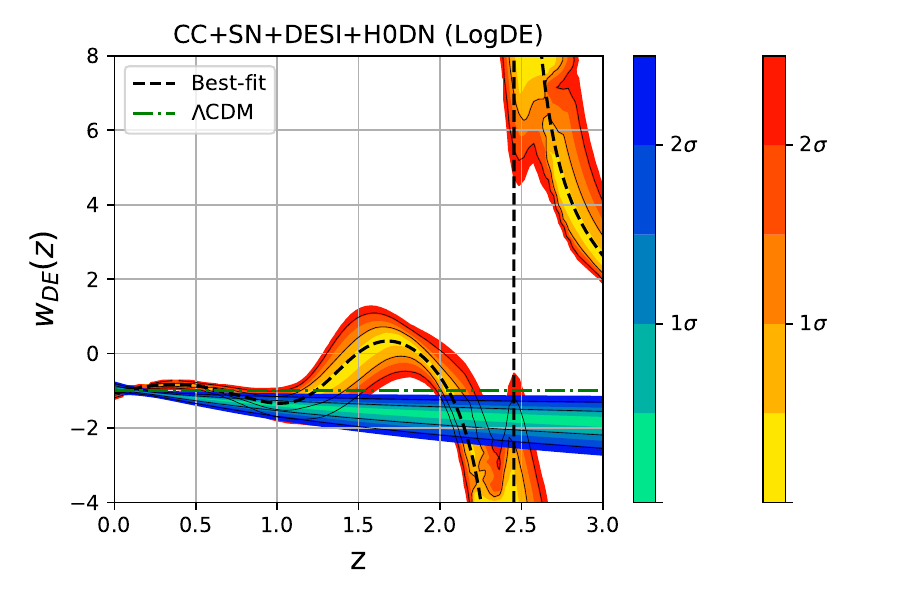}

    \includegraphics[width=0.49\linewidth]{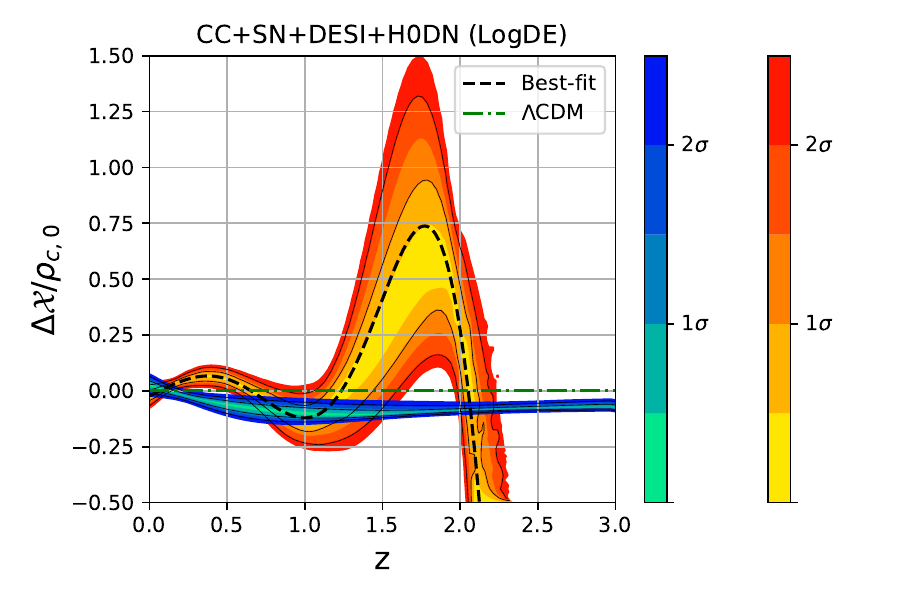}\hfill
    \includegraphics[width=0.49\linewidth]{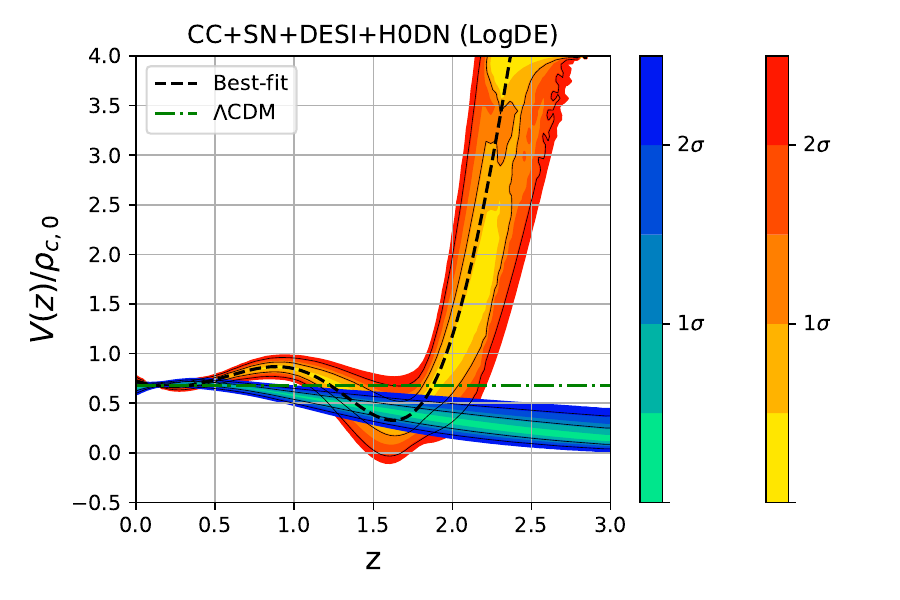}
\caption{
Posterior predictive regions for the CC+SN+DESI+H0DN dataset combination obtained from the non-parametric node-based reconstruction (Rec.) and the logarithmic parametrization. The left column, from top to bottom, shows the Hubble parameter $H(z)$, the conformal Hubble parameter $H(z)/(1+z)$, the deceleration parameter $q(z)$, and the effective kinetic contribution $\Delta\mathcal{X}/\rho_{c,0}$; the right column shows the effective DE density $\rho_{\rm DE}/\rho_{c,0}$, pressure $p_{\rm DE}/\rho_{c,0}$, EoS parameter $w_{\rm DE}(z)$, and effective potential $V(z)/\rho_{c,0}$. The cool (blue--green) and warm (yellow--red) bands correspond, respectively, to the logarithmic parametrization and reconstruction constraints, with inner and outer regions indicating approximately $1\sigma$ and $2\sigma$ credible levels. The black dashed curve shows the best-fit reconstruction, while the green dash-dotted curve shows the best-fit $\Lambda$CDM baseline. As in the main figures, results near $2.4<z<3.0$ should be interpreted with caution, especially for derivative-based quantities.
}
    \label{fig:app_log}
\end{figure}

\clearpage
 
\bibliography{references}

@preprint{Xu:2026sbw,
    author = {Xu, Tengpeng and Kumar, Suresh and Chen, Yun and Capistrano, Abra{\~a}o J. S. and Akarsu, {\"O}zg{\"u}r},
    title = "{Probing Dynamical Dark Energy with Late-Time Data: Evidence, Tensions, and the Limits of the $w_0w_a$CDM Framework}",
    eprint = "2602.11936",
    archivePrefix = "arXiv",
    primaryClass = "astro-ph.CO",
    month = "2",
    year = "2026"
}

@preprint{Pantos:2026rpe,
    author = "Pantos, Ioannis and Perivolaropoulos, Leandros",
    title = "{On the origin of the BAOtr-DESI tension}",
    eprint = "2604.11106",
    archivePrefix = "arXiv",
    primaryClass = "astro-ph.CO",
    month = "4",
    year = "2026"
}

@article{Yang:2021eud,
    author = "Yang, Weiqiang and Di Valentino, Eleonora and Pan, Supriya and Shafieloo, Arman and Li, Xiaolei",
    title = "{Generalized emergent dark energy model and the Hubble constant tension}",
    eprint = "2103.03815",
    archivePrefix = "arXiv",
    primaryClass = "astro-ph.CO",
    doi = "10.1103/PhysRevD.104.063521",
    journal = "Phys. Rev. D",
    volume = "104",
    number = "6",
    pages = "063521",
    year = "2021"
}

@article{Khalife:2023qbu,
    author = {Khalife, Ali Rida and Zanjani, Maryam Bahrami and Galli, Silvia and G{\"u}nther, Sven and Lesgourgues, Julien and Benabed, Karim},
    title = "{Review of Hubble tension solutions with new SH0ES and SPT-3G data}",
    eprint = "2312.09814",
    archivePrefix = "arXiv",
    primaryClass = "astro-ph.CO",
    reportNumber = "TTK-23-36",
    doi = "10.1088/1475-7516/2024/04/059",
    journal = "JCAP",
    volume = "04",
    pages = "059",
    year = "2024"
}

@article{Schoneberg:2021qvd,
    author = {Sch{\"o}neberg, Nils and Franco Abell{\'a}n, Guillermo and P{\'e}rez S{\'a}nchez, Andrea and Witte, Samuel J. and Poulin, Vivian and Lesgourgues, Julien},
    title = "{The H0 Olympics: A fair ranking of proposed models}",
    eprint = "2107.10291",
    archivePrefix = "arXiv",
    primaryClass = "astro-ph.CO",
    doi = "10.1016/j.physrep.2022.07.001",
    journal = "Phys. Rept.",
    volume = "984",
    pages = "1--55",
    year = "2022"
}

@article{Sahni:1999gb,
    author = "Sahni, Varun and Starobinsky, Alexei A.",
    title = "{The Case for a positive cosmological Lambda term}",
    eprint = "astro-ph/9904398",
    archivePrefix = "arXiv",
    reportNumber = "IUCAA-25-2000",
    doi = "10.1142/S0218271800000542",
    journal = "Int. J. Mod. Phys. D",
    volume = "9",
    pages = "373--444",
    year = "2000"
}

@article{AlbertoVazquez:2012ofj,
    author = "Alberto Vazquez, J. and Bridges, M. and Hobson, M. P. and Lasenby, A. N.",
    title = "{Reconstruction of the Dark Energy equation of state}",
    eprint = "1205.0847",
    archivePrefix = "arXiv",
    primaryClass = "astro-ph.CO",
    doi = "10.1088/1475-7516/2012/09/020",
    journal = "JCAP",
    volume = "09",
    pages = "020",
    year = "2012"
}

@article{Keeley:2020aym,
    author = "Keeley, Ryan E. and Shafieloo, Arman and Zhao, Gong-Bo and Vazquez, Jose Alberto and Koo, Hanwool",
    title = "{Reconstructing the Universe: Testing the Mutual Consistency of the Pantheon and SDSS/eBOSS BAO Data Sets with Gaussian Processes}",
    eprint = "2010.03234",
    archivePrefix = "arXiv",
    primaryClass = "astro-ph.CO",
    doi = "10.3847/1538-3881/abdd2a",
    journal = "Astron. J.",
    volume = "161",
    number = "3",
    pages = "151",
    year = "2021"
}

@article{Zhao:2017cud,
    author = "Zhao, Gong-Bo and others",
    title = "{Dynamical dark energy in light of the latest observations}",
    eprint = "1701.08165",
    archivePrefix = "arXiv",
    primaryClass = "astro-ph.CO",
    doi = "10.1038/s41550-017-0216-z",
    journal = "Nature Astron.",
    volume = "1",
    number = "9",
    pages = "627--632",
    year = "2017"
}

@preprint{Liu:2026rot,
    author = "Liu, Qian-Mo and Zhao, Gong-Bo",
    title = "{Low-redshift-agnostic BAO Constraints on Binned Dark-energy Density Evolution from DESI DR1 and DR2}",
    eprint = "2604.06888",
    archivePrefix = "arXiv",
    primaryClass = "astro-ph.CO",
    month = "4",
    year = "2026"
}

@article{Holsclaw:2011wi,
    author = "Holsclaw, Tracy and Alam, Ujjaini and Sanso, Bruno and Lee, Herbie and Heitmann, Katrin and Habib, Salman and Higdon, David",
    title = "{Nonparametric Reconstruction of the Dark Energy Equation of State from Diverse Data Sets}",
    eprint = "1104.2041",
    archivePrefix = "arXiv",
    primaryClass = "astro-ph.CO",
    reportNumber = "LA-UR-11-01798",
    doi = "10.1103/PhysRevD.84.083501",
    journal = "Phys. Rev. D",
    volume = "84",
    pages = "083501",
    year = "2011"
}

@article{Seikel:2012uu,
    author = "Seikel, Marina and Clarkson, Chris and Smith, Mathew",
    title = "{Reconstruction of dark energy and expansion dynamics using Gaussian processes}",
    eprint = "1204.2832",
    archivePrefix = "arXiv",
    primaryClass = "astro-ph.CO",
    doi = "10.1088/1475-7516/2012/06/036",
    journal = "JCAP",
    volume = "06",
    pages = "036",
    year = "2012"
}

@article{Jesus:2021bxq,
    author = "Jesus, J. F. and Valentim, R. and Escobal, A. A. and Pereira, S. H. and Benndorf, D.",
    title = "{Gaussian processes reconstruction of the dark energy potential}",
    eprint = "2112.09722",
    archivePrefix = "arXiv",
    primaryClass = "astro-ph.CO",
    doi = "10.1088/1475-7516/2022/11/037",
    journal = "JCAP",
    volume = "11",
    pages = "037",
    year = "2022"
}

@article{Elizalde:2022rss,
    author = "Elizalde, Emilio and Khurshudyan, Martiros and Myrzakulov, K. and Bekov, S.",
    title = "{Reconstruction of the Quintessence Dark Energy Potential from a Gaussian Process}",
    eprint = "2203.06767",
    archivePrefix = "arXiv",
    primaryClass = "gr-qc",
    doi = "10.1007/s10511-024-09828-z",
    journal = "Astrophysics",
    volume = "67",
    number = "2",
    pages = "192--214",
    year = "2024"
}

@article{Dinda:2024ktd,
    author = "Dinda, Bikash R. and Maartens, Roy",
    title = "{Model-agnostic assessment of dark energy after DESI DR1 BAO}",
    eprint = "2407.17252",
    archivePrefix = "arXiv",
    primaryClass = "astro-ph.CO",
    doi = "10.1088/1475-7516/2025/01/120",
    journal = "JCAP",
    volume = "01",
    pages = "120",
    year = "2025"
}

@article{Gadbail:2024lek,
    author = "Gadbail, Gaurav N. and Mandal, Sanjay and Sahoo, P. K. and Bamba, Kazuharu",
    title = "{Reconstruction of the scalar field potential in nonmetricity gravity through Gaussian processes}",
    eprint = "2411.00051",
    archivePrefix = "arXiv",
    primaryClass = "gr-qc",
    doi = "10.1016/j.physletb.2024.139232",
    journal = "Phys. Lett. B",
    volume = "860",
    pages = "139232",
    year = "2025"
}

@article{Yang:2025kgc,
    author = "Yang, Yuhang and Wang, Qingqing and Li, Chunyu and Yuan, Peibo and Ren, Xin and Saridakis, Emmanuel N. and Cai, Yi-Fu",
    title = "{Gaussian process reconstructions and model building of quintom dark energy from latest cosmological observations}",
    eprint = "2501.18336",
    archivePrefix = "arXiv",
    primaryClass = "astro-ph.CO",
    doi = "10.1088/1475-7516/2025/08/050",
    journal = "JCAP",
    volume = "08",
    pages = "050",
    year = "2025"
}

@article{You:2025uon,
    author = "You, Changyu and Wang, Dan and Yang, Tao",
    title = "{Dynamical dark energy implies a coupled dark sector: Insights from DESI DR2 via a data-driven approach}",
    eprint = "2504.00985",
    archivePrefix = "arXiv",
    primaryClass = "astro-ph.CO",
    doi = "10.1103/f6v7-n9fr",
    journal = "Phys. Rev. D",
    volume = "112",
    number = "4",
    pages = "043503",
    year = "2025"
}

@article{Mukherjee:2025ytj,
    author = "Mukherjee, Purba and Sen, Anjan A.",
    title = "{New expansion rate anomalies at characteristic redshifts geometrically determined using DESI-DR2 BAO and DES-SN5YR observations}",
    eprint = "2505.19083",
    archivePrefix = "arXiv",
    primaryClass = "astro-ph.CO",
    doi = "10.1088/1361-6633/ae082c",
    journal = "Rept. Prog. Phys.",
    volume = "88",
    number = "9",
    pages = "098401",
    year = "2025"
}

@article{Ye:2024ywg,
    author = "Ye, Gen and Martinelli, Matteo and Hu, Bin and Silvestri, Alessandra",
    title = "{Hints of Nonminimally Coupled Gravity in DESI 2024 Baryon Acoustic Oscillation Measurements}",
    eprint = "2407.15832",
    archivePrefix = "arXiv",
    primaryClass = "astro-ph.CO",
    doi = "10.1103/PhysRevLett.134.181002",
    journal = "Phys. Rev. Lett.",
    volume = "134",
    number = "18",
    pages = "181002",
    year = "2025"
}

@article{Wang:2025vfb,
    author = "Wang, Yun and Freese, Katherine",
    title = "{Model-independent dark energy measurements from DESI DR2 and Planck 2015 data}",
    eprint = "2505.17415",
    archivePrefix = "arXiv",
    primaryClass = "astro-ph.CO",
    doi = "10.1088/1475-7516/2026/02/023",
    journal = "JCAP",
    volume = "02",
    pages = "023",
    year = "2026"
}

@article{Ormondroyd:2025exu,
    author = "Ormondroyd, A. N. and Handley, W. J. and Hobson, M. P. and Lasenby, A. N.",
    title = "{Non-parametric reconstructions of dynamical dark energy via flexknots}",
    eprint = "2503.08658",
    archivePrefix = "arXiv",
    primaryClass = "astro-ph.CO",
    doi = "10.1093/mnras/staf1144",
    journal = "Mon. Not. Roy. Astron. Soc.",
    volume = "541",
    number = "4",
    pages = "3388--3400",
    year = "2025"
}

@article{Berti:2025phi,
    author = "Berti, Maria and Bellini, Emilio and Bonvin, Camille and Kunz, Martin and Viel, Matteo and Zumalacarregui, Miguel",
    title = "{Reconstructing the dark energy density in light of DESI BAO observations}",
    eprint = "2503.13198",
    archivePrefix = "arXiv",
    primaryClass = "astro-ph.CO",
    doi = "10.1103/dj3k-84v4",
    journal = "Phys. Rev. D",
    volume = "112",
    number = "2",
    pages = "023518",
    year = "2025"
}

@article{Shafieloo:2007cs,
    author = "Shafieloo, Arman",
    title = "{Model Independent Reconstruction of the Expansion History of the Universe and the Properties of Dark Energy}",
    eprint = "astro-ph/0703034",
    archivePrefix = "arXiv",
    doi = "10.1111/j.1365-2966.2007.12175.x",
    journal = "Mon. Not. Roy. Astron. Soc.",
    volume = "380",
    pages = "1573--1580",
    year = "2007"
}

@article{Mitra:2024ahj,
    author = "Mitra, Ayan and G{\'o}mez-Vargas, Isidro and Zarikas, Vasilios",
    title = "{Dark energy reconstruction analysis with artificial neural networks: Application on simulated Supernova Ia data from Rubin Observatory}",
    eprint = "2402.18124",
    archivePrefix = "arXiv",
    primaryClass = "astro-ph.CO",
    doi = "10.1016/j.dark.2024.101706",
    journal = "Phys. Dark Univ.",
    volume = "46",
    pages = "101706",
    year = "2024"
}

@article{Huterer:2002hy,
    author = "Huterer, Dragan and Starkman, Glenn",
    title = "{Parameterization of dark-energy properties: A Principal-component approach}",
    eprint = "astro-ph/0207517",
    archivePrefix = "arXiv",
    reportNumber = "CWRU-07-02",
    doi = "10.1103/PhysRevLett.90.031301",
    journal = "Phys. Rev. Lett.",
    volume = "90",
    pages = "031301",
    year = "2003"
}

@article{Crittenden:2005wj,
    author = "Crittenden, Robert G. and Pogosian, Levon and Zhao, Gong-Bo",
    title = "{Investigating dark energy experiments with principal components}",
    eprint = "astro-ph/0510293",
    archivePrefix = "arXiv",
    doi = "10.1088/1475-7516/2009/12/025",
    journal = "JCAP",
    volume = "12",
    pages = "025",
    year = "2009"
}

@article{Albrecht:2006mqt,
    author = "Albrecht, Andreas and Bernstein, Gary",
    title = "{Evaluating dark energy probes using multidimensional dark energy parameters}",
    eprint = "astro-ph/0608269",
    archivePrefix = "arXiv",
    doi = "10.1103/PhysRevD.75.103003",
    journal = "Phys. Rev. D",
    volume = "75",
    pages = "103003",
    year = "2007"
}

@article{Zhao:2012aw,
    author = "Zhao, Gong-Bo and Crittenden, Robert G. and Pogosian, Levon and Zhang, Xinmin",
    title = "{Examining the evidence for dynamical dark energy}",
    eprint = "1207.3804",
    archivePrefix = "arXiv",
    primaryClass = "astro-ph.CO",
    doi = "10.1103/PhysRevLett.109.171301",
    journal = "Phys. Rev. Lett.",
    volume = "109",
    pages = "171301",
    year = "2012"
}

@article{Liu:2015mkm,
    author = "Liu, Zhi-E. and Qin, Hao-Feng and Zhang, Jie and Zhang, Tong-Jie and Yu, Hao-Ran",
    title = "{Reconstructing equation of state of dark energy with principal component analysis}",
    eprint = "1501.02971",
    archivePrefix = "arXiv",
    primaryClass = "astro-ph.CO",
    doi = "10.1016/j.dark.2019.100379",
    journal = "Phys. Dark Univ.",
    volume = "26",
    pages = "100379",
    year = "2019"
}

@article{Raveri:2017qvt,
    author = "Raveri, Marco and Bull, Philip and Silvestri, Alessandra and Pogosian, Levon",
    title = "{Priors on the effective Dark Energy equation of state in scalar-tensor theories}",
    eprint = "1703.05297",
    archivePrefix = "arXiv",
    primaryClass = "astro-ph.CO",
    doi = "10.1103/PhysRevD.96.083509",
    journal = "Phys. Rev. D",
    volume = "96",
    number = "8",
    pages = "083509",
    year = "2017"
}

@article{Dai:2018zwv,
    author = "Dai, Ji-Ping and Yang, Yang and Xia, Jun-Qing",
    title = "{Reconstruction of the Dark Energy Equation of State from the Latest Observations}",
    doi = "10.3847/1538-4357/aab49a",
    journal = "Astrophys. J.",
    volume = "857",
    number = "1",
    pages = "9",
    year = "2018"
}

@article{Gomez-Valent:2021cbe,
    author = "G{\'o}mez-Valent, Adri{\`a} and Zheng, Ziyang and Amendola, Luca and Pettorino, Valeria and Wetterich, Christof",
    title = "{Early dark energy in the pre- and postrecombination epochs}",
    eprint = "2107.11065",
    archivePrefix = "arXiv",
    primaryClass = "astro-ph.CO",
    doi = "10.1103/PhysRevD.104.083536",
    journal = "Phys. Rev. D",
    volume = "104",
    number = "8",
    pages = "083536",
    year = "2021"
}

@article{Sousa-Neto:2025gpj,
    author = "Sousa-Neto, Agripino and Bengaly, Carlos and Gonzalez, Javier E. and Alcaniz, Jailson",
    title = "{Symbolic regression analysis of dynamical dark energy with DESI-DR2 and SN data}",
    eprint = "2502.10506",
    archivePrefix = "arXiv",
    primaryClass = "astro-ph.CO",
    doi = "10.1016/j.dark.2025.102108",
    journal = "Phys. Dark Univ.",
    volume = "50",
    pages = "102108",
    year = "2025"
}

@preprint{Akarsu:2026anp,
    author = {Akarsu, {\"O}zg{\"u}r and Caruana, Maria and Dialektopoulos, Konstantinos F. and Escamilla, Luis A. and Kahya, Emre O. and Levi Said, Jackson},
    title = "{Hints of sign-changing scalar field energy density and a transient acceleration phase at $z\sim 2$ from model-agnostic reconstructions}",
    eprint = "2602.08928",
    archivePrefix = "arXiv",
    primaryClass = "astro-ph.CO",
    month = "2",
    year = "2026"
}

@preprint{Ibarra-Uriondo:2026zbp,
    author = "Ibarra-Uriondo, Be{\~n}at and Bouhmadi-L{\'o}pez, Mariam",
    title = "{Sign-Switching Dark Energy: Smooth Transitions with Recent DESI DR2 Observations}",
    eprint = "2602.12347",
    archivePrefix = "arXiv",
    primaryClass = "astro-ph.CO",
    month = "2",
    year = "2026"
}

@preprint{Adil:2026kfn,
    author = {Adil, Shahnawaz A. and Zapata, Miguel A. and Akarsu, {\"O}zg{\"u}r and Vazquez, J. Alberto},
    title = "{Background-level reconstruction of scalar-field potentials from dark-energy histories and comparison with analytic potential families}",
    eprint = "2603.14693",
    archivePrefix = "arXiv",
    primaryClass = "astro-ph.CO",
    month = "3",
    year = "2026"
}

@article{Peebles:2002gy,
    author = "Peebles, P. J. E. and Ratra, Bharat",
    editor = "Hsu, Jong-Ping and Fine, D.",
    title = "{The Cosmological Constant and Dark Energy}",
    eprint = "astro-ph/0207347",
    archivePrefix = "arXiv",
    reportNumber = "KSUPT-02-3",
    doi = "10.1103/RevModPhys.75.559",
    journal = "Rev. Mod. Phys.",
    volume = "75",
    pages = "559--606",
    year = "2003"
}

@article{Velazquez:2024aya,
    author = "Vel{\'a}zquez, Jos{\'e} de Jes{\'u}s and Escamilla, Luis A. and Mukherjee, Purba and V{\'a}zquez, J. Alberto",
    title = "{Non-Parametric Reconstruction of Cosmological Observables Using Gaussian Processes Regression}",
    eprint = "2410.02061",
    archivePrefix = "arXiv",
    primaryClass = "astro-ph.CO",
    doi = "10.3390/universe10120464",
    journal = "Universe",
    volume = "10",
    number = "12",
    pages = "464",
    year = "2024"
}

@article{Gerardi:2019obr,
    author = "Gerardi, Francesca and Martinelli, Matteo and Silvestri, Alessandra",
    title = "{Reconstruction of the Dark Energy equation of state from latest data: the impact of theoretical priors}",
    eprint = "1902.09423",
    archivePrefix = "arXiv",
    primaryClass = "astro-ph.CO",
    doi = "10.1088/1475-7516/2019/07/042",
    journal = "JCAP",
    volume = "07",
    pages = "042",
    year = "2019"
}

@article{DESI:2025fii,
    author = "Lodha, K. and others",
    collaboration = "DESI",
    title = "{Extended dark energy analysis using DESI DR2 BAO measurements}",
    eprint = "2503.14743",
    archivePrefix = "arXiv",
    primaryClass = "astro-ph.CO",
    reportNumber = "FERMILAB-PUB-25-0164-PPD",
    doi = "10.1103/w4c6-1r5j",
    journal = "Phys. Rev. D",
    volume = "112",
    number = "8",
    pages = "083511",
    year = "2025"
}

@article{DESI:2025qqy,
    author = "Andrade, U. and others",
    collaboration = "DESI",
    title = "{Validation of the DESI DR2 measurements of baryon acoustic oscillations from galaxies and quasars}",
    eprint = "2503.14742",
    archivePrefix = "arXiv",
    primaryClass = "astro-ph.CO",
    reportNumber = "FERMILAB-PUB-25-0162-PPD",
    doi = "10.1103/kdys-w8vl",
    journal = "Phys. Rev. D",
    volume = "112",
    number = "8",
    pages = "083512",
    year = "2025"
}

@article{Brout:2022vxf,
    author = "Brout, Dillon and others",
    title = "{The Pantheon+ Analysis: Cosmological Constraints}",
    eprint = "2202.04077",
    archivePrefix = "arXiv",
    primaryClass = "astro-ph.CO",
    doi = "10.3847/1538-4357/ac8e04",
    journal = "Astrophys. J.",
    volume = "938",
    number = "2",
    pages = "110",
    year = "2022"
}

@article{Rubin:2023jdq,
    author = "Rubin, David and others",
    title = "{Union Through UNITY: Cosmology with 2,000 SNe Using a Unified Bayesian Framework}",
    eprint = "2311.12098",
    archivePrefix = "arXiv",
    primaryClass = "astro-ph.CO",
    doi = "10.3847/1538-4357/adc0a5",
    journal = "Astrophys. J.",
    volume = "986",
    number = "2",
    pages = "231",
    year = "2025"
}

@preprint{DES:2025sig,
    author = "Popovic, B. and others",
    collaboration = "DES",
    title = "{The Dark Energy Survey Supernova Program: A Reanalysis Of Cosmology Results And Evidence For Evolving Dark Energy With An Updated Type Ia Supernova Calibration}",
    eprint = "2511.07517",
    archivePrefix = "arXiv",
    primaryClass = "astro-ph.CO",
    reportNumber = "FERMILAB-PUB-25-0842-CSAID-PPD",
    month = "11",
    year = "2025"
}

@article{Akarsu:2024qiq,
    author = {Akarsu, {\"O}zg{\"u}r and Colg{\'a}in, Eoin {\'O}. and Sen, Anjan A. and Sheikh-Jabbari, M. M.},
    title = "{{\ensuremath{\Lambda}}CDM Tensions: Localising Missing Physics through Consistency Checks}",
    eprint = "2402.04767",
    archivePrefix = "arXiv",
    primaryClass = "astro-ph.CO",
    doi = "10.3390/universe10080305",
    journal = "Universe",
    volume = "10",
    number = "8",
    pages = "305",
    year = "2024"
}

@article{DiValentino:2022fjm,
    author = "Di Valentino, Eleonora",
    title = "{Challenges of the Standard Cosmological Model}",
    doi = "10.3390/universe8080399",
    journal = "Universe",
    volume = "8",
    number = "8",
    pages = "399",
    year = "2022"
}

@article{DES:2021vln,
    author = "Secco, L. F. and others",
    collaboration = "DES",
    title = "{Dark Energy Survey Year 3 results: Cosmology from cosmic shear and robustness to modeling uncertainty}",
    eprint = "2105.13544",
    archivePrefix = "arXiv",
    primaryClass = "astro-ph.CO",
    reportNumber = "FERMILAB-PUB-21-253-AE, DES-2019-0480",
    doi = "10.1103/PhysRevD.105.023515",
    journal = "Phys. Rev. D",
    volume = "105",
    number = "2",
    pages = "023515",
    year = "2022"
}

@article{AtacamaCosmologyTelescope:2025blo,
    author = "Louis, Thibaut and others",
    collaboration = "Atacama Cosmology Telescope",
    title = "{The Atacama Cosmology Telescope: DR6 power spectra, likelihoods and {\ensuremath{\Lambda}}CDM parameters}",
    eprint = "2503.14452",
    archivePrefix = "arXiv",
    primaryClass = "astro-ph.CO",
    reportNumber = "FERMILAB-PUB-25-0071-PPD",
    doi = "10.1088/1475-7516/2025/11/062",
    journal = "JCAP",
    volume = "11",
    pages = "062",
    year = "2025"
}

@article{DESI:2025wyn,
    author = "Gu, Gan and others",
    collaboration = "DESI",
    title = "{Dynamical dark energy in light of the DESI DR2 baryonic acoustic oscillations measurements}",
    eprint = "2504.06118",
    archivePrefix = "arXiv",
    primaryClass = "astro-ph.CO",
    reportNumber = "FERMILAB-PUB-25-0235-PPD",
    doi = "10.1038/s41550-025-02669-6",
    journal = "Nature Astron.",
    volume = "9",
    number = "12",
    pages = "1879--1889",
    year = "2025",
    note = "[Erratum: Nature Astron. 9, 1898 (2025)]"
}

@article{KiDS:2020suj,
    author = "Asgari, Marika and others",
    collaboration = "KiDS",
    title = "{KiDS-1000 Cosmology: Cosmic shear constraints and comparison between two point statistics}",
    eprint = "2007.15633",
    archivePrefix = "arXiv",
    primaryClass = "astro-ph.CO",
    doi = "10.1051/0004-6361/202039070",
    journal = "Astron. Astrophys.",
    volume = "645",
    pages = "A104",
    year = "2021"
}

@article{Wright:2025xka,
    author = "Wright, Angus H. and others",
    title = "{KiDS-Legacy: Cosmological constraints from cosmic shear with the complete Kilo-Degree Survey}",
    eprint = "2503.19441",
    archivePrefix = "arXiv",
    primaryClass = "astro-ph.CO",
    doi = "10.1051/0004-6361/202554908",
    journal = "Astron. Astrophys.",
    volume = "703",
    pages = "A158",
    year = "2025"
}

@misc{SPT-3G:2025bzu,
    author = "Camphuis, E. and others",
    collaboration = "SPT-3G",
    title = "{SPT-3G D1: CMB temperature and polarization power spectra and cosmology from 2019 and 2020 observations of the SPT-3G Main field}",
    eprint = "2506.20707",
    archivePrefix = "arXiv",
    primaryClass = "astro-ph.CO",
    reportNumber = "FERMILAB-PUB-25-0144-PPD",
    month = "6",
    year = "2025"
}

@article{Riess:2021jrx,
    author = "Riess, Adam G. and others",
    title = "{A Comprehensive Measurement of the Local Value of the Hubble Constant with 1 km s$^{-1}$ Mpc$^{-1}$ Uncertainty from the Hubble Space Telescope and the SH0ES Team}",
    eprint = "2112.04510",
    archivePrefix = "arXiv",
    primaryClass = "astro-ph.CO",
    doi = "10.3847/2041-8213/ac5c5b",
    journal = "Astrophys. J. Lett.",
    volume = "934",
    number = "1",
    pages = "L7",
    year = "2022"
}

@article{eBOSS:2020yzd,
    author = "Alam, Shadab and others",
    collaboration = "eBOSS",
    title = "{Completed SDSS-IV extended Baryon Oscillation Spectroscopic Survey: Cosmological implications from two decades of spectroscopic surveys at the Apache Point Observatory}",
    eprint = "2007.08991",
    archivePrefix = "arXiv",
    primaryClass = "astro-ph.CO",
    doi = "10.1103/PhysRevD.103.083533",
    journal = "Phys. Rev. D",
    volume = "103",
    number = "8",
    pages = "083533",
    year = "2021"
}

@article{Zhang:2012mp,
    author = "Zhang, Cong and Zhang, Han and Yuan, Shuo and Zhang, Tong-Jie and Sun, Yan-Chun",
    title = "{Four new observational $H(z)$ data from luminous red galaxies in the Sloan Digital Sky Survey data release seven}",
    eprint = "1207.4541",
    archivePrefix = "arXiv",
    primaryClass = "astro-ph.CO",
    doi = "10.1088/1674-4527/14/10/002",
    journal = "Res. Astron. Astrophys.",
    volume = "14",
    number = "10",
    pages = "1221--1233",
    year = "2014"
}

@article{Jimenez:2003iv,
    author = "Jimenez, Raul and Verde, Licia and Treu, Tommaso and Stern, Daniel",
    title = "{Constraints on the equation of state of dark energy and the Hubble constant from stellar ages and the CMB}",
    eprint = "astro-ph/0302560",
    archivePrefix = "arXiv",
    doi = "10.1086/376595",
    journal = "Astrophys. J.",
    volume = "593",
    pages = "622--629",
    year = "2003"
}

@article{Simon:2004tf,
    author = "Simon, Joan and Verde, Licia and Jimenez, Raul",
    title = "{Constraints on the redshift dependence of the dark energy potential}",
    eprint = "astro-ph/0412269",
    archivePrefix = "arXiv",
    doi = "10.1103/PhysRevD.71.123001",
    journal = "Phys. Rev. D",
    volume = "71",
    pages = "123001",
    year = "2005"
}

@article{Moresco:2012by,
    author = "Moresco, Michele and Verde, Licia and Pozzetti, Lucia and Jimenez, Raul and Cimatti, Andrea",
    title = "{New constraints on cosmological parameters and neutrino properties using the expansion rate of the Universe to z{\textasciitilde}1.75}",
    eprint = "1201.6658",
    archivePrefix = "arXiv",
    primaryClass = "astro-ph.CO",
    doi = "10.1088/1475-7516/2012/07/053",
    journal = "JCAP",
    volume = "07",
    pages = "053",
    year = "2012"
}

@article{Moresco:2016mzx,
    author = "Moresco, Michele and Pozzetti, Lucia and Cimatti, Andrea and Jimenez, Raul and Maraston, Claudia and Verde, Licia and Thomas, Daniel and Citro, Annalisa and Tojeiro, Rita and Wilkinson, David",
    title = "{A 6{\%} measurement of the Hubble parameter at $z\sim0.45$: direct evidence of the epoch of cosmic re-acceleration}",
    eprint = "1601.01701",
    archivePrefix = "arXiv",
    primaryClass = "astro-ph.CO",
    doi = "10.1088/1475-7516/2016/05/014",
    journal = "JCAP",
    volume = "05",
    pages = "014",
    year = "2016"
}

@article{Ratsimbazafy:2017vga,
    author = {Ratsimbazafy, A. L. and Loubser, S. I. and Crawford, S. M. and Cress, C. M. and Bassett, B. A. and Nichol, R. C. and V{\"a}is{\"a}nen, P.},
    title = "{Age-dating Luminous Red Galaxies observed with the Southern African Large Telescope}",
    eprint = "1702.00418",
    archivePrefix = "arXiv",
    primaryClass = "astro-ph.CO",
    doi = "10.1093/mnras/stx301",
    journal = "Mon. Not. Roy. Astron. Soc.",
    volume = "467",
    number = "3",
    pages = "3239--3254",
    year = "2017"
}

@article{Moresco:2015cya,
    author = "Moresco, Michele",
    title = "{Raising the bar: new constraints on the Hubble parameter with cosmic chronometers at z {\ensuremath{\sim}} 2}",
    eprint = "1503.01116",
    archivePrefix = "arXiv",
    primaryClass = "astro-ph.CO",
    doi = "10.1093/mnrasl/slv037",
    journal = "Mon. Not. Roy. Astron. Soc.",
    volume = "450",
    number = "1",
    pages = "L16--L20",
    year = "2015"
}

@misc{H0DN:2025lyy,
    author = "Casertano, Stefano and others",
    collaboration = "H0DN",
    title = "{The Local Distance Network: a community consensus report on the measurement of the Hubble constant at 1{\%} precision}",
    eprint = "2510.23823",
    archivePrefix = "arXiv",
    primaryClass = "astro-ph.CO",
    month = "10",
    year = "2025"
}

@article{Akarsu:2021fol,
    author = {Akarsu, {\"O}zg{\"u}r and Kumar, Suresh and {\"O}z{\"u}lker, Emre and Vazquez, J. Alberto},
    title = "{Relaxing cosmological tensions with a sign switching cosmological constant}",
    eprint = "2108.09239",
    archivePrefix = "arXiv",
    primaryClass = "astro-ph.CO",
    doi = "10.1103/PhysRevD.104.123512",
    journal = "Phys. Rev. D",
    volume = "104",
    number = "12",
    pages = "123512",
    year = "2021"
}

@article{Akarsu:2022typ,
    author = {Akarsu, Ozgur and Kumar, Suresh and {\"O}z{\"u}lker, Emre and Vazquez, J. Alberto and Yadav, Anita},
    title = "{Relaxing cosmological tensions with a sign switching cosmological constant: Improved results with Planck, BAO, and Pantheon data}",
    eprint = "2211.05742",
    archivePrefix = "arXiv",
    primaryClass = "astro-ph.CO",
    doi = "10.1103/PhysRevD.108.023513",
    journal = "Phys. Rev. D",
    volume = "108",
    number = "2",
    pages = "023513",
    year = "2023"
}

@article{Akarsu:2022lhx,
    author = {Akarsu, Ozgur and Colgain, Eoin O. and {\"O}zulker, Emre and Thakur, Somyadip and Yin, Lu},
    title = "{Inevitable manifestation of wiggles in the expansion of the late Universe}",
    eprint = "2207.10609",
    archivePrefix = "arXiv",
    primaryClass = "astro-ph.CO",
    doi = "10.1103/PhysRevD.107.123526",
    journal = "Phys. Rev. D",
    volume = "107",
    number = "12",
    pages = "123526",
    year = "2023"
}

@article{DESI:2025zgx,
    author = "Abdul Karim, M. and others",
    collaboration = "DESI",
    title = "{DESI DR2 results. II. Measurements of baryon acoustic oscillations and cosmological constraints}",
    eprint = "2503.14738",
    archivePrefix = "arXiv",
    primaryClass = "astro-ph.CO",
    reportNumber = "FERMILAB-PUB-25-0169-PPD",
    doi = "10.1103/tr6y-kpc6",
    journal = "Phys. Rev. D",
    volume = "112",
    number = "8",
    pages = "083515",
    year = "2025"
}

@article{DESI:2024mwx,
    author = "Adame, A. G. and others",
    collaboration = "DESI",
    title = "{DESI 2024 VI: cosmological constraints from the measurements of baryon acoustic oscillations}",
    eprint = "2404.03002",
    archivePrefix = "arXiv",
    primaryClass = "astro-ph.CO",
    reportNumber = "FERMILAB-PUB-24-0154-PPD",
    doi = "10.1088/1475-7516/2025/02/021",
    journal = "JCAP",
    volume = "02",
    pages = "021",
    year = "2025"
}

@article{DESI:2024aqx,
    author = "Calderon, R. and others",
    collaboration = "DESI",
    title = "{DESI 2024: reconstructing dark energy using crossing statistics with DESI DR1 BAO data}",
    eprint = "2405.04216",
    archivePrefix = "arXiv",
    primaryClass = "astro-ph.CO",
    doi = "10.1088/1475-7516/2024/10/048",
    journal = "JCAP",
    volume = "10",
    pages = "048",
    year = "2024"
}

@misc{Akarsu:2023mfb,
    author = "Akarsu, Ozgur and Di Valentino, Eleonora and Kumar, Suresh and Nunes, Rafael C. and Vazquez, J. Alberto and Yadav, Anita",
    title = "{$\Lambda_{\rm s}$CDM model: A promising scenario for alleviation of cosmological tensions}",
    eprint = "2307.10899",
    archivePrefix = "arXiv",
    primaryClass = "astro-ph.CO",
    month = "7",
    year = "2023"
}

@article{Akarsu:2024eoo,
    author = {Akarsu, {\"O}zg{\"u}r and De Felice, Antonio and Di Valentino, Eleonora and Kumar, Suresh and Nunes, Rafael C. and {\"O}z{\"u}lker, Emre and Vazquez, J. Alberto and Yadav, Anita},
    title = "{Cosmological constraints on {\ensuremath{\Lambda}}sCDM scenario in a type II minimally modified gravity}",
    eprint = "2406.07526",
    archivePrefix = "arXiv",
    primaryClass = "astro-ph.CO",
    reportNumber = "YITP-24-57",
    doi = "10.1103/PhysRevD.110.103527",
    journal = "Phys. Rev. D",
    volume = "110",
    number = "10",
    pages = "103527",
    year = "2024"
}

@article{Akarsu:2019hmw,
    author = {Akarsu, {\"O}zg{\"u}r and Barrow, John D. and Escamilla, Luis A. and Vazquez, J. Alberto},
    title = "{Graduated dark energy: Observational hints of a spontaneous sign switch in the cosmological constant}",
    eprint = "1912.08751",
    archivePrefix = "arXiv",
    primaryClass = "astro-ph.CO",
    doi = "10.1103/PhysRevD.101.063528",
    journal = "Phys. Rev. D",
    volume = "101",
    number = "6",
    pages = "063528",
    year = "2020"
}

@misc{simplemc,
  author = "A. Slosar and J. A. Vazquez",
  year = "2019",
  howpublished = "\url{https://github.com/ja-vazquez/SimpleMC}"
}

@article{speagle2020dynesty,
 author = "Speagle, Joshua S.",
    title = "{dynesty: a dynamic nested sampling package for estimating Bayesian posteriors and evidences}",
    eprint = "1904.02180",
    archivePrefix = "arXiv",
    primaryClass = "astro-ph.IM",
    doi = "10.1093/mnras/staa278",
    journal = "Mon. Not. Roy. Astron. Soc.",
    volume = "493",
    number = "3",
    pages = "3132--3158",
    year = "2020"
}

@article{skilling,
    author = {Skilling, John},
    title = {Nested Sampling},
    journal = {AIP Conference Proceedings},
    volume = {735},
    number = {1},
    pages = {395-405},
    year = {2004},
    month = {11},
    doi = {10.1063/1.1835238},
    url = {https://doi.org/10.1063/1.1835238},
}

@article{Scolnic:2021amr,
    author = "Scolnic, Dan and others",
    title = "{The Pantheon+ Analysis: The Full Data Set and Light-curve Release}",
    eprint = "2112.03863",
    archivePrefix = "arXiv",
    primaryClass = "astro-ph.CO",
    doi = "10.3847/1538-4357/ac8b7a",
    journal = "Astrophys. J.",
    volume = "938",
    number = "2",
    pages = "113",
    year = "2022"
}

@article{Sabogal:2024qxs,
    author = {Sabogal, Miguel A. and Akarsu, {\"O}zg{\"u}r and Bonilla, Alexander and Di Valentino, Eleonora and Nunes, Rafael C.},
    title = "{Exploring new physics in the late Universe{\textquoteright}s expansion through non-parametric inference}",
    eprint = "2407.04223",
    archivePrefix = "arXiv",
    primaryClass = "astro-ph.CO",
    doi = "10.1140/epjc/s10052-024-13081-1",
    journal = "Eur. Phys. J. C",
    volume = "84",
    number = "7",
    pages = "703",
    year = "2024"
}

@article{Escamilla:2023shf,
    author = "Escamilla, Luis A. and Akarsu, Ozgur and Di Valentino, Eleonora and Vazquez, J. Alberto",
    title = "{Model-independent reconstruction of the interacting dark energy kernel: Binned and Gaussian process}",
    eprint = "2305.16290",
    archivePrefix = "arXiv",
    primaryClass = "astro-ph.CO",
    doi = "10.1088/1475-7516/2023/11/051",
    journal = "JCAP",
    volume = "11",
    pages = "051",
    year = "2023"
}

@article{Giare:2024gpk,
    author = "Giar{\`e}, William and Najafi, Mahdi and Pan, Supriya and Di Valentino, Eleonora and Firouzjaee, Javad T.",
    title = "{Robust preference for Dynamical Dark Energy in DESI BAO and SN measurements}",
    eprint = "2407.16689",
    archivePrefix = "arXiv",
    primaryClass = "astro-ph.CO",
    doi = "10.1088/1475-7516/2024/10/035",
    journal = "JCAP",
    volume = "10",
    pages = "035",
    year = "2024"
}

@article{Giare:2024oil,
    author = "Giar{\`e}, William",
    title = "{Dynamical dark energy beyond Planck? Constraints from multiple CMB probes, DESI BAO, and type-Ia supernovae}",
    eprint = "2409.17074",
    archivePrefix = "arXiv",
    primaryClass = "astro-ph.CO",
    doi = "10.1103/ss37-cxhn",
    journal = "Phys. Rev. D",
    volume = "112",
    number = "2",
    pages = "023508",
    year = "2025"
}

@article{Najafi:2024qzm,
    author = "Najafi, Mahdi and Pan, Supriya and Di Valentino, Eleonora and Firouzjaee, Javad T.",
    title = "{Dynamical dark energy confronted with multiple CMB missions}",
    eprint = "2407.14939",
    archivePrefix = "arXiv",
    primaryClass = "astro-ph.CO",
    doi = "10.1016/j.dark.2024.101539",
    journal = "Phys. Dark Univ.",
    volume = "45",
    pages = "101539",
    year = "2024"
}

@article{Giare:2025pzu,
    author = "Giar{\`e}, William and Mahassen, Tariq and Di Valentino, Eleonora and Pan, Supriya",
    title = "{An overview of what current data can (and cannot yet) say about evolving dark energy}",
    eprint = "2502.10264",
    archivePrefix = "arXiv",
    primaryClass = "astro-ph.CO",
    doi = "10.1016/j.dark.2025.101906",
    journal = "Phys. Dark Univ.",
    volume = "48",
    pages = "101906",
    year = "2025"
}

@article{Fazzari:2025lzd,
    author = "Fazzari, Elisa and Giar{\`e}, William and Di Valentino, Eleonora",
    title = "{Cosmographic Footprints of Dynamical Dark Energy}",
    eprint = "2509.16196",
    archivePrefix = "arXiv",
    primaryClass = "astro-ph.CO",
    doi = "10.3847/2041-8213/ae2917",
    journal = "Astrophys. J. Lett.",
    volume = "996",
    number = "1",
    pages = "L5",
    year = "2026"
}

@article{Perivolaropoulos:2021jda,
    author = "Perivolaropoulos, Leandros and Skara, Foteini",
    title = "{Challenges for {\ensuremath{\Lambda}}CDM: An update}",
    eprint = "2105.05208",
    archivePrefix = "arXiv",
    primaryClass = "astro-ph.CO",
    doi = "10.1016/j.newar.2022.101659",
    journal = "New Astron. Rev.",
    volume = "95",
    pages = "101659",
    year = "2022"
}

@article{Verde:2019ivm,
    author = "Verde, L. and Treu, T. and Riess, A. G.",
    title = "{Tensions between the Early and the Late Universe}",
    eprint = "1907.10625",
    archivePrefix = "arXiv",
    primaryClass = "astro-ph.CO",
    doi = "10.1038/s41550-019-0902-0",
    journal = "Nature Astron.",
    volume = "3",
    pages = "891",
    year = "2019"
}

@article{Breuval:2024lsv,
    author = "Breuval, Louise and Riess, Adam G. and Casertano, Stefano and Yuan, Wenlong and Macri, Lucas M. and Romaniello, Martino and Murakami, Yukei S. and Scolnic, Daniel and Anand, Gagandeep S. and Soszy{\'n}ski, Igor",
    title = "{Small Magellanic Cloud Cepheids Observed with the Hubble Space Telescope Provide a New Anchor for the SH0ES Distance Ladder}",
    eprint = "2404.08038",
    archivePrefix = "arXiv",
    primaryClass = "astro-ph.CO",
    doi = "10.3847/1538-4357/ad630e",
    journal = "Astrophys. J.",
    volume = "973",
    number = "1",
    pages = "30",
    year = "2024"
}

@article{DiValentino:2021izs,
    author = "Di Valentino, Eleonora and Mena, Olga and Pan, Supriya and Visinelli, Luca and Yang, Weiqiang and Melchiorri, Alessandro and Mota, David F. and Riess, Adam G. and Silk, Joseph",
    title = "{In the realm of the Hubble tension{\textemdash}a review of solutions}",
    eprint = "2103.01183",
    archivePrefix = "arXiv",
    primaryClass = "astro-ph.CO",
    reportNumber = "IPPP/20/108",
    doi = "10.1088/1361-6382/ac086d",
    journal = "Class. Quant. Grav.",
    volume = "38",
    number = "15",
    pages = "153001",
    year = "2021"
}

@article{Abdalla:2022yfr,
    author = "Abdalla, Elcio and others",
    title = "{Cosmology intertwined: A review of the particle physics, astrophysics, and cosmology associated with the cosmological tensions and anomalies}",
    eprint = "2203.06142",
    archivePrefix = "arXiv",
    primaryClass = "astro-ph.CO",
    reportNumber = "FERMILAB-CONF-22-192-SCD",
    doi = "10.1016/j.jheap.2022.04.002",
    journal = "JHEAp",
    volume = "34",
    pages = "49--211",
    year = "2022"
}

@article{Planck:2018vyg,
    author = "Aghanim, N. and others",
    collaboration = "Planck",
    title = "{Planck 2018 results. VI. Cosmological parameters}",
    eprint = "1807.06209",
    archivePrefix = "arXiv",
    primaryClass = "astro-ph.CO",
    doi = "10.1051/0004-6361/201833910",
    journal = "Astron. Astrophys.",
    volume = "641",
    pages = "A6",
    year = "2020",
    note = "[Erratum: Astron.Astrophys. 652, C4 (2021)]"
}

@article{Copeland:2006wr,
    author = "Copeland, Edmund J. and Sami, M. and Tsujikawa, Shinji",
    title = "{Dynamics of dark energy}",
    eprint = "hep-th/0603057",
    archivePrefix = "arXiv",
    doi = "10.1142/S021827180600942X",
    journal = "Int. J. Mod. Phys. D",
    volume = "15",
    pages = "1753--1936",
    year = "2006"
}

@article{Weinberg:1988cp,
    author = "Weinberg, Steven",
    editor = "Hsu, Jong-Ping and Fine, D.",
    title = "{The Cosmological Constant Problem}",
    reportNumber = "UTTG-12-88",
    doi = "10.1103/RevModPhys.61.1",
    journal = "Rev. Mod. Phys.",
    volume = "61",
    pages = "1--23",
    year = "1989"
}

@article{CosmoVerseNetwork:2025alb,
    author = "Di Valentino, Eleonora and others",
    collaboration = "CosmoVerse Network",
    title = "{The CosmoVerse White Paper: Addressing observational tensions in cosmology with systematics and fundamental physics}",
    eprint = "2504.01669",
    archivePrefix = "arXiv",
    primaryClass = "astro-ph.CO",
    doi = "10.1016/j.dark.2025.101965",
    journal = "Phys. Dark Univ.",
    volume = "49",
    pages = "101965",
    year = "2025"
}

@article{DiValentino:2020zio,
    author = "Di Valentino, Eleonora and others",
    title = "{Snowmass2021 - Letter of interest cosmology intertwined II: The hubble constant tension}",
    eprint = "2008.11284",
    archivePrefix = "arXiv",
    primaryClass = "astro-ph.CO",
    reportNumber = "FERMILAB-PUB-21-590-PPD",
    doi = "10.1016/j.astropartphys.2021.102605",
    journal = "Astropart. Phys.",
    volume = "131",
    pages = "102605",
    year = "2021"
}

@article{DiValentino:2020vvd,
    author = "Di Valentino, Eleonora and others",
    title = "{Cosmology Intertwined III: $f \sigma_8$ and $S_8$}",
    eprint = "2008.11285",
    archivePrefix = "arXiv",
    primaryClass = "astro-ph.CO",
    reportNumber = "FERMILAB-PUB-20-495-AE",
    doi = "10.1016/j.astropartphys.2021.102604",
    journal = "Astropart. Phys.",
    volume = "131",
    pages = "102604",
    year = "2021"
}

@article{Vagnozzi:2023nrq,
    author = "Vagnozzi, Sunny",
    title = "{Seven Hints That Early-Time New Physics Alone Is Not Sufficient to Solve the Hubble Tension}",
    eprint = "2308.16628",
    archivePrefix = "arXiv",
    primaryClass = "astro-ph.CO",
    doi = "10.3390/universe9090393",
    journal = "Universe",
    volume = "9",
    number = "9",
    pages = "393",
    year = "2023"
}

@inproceedings{Akarsu:2025dmj,
    author = {Akarsu, {\"O}zg{\"u}r and Eingorn, Maxim and Perivolaropoulos, Leandros and Y{\"u}kselci, A. Emrah and Zhuk, Alexander},
    title = "{Dynamical dark energy with AdS-dS transitions vs. Baryon Acoustic Oscillations at $z =$ 2.3-2.4}",
    eprint = "2504.07299",
    archivePrefix = "arXiv",
    primaryClass = "astro-ph.CO",
    month = "4",
    year = "2025"
}

@article{Tan:2025xas,
    author = "Tan, Hai Siong",
    title = "{Inferring Cosmological Parameters with Evidential Physics-Informed Neural Networks}",
    eprint = "2509.24327",
    archivePrefix = "arXiv",
    primaryClass = "astro-ph.CO",
    doi = "10.3390/universe11120403",
    journal = "Universe",
    volume = "11",
    number = "12",
    pages = "403",
    year = "2025"
}

@article{Bouhmadi-Lopez:2025spo,
    author = "Bouhmadi-L{\'o}pez, Mariam and Ibarra-Uriondo, Be{\~n}at",
    title = "{Cosmological perturbations for smooth sign-switching dark energy models}",
    eprint = "2506.18992",
    archivePrefix = "arXiv",
    primaryClass = "gr-qc",
    doi = "10.1016/j.dark.2025.102129",
    journal = "Phys. Dark Univ.",
    volume = "50",
    pages = "102129",
    year = "2025"
}

@article{Gonzalez-Fuentes:2025lei,
    author = "Gonz{\'a}lez-Fuentes, Alex and G{\'o}mez-Valent, Adri{\`a}",
    title = "{Reconstruction of dark energy and late-time cosmic expansion using the Weighted Function Regression method}",
    eprint = "2506.11758",
    archivePrefix = "arXiv",
    primaryClass = "astro-ph.CO",
    doi = "10.1088/1475-7516/2025/12/049",
    journal = "JCAP",
    volume = "12",
    pages = "049",
    year = "2025"
}

@article{Bouhmadi-Lopez:2025ggl,
    author = "Bouhmadi-L{\'o}pez, Mariam and Ibarra-Uriondo, Be{\~n}at",
    title = "{Cosmographic analysis of sign-switching dark energy}",
    eprint = "2506.12139",
    archivePrefix = "arXiv",
    primaryClass = "gr-qc",
    doi = "10.1103/v1cl-pr54",
    journal = "Phys. Rev. D",
    volume = "112",
    number = "6",
    pages = "063559",
    year = "2025"
}

@misc{Akarsu:2025gwi,
    author = {Akarsu, {\"O}zg{\"u}r and Perivolaropoulos, Leandros and Tsikoundoura, Anna and Y{\"u}kselci, A. Emrah and Zhuk, Alexander},
    title = "{Dynamical dark energy with AdS-to-dS and dS-to-dS transitions: Implications for the $H_0$ tension}",
    eprint = "2502.14667",
    archivePrefix = "arXiv",
    primaryClass = "astro-ph.CO",
    month = "2",
    year = "2025"
}

@article{Souza:2024qwd,
    author = {Souza, Mateus S. and Barcelos, Ana M. and Nunes, Rafael C. and Akarsu, {\"O}zg{\"u}r and Kumar, Suresh},
    title = "{Mapping the {\ensuremath{\Lambda}}$_{s}$CDM Scenario to f(T) Modified Gravity: Effects on Structure Growth Rate}",
    eprint = "2501.18031",
    archivePrefix = "arXiv",
    primaryClass = "astro-ph.CO",
    doi = "10.3390/universe11010002",
    journal = "Universe",
    volume = "11",
    number = "1",
    pages = "2",
    year = "2025"
}

@article{Gomez-Valent:2024ejh,
    author = "G{\'o}mez-Valent, Adria and Sol{\`a} Peracaula, Joan",
    title = "{Composite dark energy and the cosmological tensions}",
    eprint = "2412.15124",
    archivePrefix = "arXiv",
    primaryClass = "astro-ph.CO",
    doi = "10.1016/j.physletb.2025.139391",
    journal = "Phys. Lett. B",
    volume = "864",
    pages = "139391",
    year = "2025"
}

@article{Akarsu:2024nas,
    author = "Akarsu, Ozgur and Bulduk, Bilal and De Felice, Antonio and Kat{\i}rc{\i}, Nihan and Uzun, N. Merve",
    title = "{Unexplored regions in teleparallel f(T) gravity: Sign-changing dark energy density}",
    eprint = "2410.23068",
    archivePrefix = "arXiv",
    primaryClass = "gr-qc",
    reportNumber = "YITP-24-119",
    doi = "10.1103/1xd4-k91h",
    journal = "Phys. Rev. D",
    volume = "112",
    number = "8",
    pages = "083532",
    year = "2025"
}

@article{Anchordoqui:2024dqc,
    author = "Anchordoqui, Luis A. and Antoniadis, Ignatios and Bielli, Daniele and Chatrabhuti, Auttakit and Isono, Hiroshi",
    title = "{Thin-wall vacuum decay in the presence of a compact dimension meets the H$_{0}$ and S$_{8}$ tensions}",
    eprint = "2410.18649",
    archivePrefix = "arXiv",
    primaryClass = "hep-th",
    doi = "10.1007/JHEP07(2025)021",
    journal = "JHEP",
    volume = "07",
    pages = "021",
    year = "2025"
}

@article{Escamilla:2024ahl,
    author = {Escamilla, Luis A. and {\"O}z{\"u}lker, Emre and Akarsu, {\"O}zg{\"u}r and Di Valentino, Eleonora and V{\'a}zquez, J. A.},
    title = "{Improved late-time fits with wavelet extensions of {\ensuremath{\Lambda}}CDM}",
    eprint = "2408.12516",
    archivePrefix = "arXiv",
    primaryClass = "astro-ph.CO",
    doi = "10.1093/mnras/staf1732",
    journal = "Mon. Not. Roy. Astron. Soc.",
    volume = "544",
    number = "1",
    pages = "836--854",
    year = "2025"
}

@article{Anchordoqui:2023woo,
    author = "Anchordoqui, Luis A. and Antoniadis, Ignatios and Lust, Dieter",
    title = "{Anti-de Sitter {\textrightarrow} de Sitter transition driven by Casimir forces and mitigating tensions in cosmological parameters}",
    eprint = "2312.12352",
    archivePrefix = "arXiv",
    primaryClass = "hep-th",
    reportNumber = "MPP-2023-288; LMU-ASC 39/23",
    doi = "10.1016/j.physletb.2024.138775",
    journal = "Phys. Lett. B",
    volume = "855",
    pages = "138775",
    year = "2024"
}

@article{Alexandre:2023nmh,
    author = "Alexandre, Bruno and Gielen, Steffen and Magueijo, Jo{\~a}o",
    title = "{Overall signature of the metric and the cosmological constant}",
    eprint = "2306.11502",
    archivePrefix = "arXiv",
    primaryClass = "hep-th",
    doi = "10.1088/1475-7516/2024/02/036",
    journal = "JCAP",
    volume = "02",
    pages = "036",
    year = "2024"
}

@article{Bernardo:2021cxi,
    author = "Bernardo, Reginald Christian and Grand{\'o}n, Daniela and Said Levi, Jackson and C{\'a}rdenas, V{\'\i}ctor H.",
    title = "{Parametric and nonparametric methods hint dark energy evolution}",
    eprint = "2111.08289",
    archivePrefix = "arXiv",
    primaryClass = "astro-ph.CO",
    doi = "10.1016/j.dark.2022.101017",
    journal = "Phys. Dark Univ.",
    volume = "36",
    pages = "101017",
    year = "2022"
}

@article{Bonilla:2020wbn,
    author = "Bonilla, Alexander and Kumar, Suresh and Nunes, Rafael C.",
    title = "{Measurements of $H_0$ and reconstruction of the dark energy properties from a model-independent joint analysis}",
    eprint = "2011.07140",
    archivePrefix = "arXiv",
    primaryClass = "astro-ph.CO",
    doi = "10.1140/epjc/s10052-021-08925-z",
    journal = "Eur. Phys. J. C",
    volume = "81",
    number = "2",
    pages = "127",
    year = "2021"
}

@article{DeFelice:2020cpt,
    author = "De Felice, Antonio and Mukohyama, Shinji and Pookkillath, Masroor C.",
    title = "{Addressing $H_0$ tension by means of VCDM}",
    eprint = "2009.08718",
    archivePrefix = "arXiv",
    primaryClass = "astro-ph.CO",
    reportNumber = "YITP-20-117, IPMU 20-0098",
    doi = "10.1016/j.physletb.2021.136201",
    journal = "Phys. Lett. B",
    volume = "816",
    pages = "136201",
    year = "2021",
    note = "[Erratum: Phys.Lett.B 818, 136364 (2021)]"
}

@article{Wang:2018fng,
    author = "Wang, Yuting and Pogosian, Levon and Zhao, Gong-Bo and Zucca, Alex",
    title = "{Evolution of dark energy reconstructed from the latest observations}",
    eprint = "1807.03772",
    archivePrefix = "arXiv",
    primaryClass = "astro-ph.CO",
    doi = "10.3847/2041-8213/aaf238",
    journal = "Astrophys. J. Lett.",
    volume = "869",
    pages = "L8",
    year = "2018"
}

@article{Escamilla:2021uoj,
    author = "Escamilla, Luis A. and Vazquez, J. Alberto",
    title = "{Model selection applied to reconstructions of the Dark Energy}",
    eprint = "2111.10457",
    archivePrefix = "arXiv",
    primaryClass = "astro-ph.CO",
    doi = "10.1140/epjc/s10052-023-11404-2",
    journal = "Eur. Phys. J. C",
    volume = "83",
    number = "3",
    pages = "251",
    year = "2023"
}

@article{Li:2019yem,
    author = "Li, Xiaolei and Shafieloo, Arman",
    title = "{A Simple Phenomenological Emergent Dark Energy Model can Resolve the Hubble Tension}",
    eprint = "1906.08275",
    archivePrefix = "arXiv",
    primaryClass = "astro-ph.CO",
    doi = "10.3847/2041-8213/ab3e09",
    journal = "Astrophys. J. Lett.",
    volume = "883",
    number = "1",
    pages = "L3",
    year = "2019"
}

@article{Anchordoqui:2024gfa,
    author = "Anchordoqui, Luis A. and Antoniadis, Ignatios and Lust, Dieter and Noble, Neena T. and Soriano, Jorge F.",
    title = "{From infinite to infinitesimal: Using the universe as a dataset to probe Casimir corrections to the vacuum energy from fields inhabiting the dark dimension}",
    eprint = "2404.17334",
    archivePrefix = "arXiv",
    primaryClass = "astro-ph.CO",
    reportNumber = "MPP-2024-91; LMU-ASC 05/24",
    doi = "10.1016/j.dark.2024.101715",
    journal = "Phys. Dark Univ.",
    volume = "46",
    pages = "101715",
    year = "2024"
}

@article{Gomez-Valent:2024tdb,
    author = "Gomez-Valent, Adria and Sol{\`a} Peracaula, Joan",
    title = "{Phantom Matter: A Challenging Solution to the Cosmological Tensions}",
    eprint = "2404.18845",
    archivePrefix = "arXiv",
    primaryClass = "astro-ph.CO",
    doi = "10.3847/1538-4357/ad7a62",
    journal = "Astrophys. J.",
    volume = "975",
    number = "1",
    pages = "64",
    year = "2024"
}

@article{Soriano:2025gxd,
    author = "Soriano, Jorge F. and Wohlberg, Shimon and Anchordoqui, Luis A.",
    title = "{New insights on a sign-switching {\ensuremath{\Lambda}}}",
    eprint = "2502.19239",
    archivePrefix = "arXiv",
    primaryClass = "astro-ph.CO",
    doi = "10.1016/j.dark.2025.101911",
    journal = "Phys. Dark Univ.",
    volume = "48",
    pages = "101911",
    year = "2025"
}

@article{Chevallier:2000qy,
    author = "Chevallier, Michel and Polarski, David",
    title = "{Accelerating universes with scaling dark matter}",
    eprint = "gr-qc/0009008",
    archivePrefix = "arXiv",
    doi = "10.1142/S0218271801000822",
    journal = "Int. J. Mod. Phys. D",
    volume = "10",
    pages = "213--224",
    year = "2001"
}

@article{Linder:2002et,
    author = "Linder, Eric V.",
    title = "{Exploring the expansion history of the universe}",
    eprint = "astro-ph/0208512",
    archivePrefix = "arXiv",
    doi = "10.1103/PhysRevLett.90.091301",
    journal = "Phys. Rev. Lett.",
    volume = "90",
    pages = "091301",
    year = "2003"
}

@article{Barboza:2008rh,
    author = "Barboza, Jr., E. M. and Alcaniz, J. S.",
    title = "{A parametric model for dark energy}",
    eprint = "0805.1713",
    archivePrefix = "arXiv",
    primaryClass = "astro-ph",
    doi = "10.1016/j.physletb.2008.08.012",
    journal = "Phys. Lett. B",
    volume = "666",
    pages = "415--419",
    year = "2008"
}

@article{Jassal:2004ej,
    author = "Jassal, H. K. and Bagla, J. S. and Padmanabhan, T.",
    title = "{WMAP constraints on low redshift evolution of dark energy}",
    eprint = "astro-ph/0404378",
    archivePrefix = "arXiv",
    doi = "10.1111/j.1745-3933.2005.08577.x",
    journal = "Mon. Not. Roy. Astron. Soc.",
    volume = "356",
    pages = "L11--L16",
    year = "2005"
}

@article{Jassal:2005qc,
    author = "Jassal, Harvinder Kaur and Bagla, J. S. and Padmanabhan, T.",
    title = "{Observational constraints on low redshift evolution of dark energy: How consistent are different observations?}",
    eprint = "astro-ph/0506748",
    archivePrefix = "arXiv",
    doi = "10.1103/PhysRevD.72.103503",
    journal = "Phys. Rev. D",
    volume = "72",
    pages = "103503",
    year = "2005"
}

@article{Akarsu:2024qsi,
    author = {Akarsu, {\"O}zg{\"u}r and De Felice, Antonio and Di Valentino, Eleonora and Kumar, Suresh and Nunes, Rafael C. and {\"O}z{\"u}lker, Emre and Vazquez, J. Alberto and Yadav, Anita},
    title = "{{\ensuremath{\Lambda}}sCDM cosmology from a type-II minimally modified gravity}",
    eprint = "2402.07716",
    archivePrefix = "arXiv",
    primaryClass = "astro-ph.CO",
    reportNumber = "YITP-24-18",
    doi = "10.1093/mnras/staf2276",
    journal = "Mon. Not. Roy. Astron. Soc.",
    volume = "546",
    number = "1",
    pages = "staf2276",
    year = "2026"
}

@article{Pan:2019brc,
    author = "Pan, Supriya and Yang, Weiqiang and Paliathanasis, Andronikos",
    title = "{Imprints of an extended Chevallier{\textendash}Polarski{\textendash}Linder parametrization on the large scale of our universe}",
    eprint = "1902.07108",
    archivePrefix = "arXiv",
    primaryClass = "astro-ph.CO",
    doi = "10.1140/epjc/s10052-020-7832-y",
    journal = "Eur. Phys. J. C",
    volume = "80",
    number = "3",
    pages = "274",
    year = "2020"
}

@article{Tripathi:2016slv,
    author = "Tripathi, Ashutosh and Sangwan, Archana and Jassal, H. K.",
    title = "{Dark energy equation of state parameter and its evolution at low redshift}",
    eprint = "1611.01899",
    archivePrefix = "arXiv",
    primaryClass = "astro-ph.CO",
    doi = "10.1088/1475-7516/2017/06/012",
    journal = "JCAP",
    volume = "06",
    pages = "012",
    year = "2017"
}

@article{Gokcen:2026pkq,
    author = {G{\"o}k{\c{c}}en, Mine and Akarsu, {\"O}zg{\"u}r and Di Valentino, Eleonora},
    title = "{Revisiting CPL with sign-switching density: To cross or not to cross the NECB}",
    eprint = "2602.21169",
    archivePrefix = "arXiv",
    primaryClass = "astro-ph.CO",
    doi = "10.1016/j.dark.2026.102273",
    journal = "Phys. Dark Univ.",
    volume = "52",
    pages = "102273",
    year = "2026"
}

\end{document}